\documentclass[prc,amsmath,amsfonts,showpacs,eqsecnum,floatfix,multicol]{revtex4}

 \usepackage{axodraw}   
 \usepackage{graphicx}  
 \usepackage{pstcol}  
 \usepackage{bm}    

\begin{document}
\hyphenation{Rijken}
\hyphenation{Nijmegen}
 
\title{
    Extended-soft-core Baryon-Baryon Model ESC08\\   
    I. Nucleon-Nucleon Scattering }
\author{M.M.\ Nagels and Th.A.\ Rijken}
\affiliation{ Institute of Mathematics, Astrophysics, and Particle Physics \\
 University of Nijmegen, Nijmegen, The Netherlands}               
\author{Y.\ Yamamoto}  
\affiliation{Nishina Center for Accelerator-Based Science, Insitute for 
 Physical and Chemical Research (RIKEN). Wako, Saitama, 351-0198, Japan}

\date{version of: \today}
 
\begin{abstract}                                       
The Nijmegen extended-soft-core ESC08c model for the baryon-baryon (BB) 
interactions of the SU(3) flavor-octet of baryons (N, $\Lambda$, $\Sigma$, and $\Xi$)
is presented. In this first of a series of papers, the NN results are reported in
detail.
In the spirit of the Yukawa-approach to the nuclear force problem, 
the interactions are studied from the meson-exchange picture viewpoint,       
using generalized soft-core Yukawa-functions. 
These interactions are supplemented with (i) multiple-gluon-exchange, 
and (ii) structural effects due to the quark-core of the baryons.
The extended-soft-core (ESC) meson-exchange interactions 
consist of local- and non-local-potentials due to (i) One-boson-exchanges 
(OBE), which are the members of nonets of pseudoscalar, vector, scalar, and
axial-vector mesons, (ii) diffractive (i.e. multiple-gluon) exchanges, 
(iii) two pseudo-scalar exchange (PS-PS),
and (iv) meson-pair-exchange (MPE). The OBE- and MPE-vertices are regulated 
by gaussian form factors producing potentials with a soft behavior near the origin.
The assignment of the cut-off masses for the BBM-vertices is dependent on the 
$SU(3)$-classification of the exchanged mesons for OBE, and a similar scheme 
for MPE.        

The ESC-models ESC08, as well as its predecessor ESC04, describe the 
nucleon-nucleon (NN), hyperon-nucleon (YN), and hyperon-hyperon (YY)  
interactions in a unified way using broken $SU(3)$-symmetry. 
Important non-standard ingredients in the OBE-sector in the ESC-models 
are (i) the axial-vector
meson potentials, and (ii) a zero in the scalar- and axial-vector meson form factors. 
These innovations
make it possible to keep the meson coupling parameters of the model 
qualitatively in accordance with  
the predictions of the $^3P_0$-dominated quark-antiquark pair creation 
(QPC) model. 
SU(3)-symmetry serves to connect the $NN$ with the $YN$ and the $YY$
channels.
In the fit to NN and YN many parameters are essentially fixed by the NN-data.
A few, but severely constrained e.g. $F/(F+D)$-ratio's, parameters are left
for determination of the $YN$-interactions and the $YY$ experimental
indications.
In particular,
the meson-baryon coupling constants are calculated via $SU(3)$
using the coupling constants of the $NN \oplus YN$-analysis as input.
In ESC08 the couplings are kept SU(3)-symmetric.

In establishing the parameters of the model a simultaneous fit to NN- and YN-channels 
has been performed. Here the information about $\Lambda\Lambda$, $\Xi N$, and
hypernuclei played an important role in the form of using constraints.
In particular, the experimental indications 
for the $\Lambda\Lambda$-attraction, the $\Xi$-nuclear and the $\Sigma$-nuclear
well-depth were directive.            
About 25 physical coupling parameters and 8 cut-off and diffractive masses, were
searched. 
The obtained OBE-couplings and the $F/(F+D)$-ratio's can be well understood in the context 
of the QPC-model. 

The simultaneous fit of the ESC-models to the NN- and YN- scattering data 
with a single set of parameters has achieved 
excellent results for the NN- and YN-data, and for the YY-data in accordance 
with the experimental indications for $\Lambda\Lambda$ and $\Xi N$.
In the case of ESC08c, the version discussed here, the achievements are: 
(i) For the selected 4313 pp and np scattering data with energies 
$0 \leq T_{lab} \leq 350$ MeV, the model reaches a fit 
having $\chi^2/N_{data}=1.08$.       
(ii) The deuteron binding energy and all the NN scattering lengths are fitted
 very nicely.
(iii) The YN-data are described very well with $\chi^2/N_{data}=1.08$, 
giving at the same time good descriptions of the $\Lambda$, $\Sigma$, and $\Xi$ 
nuclear well-depths. 
(iv) The model predicts a bound $\Xi N(^3S_1,I=1)$ state with
binding energy 1.56 MeV.
 \end{abstract}

\maketitle

 --------------------------------------------------------------------\\
\twocolumngrid
\section{Introduction}                                     
\label{sec:1}
 
In a new series of papers we present the results obtained with 
the recent ESC08c-version of the 
Extended-Soft-Core (ESC) model \cite{Rij93} for nucleon-nucleon (NN), 
hyperon-nucleon (YN), and hyperon-hyperon (YY) interactions with $S=0,-1,-2$. 
Moreover, we present predictions for the YY-channels with $S=-3,-4$.

The combined study of all baryon-baryon (BB) interactions, 
exploiting all experimental information hitherto available, both on BB-scattering
and (hyper-)nuclear systems, might throw light on the basic mechanisms of
these interactions. The program, which in its original form was formulated in
Refs.~\cite{NRS77,MRS89}, pursuits the aims:
\begin{itemize}
\item  To study the assumption of broken $SU(3)$-symmetry. For example we
       investigate the properties of the scalar mesons
       ($\varepsilon(760)$, $f_{0}(975)$, $a_{0}(980)$, $\kappa(800)$).
\item  To determine the $F/(F+D)$-ratio's.
\item  To study the connection between QCD, the quark-model, 
       and low energy physics.
\item  To extract, in spite of the scarce experimental $YN$- and $YY$-data,
 information about scattering lengths, effective
 ranges, the existence of resonances and bound states, etc.
\item  To provide realistic baryon-baryon potentials, which can be 
       applied in few-body calculations, nuclear- and hyperonic matter
       studies, neutron-stars;
\item  To extend the theoretical description to the baryon-baryon channels with
       strangeness S=-2. This in particular for the $\Lambda\Lambda$
       and $\Xi N$ channels, where some data already exist, and for which 
       experiments will be realized in the near future.
\item  Finally, to extend the theoretical description to all baryon-baryon channels with
       strangeness S=-3,-4. These will be parameter free predictions, 
       and have, like the other BB-channels, relevance for the study of 
       hyperonic matter and compact stars.
\end{itemize}
With this series of papers this program nears essentially its completion.

As has been amply demonstrated, see Ref.'s \cite{Rij04a,Rij04b,HYPX,PTP185.a}, 
the ESC-model interactions give excellent simultaneous descriptions of the NN and YN data. 
Also it turned out that the
ESC-approach gives great improvements for the NN description as compared to the 
One-Boson-Exchange (OBE) models, e.g. \cite{MRS89,RSY99}, and other existing models
in the literature. The ESC08c-model presents the culmination in this respect: 
the NN-model has a quality on equal par with the energy-dependent 
partial-wave analysis (PWA) \cite{Sto93,Klo93}.

The ESC04-model papers \cite{Rij04a,Rij04b,Rij04c} 
contain the first rather extensive exposition of the ESC-approach. 
As compared to the earlier versions of the ESC-model, we introduced in ESC04-models 
\cite{Rij04a,Rij04b,Rij04c} several innovations: 
Firstly, we introduced a zero in the form factor of 
the mesons with P-wave quark-antiquark contents, which applies to the scalar and
axial-vector mesons.
Secondly, we exploited the exchange of the axial-vector mesons with $J^{PC}=1^{++}$ and
$J^{PC}=1^{+-}$. Thirdly, we employed some $\Lambda\Lambda, \Xi N$ information.\\

In the ESC08-models on top of these improvements, we introduce 
in the ESC-approach for the first time: (i) Odderon-exchange $J^{PC}=1^{-+}$. 
Odderon-exchange represents the exchange of an
odd-number of gluons at short-distance. This to complement pomeron-exchange which
stands for the exchange of an even-number of gluons. 
(ii) Quark-core effects. The quark-core effects represent
structural effects caused by the occurrence of Pauli-blocked configurations 
in two-baryon systems.
These structural effects depend on the BB-channel and cannot be described by
t-channel exchanges.\\
 Furthermore, (iii) the
axial-vector ($J^{PC}=1^{++}$) mesons are treated with the most general vertices,
and the  
$(\mbox{\boldmath $\sigma$}_1\cdot{\bf q})(\mbox{\boldmath $\sigma$}_2\cdot{\bf q})$-operator 
is evaluated in a superior mannner compared to ESC04.
Not included are the potentials from the tensor ($J^{PC}=2^{++}$) mesons.
Attempts including the latter mesons did not lead to substantial potentials from these
mesons or qualitative changes in the other contributions to the potentials.
The first results with the ESC08-model are reported in \cite{HYPX,PTP185.a}.

In this first paper of the series, we display and discuss the NN results of 
the simultaneous fit to the NN- and YN-data, including some $\Lambda\Lambda, \Xi N$ 
and $\Sigma N$ information from hypernuclei, using a single set of parameters.
In the second paper, henceforth referred to as II \cite{RNY10b},  
we report on the results for strangeness S=-1 YN-channels,
using the same simultaneous fit of the NN- and YN-data. 
This simultaneous fitting procedure was first introduced in \cite{Rij04b}, and its 
importance and advantages will be discussed in II.
In the third paper, henceforth referred to as III \cite{RNY10c},  
we report on the results and predictions for YY with strangeness $S=-2$.             
Finally, in the fourth paper (IV), we describe the predictions for YY 
with strangeness $S=-3,-4$.             
 
The contents of this paper are as follows. 
In section~\ref{sec:21} a description of the physical background and dynamical
contents of the ESC08-model is given.
In section~\ref{sec:2} the two-body integral equations in momentum space
are discussed. Also, the expansion into Pauli-spinor invariants is reviewed.
In section~\ref{sec:3} the ESC-potentials in momentum and 
configuration space for non-strange mesons are discussed in detail. 
In particular the new potentials are given. 
Section~\ref{sec:4} contains some brief remarks on the ESC-couplings 
and the QPC-model.
In section~\ref{sec:5} the simultaneous $NN \oplus YN \oplus YY$ fitting 
procedure is reviewed.
Here, also the results for the coupling constants and $F/(F+D)$-ratios
for OBE and MPE are given. 
In section~\ref{sec:6} the NN-results for the ESC08c-model are displayed.               
In section~\ref{sec:7} a solution for the the nuclear saturation and 
neutron star mass is described.                

In section~\ref{sec:8} we discuss the results and draw some conclusions.
In appendix~\ref{app:C} the B-field formalism for vector- and axial-vector mesons is
described. The exact treatment of the non-local-tensor operator is explained in
appendix~\ref{app:B}. 
In appendix~\ref{app:NTC} the treatment of the non-local tensor potential is reviewed.
In appendix~\ref{app:A} the basic formulas for the 
configuration space gaussian-yukawa functions are given.

 --------------------------------------------------------------------\\
\twocolumngrid
\section{Physical Content of the ESC-model}             
\label{sec:21}
 
The general physical basis, within the context of QCD, for the Nijmegen 
soft-core models has been outlined in the introduction of \cite{Rij04a}.
The description of baryon-interactions at low energies 
in terms of baryons and mesons 
can be reached through the following stages:
(i) The strongly interacting sector of the standard-model (SM) contains 
three families of quarks: (ud), (cs), (tb). (ii) Integrating out the
heavy quarks (c,b,t) leads to a QCD-world with effective interactions 
for the (u,d,s) quarks. (iii) This QCD-world is characterized by a phase
transition of the vacuum. Thereby the quarks gets dressed 
and become the so-called constituent quarks. The emerging picture
is that of the constituent-quark-model (CQM) \cite{Man84}.
The phase transition has transformed the effective QCD-world into
an complex hadronic-world.
(iv) The strong coupling lattice QCD (SCQCD) seems to be a proper model
to study the low energy meson-baryon and baryon-baryon physics, 
see \cite{Mil89} for applications and references.
Here the lattice spacing $a \geq 0.11$ fm provides a momentum scale for 
which the QCD coupling $g \geq 1.1$. Emerging is a picture where the 
meson-baryon coupling constants get large, and quark-exchange 
effects are rather small. The latter is due to the suppression due to the
gluonic overlaps involved. For a similar reason it has been argued \cite{LN87}
that the pomeron is exchanged between the individual quarks of the baryons.
In this picture the Nijmegen soft-core approach
to baryon-baryon interactions has a natural motivation.
(v) For the mesons we restrict ourselves to
mesons with $M \leq 1.5$ GeV$/c^2$, arriving at a so-called 
{\it effective field theory} as the arena for our description of the  
low energy baryon-baryon scattering.  

In view of the success of QCD, pseudo-scalar dominance of the divergence of
the axial-vector current (PCAC) leading to small light ("current") quark masses 
\cite{Ynd80,Gas75}, the spectroscopic success of the CQM, where the quarks 
have definite color charges, in generating       
the masses of the pseudo-scalar and vector nonets, and the masses and
magnetic moments of the baryon octet is rather surprising \cite{Gas81,Pov95}.
The transition from "current" to "constituent" quarks comes from dressing the
quark fields in the original QCD Lagrangian, see e.g. 
Ref.~\cite{Pol76,Man84,Lav97}.

In all works of the Nijmegen group on the baryon-baryon models,  
(broken) $SU(3)$ flavor-symmetry is explored to connect the $NN$, $YN$, and $YY$
channels, making possible a simultaneous fitting of all 
the available BB-data using a single set of model-parameters. 
The dynamical basis is the (approximate)  
permutation symmetry w.r.t. the constituent (u,d,s)-quarks. This has its roots in the 
approximate equality of the quark-masses, and more importantly that the gluons have no
flavor.
This enables the calculation of the baryon-baryon-meson coupling constants 
using as parameters the nucleon-nucleon-meson couplings and the $F/(F+D)$-ratio's.
This provides a strong correlation between the (rich) nucleon-nucleon- and the 
(scarce) hyperon-nucleon-data. 

The obtained coupling constants of the $BBM$-vertices are interpreted studying the
predictions of the constituent quark-model (CQM) in the form of 
the quark-antiquark pair creation model (QPC).
It has been argued that the $^3P_0$-mechanism \cite{Mic69,LeY73}
is dominant over the $^3S_1$-mechanism in lattice QCD \cite{Isg85}.
It turned out that the fitted coupling constants in ESC04 and ESC08
indeed follow mainly the pattern of couplings set by the $^3P_0$-model. 
Also, all $\alpha=F/(F+D)$-ratios are required to deviate no more than 0.1 
from the QPC-model predictions for the $BBM$- and the $BB-Pair$-vertices. 
Although it is in principle attractive to study the SU(3)-breaking of the $BBM$-couplings
using the QPC-model, as has been explored in ESC04 \cite{Rij04b}, in ESC08 the couplings
are treated as SU(3)-symmetric.
In the Nijmegen soft-core OBE- and ESC-models the BBM-vertices 
are described by coupling constants and gaussian form factors. 
Given the fact that in the CQM the quark wave functions for the
baryons are very much like ground state harmonic oscillator
functions, a gaussian behavior of the form factors is most natural.
These form factors guarantee a soft behavior of the potentials 
in configuration space at small distances.
The cut-off parameters in the form factors depend only on the type of meson 
(pseudoscalar, vector, etc.).
Within a meson SU(3)-multiplet we distinguish between octet and singlet form
factors.  Since there is
singlet-octet mixing for the I=0 mesons, we attribute the singlet and octet
cut-off to the dominant singlet or octet particle respectively.
For the considered nonets the singlet and octet cut-off are the same or close. 

In this way we have full predictive power for the $S=-2,-3,-4$ baryon-baryon channels, 
e.g. $\Lambda\Lambda, \Xi N$-channels which involve the singlet 
$\{1\}$-irrep that does not occur in the $N\!N$ and $Y\!N$ channels. 

Field theory allows both linear and non-linear realizations of chiral-symmetry (CS)
\cite{Schw67,Wei68,DeAlf73}. At low-energy phenomenologically the non-linear
realization is the most economical and natural. 
Therefore, we have chosen the pv-coupling and not the ps-coupling  
for the pseudoscalar mesons. This choice affects some $1/M^2$-terms in the 
ps-ps-exchange potential, In ESC04 we tested mixtures of the pv- and 
ps-coupling, but in ESC08 we use only the pv-coupling. In the non-linear
realization chiral-symmetry for the couplings of the scalar-, vector-, 
axial-vector-, etc. mesons is realized through isospin-symmetry SU(2,I) 
\cite{Wei68,DeAlf73}.

The potentials of the ESC-model are generated by (i) One-Boson-Exchange (OBE),
(ii) uncorrelated Two-Meson-Exchange (TME), (iii) Meson-Pair-Exchange (MPE),
(iv) Diffractive/Multi-gluon Exchange, (v) Quark-Core Effects (QCE).

\begin{enumerate}
\item [(i)]  
The OBE-part of the dynamical contents of the ESC08-models is determined 
by the following meson-exchanges:
\begin{enumerate}
\item   $J^{PC}=0^{--}$: The pseudoscalar-meson nonet $\pi,\ \eta,\ \eta',\ K$ with the
        $\eta-\eta'$ mixing angle $\theta_{P}=-13^{0}$ \cite{KLOE09},
        close to the \mbox{Gell-Mann-Okubo} quadratic mass formula \cite{GMO62}.
\item   $J^{PC}=1^{--}$: The vector-meson nonet $\rho,\ \phi,\ K^{\star},\ \omega$ with
        the $\phi-\omega$ mixing angle $\theta_{V}= 38.70^{0}$ \cite{KLOE09}.
        This follows from the quadratic GMO mass-formula, and is close to ideal mixing.
\item   $J^{PC}=1^{++}$: The axial-vector-meson nonet $a_1, f_1\ K_1, f_1'$ with
        the $f_1-f_1'$ mixing angle $\theta_{A}=  50.0^{0}$ \cite{SR97}.
\item   $J^{PC}=0^{++}$: The scalar-meson nonet 
        $a_0(962)=\delta,f_0(993)=S^{\star},\kappa(800),f_0(760)=\varepsilon$
        with the ideal $S^{\star}-\varepsilon$ mixing angle $\theta_{S}=35.26^{0}$.
\item   $J^{PC}=1^{+-}$: The axial-vector-meson nonet $b_1, f_1\ K_1, f_1'$ with
        the $h_1-h_1'$ ideal mixing angle $\theta_{B}= 35.26^{0}$.
\end{enumerate}
The soft-core approach of the OBE has been given originally for 
 $NN$ in \cite{NRS78}, and for $YN$ in \cite{MRS89}. With respect to these
OBE-interactions the ESC-models contain the modification of the form factor   
by introducing a zero for the mesons being P-wave quark-antiquark states in the 
CQM: the scalar- and axial-vector-mesons.  Such a zero is natural 
in the $^3P_0$-quark-pair-creation (QPC) \cite{Mic69,LeY73}  model for the coupling 
of the mesonic quark-antiquark ($Q\bar{Q}$) system to baryons. 
A consequence of such a zero is that a bound state in $\Lambda p$-scattering is less 
likely to occur.


\item [(ii)]  
The configuration space soft-core uncorrelated two-meson exchange for $NN$ has been derived 
in \cite{Rij91,RS96a}. Similarly to ESC04, also in ESC08 we use these potentials for 
ps-ps exchange with a complete $SU(3)_f$-symmetric treatment in NN, YN and YY. 
For example, we include double $K$-exchange in $NN$-scattering.
Since this includes two-pion exchange (TPE) the long-range part of the
potentials are represented. Here it is tacitly assumed that other TME potentials,
like ps-vc, ps-sc, etc., are either small due to cancellations, or 
can be described adequately by using effective couplings in the OBE-potentials. 
When these effective couplings do not deviate from
experimentally determined couplings it may be assumed that the corrections
from these other SU(3) meson-nonets in the TME potentials are small. 
This is our working hypothesis for the TME-potentials.
From the point of view of SU(3), since OBE contains only $\{8\}$- and $\{1\}$-exchange,
TME can not be represented completely in terms of OBE. This because TME also has
$\{27\}-, \{10\}$-, and $\{10^*\}$-exchange components.
Therefore, the predictions made by the ESC-models could be sensitive 
to this incompleteness of TME in the ESC-models.
At present the BB-data and the hypernuclear-data do not give information at this point.
\item [(iii)]  
Meson-pair exchanges (MPE) have been introduced in \cite{Rij93} for $NN$ and described 
in detail in \cite{RS96b}. The two-meson-baryon-baryon vertices are the low energy
approximations of (a) the heavy-meson and their two-meson decays, and (b) baryon-resonance
contributions $\Delta_{33}$ etc \cite{RS96b,SR97}.


\item [(iv)]  
Diffractive contributions to the soft-core potential have been 
introduced from the beginning, cfr. \cite{NRS78}. The pomeron
is thought of being related to an even number of gluon-exchanges. 
Here we introduce the Odderon-potential, 
which is related to an odd number of gluon exchanges.
\begin{enumerate}
\item   $J^{PC}=0^{++}$: The `diffractive' contribution from the pomeron P,
        which is a unitary singlet.
        These interactions give a repulsive contribution to the
        potentials in all channels of a gaussian type. 
\item   $J^{PC}=1^{--}$: The `diffractive' contribution from the odderon O.
        The origin of the odderon is assumed to be purely the exchange of
        the color-singlets with an odd number of gluons. Similarly to the
        pomeron, the odderon potential is taken to be 
        an SU(3)$_F$ singlet and of the gaussian form.
\end{enumerate}
\noindent As an explanation of the repulsive character of the pomeron-potential the following: 
The $J^{PC}$ is identical to that for the scalar-mesons. Naively, one would expect an 
attractive central potential. However, considering the two-gluon model for the
pomeron \cite{Low75,Nus75} the two-gluon parallel and crossed diagram contributions
to the BB-interaction can be shown to cancel adiabatically. The remaining 
non-adiabatic contribution is repulsive \cite{Pad89}.
\item [(v)]  
Quark-Core-Effects in the soft-core model can supply extra repulsion,
which may be required in some BB-channels.
Baryon-baryon studies with the soft-core OBE and ESC-models
thus far show that it is difficult to achieve a strongly enough repulsive short-range
interactions in (i) the $\Sigma^+ p(I=3/2,^3S_1)$- and (ii) the
$\Sigma N(I=1/2,^1S_0)$-channel. The short-range repulsion in baryon-baryon may  
in principle come from: (a) meson- and multi-gluon-exchange \cite{Rij04a,Rij04b}, 
and/or (b) the occurrence of forbidden six-quark SU(6)-states 
by the Pauli-principle \cite{Ots65,Oka00,Fuj07}. In view of the mentioned difficulties,
we have developed a phenomenological method for the ESC-model, which enables us 
to incorporate this quark-structural effect. This is an important new ingredient
of the here presented ESC08-model. This structural effect we describe phenomenologically
by gaussian repulsions, similar to the pomeron.
In the ESC08c-model we take the strength of this repulsion proportional to the 
weights of the SU(6)-forbidden [51]-configuration in the various BB-channels.
This in contrast to ESC08a,b \cite{HYPX,PTP185.a} where the quark-core effect is 
only included in the BB-channels with a dominant occurrence of the [51]-configuration.
\end{enumerate}
The different sources of $SU(3)$-breaking are discussed in paper II of this series.
With this simultaneous treatment of the $N\!N$, $Y\!N$, and $Y\!Y$ channels we have 
achieved 
a high quality description of the baryon-baryon interactions. The results,
using a single set of meson and quark-core parameters, include:
(a) a description of the NN-data with a $\chi^2_{pdp} = 1.081$
and good low energy parameters for the NN-channels including the binding energy
$E_B$ of the deuteron,                  
(b) a very good fit to the YN-scattering data.
(c) the fitting parameters with a clear physical significance, like e.g. the
 $NN\pi$-, $NN\rho$-couplings etc. and with realistic values of the
$F/(F+D)$-ratio's $\alpha_{PV}$ and $\alpha_{V}^{m}$. 
The fitting has been done under the constraints of the G-matrix results for the
ESC08-interactions. These show (i) satisfactory well-depth values for
$U_\Lambda, U_\Sigma > 0$, and $U_\Xi < 0$,
(ii) proper spin-spin ($U_{\sigma\sigma} \geq 1$, 
 and small spin-orbit interactions for $\Lambda N$.
All these features are in agreement with the Hyperball-data \cite{Has06} and the
NAGARA-event \cite{Tak01}. 
 
As in all Nijmegen models, the Coulomb interaction is included exactly, 
for which we solve the multichannel
Schr\"{o}dinger equation on the physical particle basis.
The nuclear potentials are calculated on the isospin basis.
This means that we include only the so-called 'medium strong' 
SU(3)-breaking and the charge symmetry breaking (CSB) in the potentials.
 
 

\section{Two-Body Integral Equations in Momentum Space}
\label{sec:2}

\subsection{Three-dimensional Two-Body Equations}
We consider the baryon-baryon reactions 
\begin{eqnarray}
 B(p_{a},s_{a})+B(p_{b},s_{b}) \rightarrow
    B(p_{a'},s_{a'})+B(p_{b'},s_{b'}) &&
\label{eq:30.1} \end{eqnarray}
In the following we also refer to a and a' as particles 1 and 1' (or 3),
and to b and b' as particles 2 and 2' (or 4).
The total four-momenta for the initial and the final states are denoted
as $P = p_{a} + p_{b}, P' = p_{a'} + p_{b'}$, and similarly the relative momenta
by $p = \frac{1}{2}(p_{a}-p_{b}), p' = \frac{1}{2}(p_{a'}-p_{b'})$. 
In the center-of-mass system (CM-system) for a and b
on-mass-shell one has 
$P = ( W , {\bf 0}) \hspace{0.2cm} , \hspace{0.2cm} p = ( 0 , {\bf p})
 \hspace{0.2cm} , \hspace{0.2cm} p' = ( 0 , {\bf p}')$.   
In the following, the on-mass-shell CM-momenta for the initial
and final states are denoted respectively by ${\bf p}$ and ${\bf p}'$.
So, $p_{a}^{0}=E_{a}({\bf p})=\sqrt{{\bf p}^{2}+M_{a}^{2}}$ and
$p_{a'}^{0}=E_{a'}({\bf p}')=\sqrt{{\bf p'}^{2}+M_{a'}^{2}}$, and
similarly for b(2) and b'(4). 
Because of translation-invariance $P=P'$ and
$W=W'=E_{a}({\bf p})+E_{b}({\bf p})=E_{a'}({\bf p}')+E_{b'}({\bf p}')$.
The transition amplitude matrix $M$ is related to the $S$-matrix via
\begin{equation}
 \langle f|S|i\rangle = \langle f|i\rangle -i(2\pi)^4\delta^4(P_f-P_i)
 \langle f| M | i \rangle.
\label{eq:30.2} \end{equation}
The two-particle states we normalize in the following way
\begin{eqnarray}
  \langle {\bf p}_{1}',{\bf p}_{2}'|{\bf p}_{1},{\bf p}_{2}\rangle
  &=& (2\pi)^{3}2E({\bf p}_{1}) \delta^{3}({\bf p}_{1}'-{\bf p}_{1})\cdot 
\nonumber\\ && \times 
  (2\pi)^{3}2E({\bf p}_{2}) \delta^{3}({\bf p}_{2}'-{\bf p}_{2}).
\label{eq:30.3} \end{eqnarray}
Three-dimensional integral equations for the amplitudes $\langle f|M|i\rangle$
have been derived in various ways, see e.g. \cite{Log63,Thom70,Ger75,NRS77,Rij85}.
Here, we follow Ref.~\cite{Rij04a} which employs the Macke-Klein procedure \cite{Klein53}.
After redefining the CM-amplitude $M({\bf p}',{\bf p} |W)$ by 

\onecolumngrid 

\begin{equation}
 M({\bf p}',{\bf p} |W) \rightarrow 
 \sqrt{\frac{M_{a}M_{b}}{E_{a}({\bf p}') E_{b}({\bf p}')} }
 M({\bf p}',{\bf p} |W)  
 \sqrt{\frac{M_{a}M_{b}}{E_{a}({\bf p}') E_{b}({\bf p}')} }
\label{eq:30.4} \end{equation}
 one arrives, see for details Ref.~\cite{Rij04a}, at the Thompson equation \cite{Thom70}
\begin{eqnarray}
 M({\bf p}',{\bf p}| W) &=& K^{irr}({\bf p}',{\bf p}|W) + 
 \int\!\frac{d^{3}p^{\prime\prime}}{(2\pi)^3} 
 K^{irr}({\bf p}',{\bf p}^{\prime\prime}|W)\ E_{2}^{(+)}({\bf p}^{\prime\prime}; W)\
 M({\bf p}^{\prime\prime},{\bf p}|W),  
 \nonumber\\
\label{eq:30.22} \end{eqnarray}
where 
$ E_{2}^{(+)}({\bf p}^{\prime\prime}; W)=  
 \left( W-{\cal W}({\bf p}^{\prime\prime})+i\delta\right)^{-1} $,  
and the two-nucleon irreducible kernel is given by
\begin{eqnarray}
  K^{{\it irr}}({\bf p}',{\bf p}| W)&=& -\frac{1}{(2\pi)^{2}}
 \sqrt{\frac{M_{a}M_{b}}{E_{a}({\bf p}') E_{b}({\bf p}')} }
 \sqrt{\frac{M_{a}M_{b}}{E_{a}({\bf p}) E_{b}({\bf p})} }
 \left(W-{\cal W}({\bf p}')\right)\left(W-{\cal W}({\bf p})\right)
 \nonumber \\[0.2cm] &\times&
  \int_{-\infty}^{+\infty} dp_{0}'
   \int_{-\infty}^{+\infty} dp_{0} \left[ \vphantom{\frac{A}{A}}
 \left\{F_{W}^{(a)}({\bf p}',p_{0}')
 F_{W}^{(b)}(-{\bf p}',-p_{0}')\right\}^{-1}
 \right.\nonumber \\[0.2cm] &\times& \left.
 \left[ I(p_{0}',{\bf p}'; p_{0},{\bf p}) \right]_{++,++}
    \left\{F_{W}^{(a)}({\bf p},p_{0})
    F_{W}^{(b)}(-{\bf p},-p_{0})\right\}^{-1}
\vphantom{\frac{A}{A}}\right], 
\label{Thomp2}  \end{eqnarray}
where $F_W({\bf p},p_0)= p_0-E({\bf p})+W/2+i\delta$.
This same expression for the kernel was exploited in \cite{Rij91,RS96a,RS96b}. 

In case one does not assume the strong pair-suppression, one must study instead 
of equation (\ref{eq:30.22}) a more general equation with couplings between the 
positive and negative energy spinorial amplitudes. Also to this more general case
one can apply the described three-dimensional reduction, and we refer the reader to 
Ref.~\cite{Klein74} for a treatment of this case.

 The $M/E$-factors in (\ref{Thomp2}) are due to the difference
between the relativistic and the non-relativistic normalization of
the two-particle states. In the following we simply put
$M/E({\bf p})=1$ in the kernel $K^{irr}$ Eq.~(\ref{Thomp2}). The corrections
to this approximation would give $(1/M)^{2}$-corrections
to the potentials, which we neglect in this paper. In the same approximation
there is no difference between the Thompson \cite{Thom70}
and the Lippmann-Schwinger equation, when the connection between these 
equations is made using multiplication factors. Henceforth, we will not
distinguish between the two.
 
The contributions to the two-particle irreducible kernel
$K^{{\it irr}}$ up to second order in the meson-exchange
are given in detail in \cite{RS96a,RS96b}.\\


\twocolumngrid
 
\subsection{Lippmann-Schwinger Equation}
\label{sec:2b}

  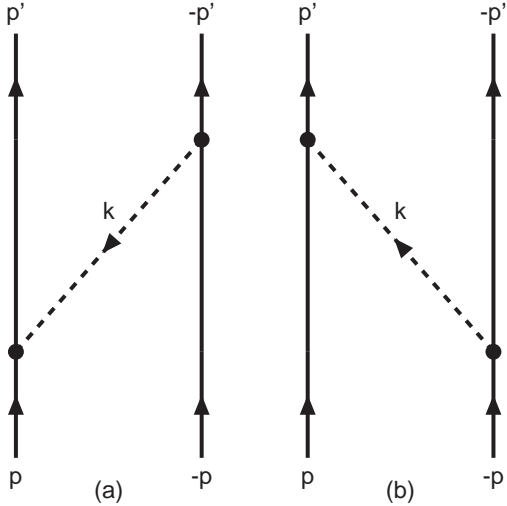
\begin{figure}[hbt]
 \begin{center} \begin{picture}(200,180)(0,0)
 \SetPFont{Helvetica}{9}
 \SetScale{1.0} \SetWidth{1.5}
            
 \ArrowLine(15,10)(15,50)  
 \Line(15,50)(15,130)   
 \ArrowLine(15,130)(15,170)  
 \ArrowLine(85,10)(85,50)   
 \Line(85,50)(85,130)   
 \ArrowLine(85,130)(85,170)  
 \Vertex(15,50){3}
 \Vertex(85,130){3}
 \DashArrowLine(85,130)(15,50){3}           
 \PText(15, 00)(0)[b]{p}
 \PText(15,175)(0)[b]{p'}
 \PText(85, 00)(0)[b]{-p}
 \PText(85,175)(0)[b]{-p'}
 \PText(50,100)(0)[b]{k}
 \PText(50,-5 )(0)[b]{(a)}

\SetOffset(10,0)

 \ArrowLine(115,10)(115,50)   
 \Line(115,50)(115,130)   
 \ArrowLine(115,130)(115,170)  
 \ArrowLine(185,10)(185,50)   
 \Line(185,50)(185,130)   
 \ArrowLine(185,130)(185,170)  
 \Vertex(115,130){3}
 \Vertex(185, 50){3}
 \DashArrowLine(185,50)(115,130){3}           
 \PText(115, 00)(0)[b]{p}
 \PText(115,175)(0)[b]{p'}
 \PText(185, 00)(0)[b]{-p}
 \PText(185,175)(0)[b]{-p'}
 \PText(150,100)(0)[b]{k}
 \PText(150,-5 )(0)[b]{(b)}
 \end{picture} 
  \end{center}
\caption{One-boson-exchange graphs: 
        The dashed lines with momentum ${\bf k}$ refers to the
        bosons: pseudo-scalar, vector, axial-vector, or scalar mesons.}
\label{obefig}
 \end{figure}

  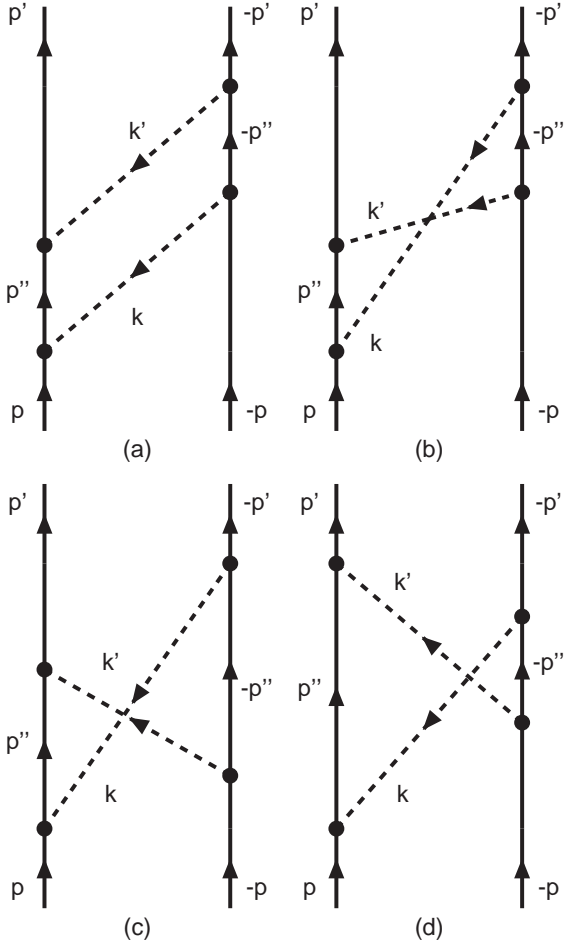
\begin{figure}[hbt]
 \begin{center} \begin{picture}(220,350)(0,0)
 \SetPFont{Helvetica}{9}
 \SetScale{1.0} \SetWidth{1.5}

 \Vertex(15,220){3}
 \Vertex(15,260){3}
 \Vertex(85,280){3}
 \Vertex(85,320){3}

 \ArrowLine(15,190)(15,220)  
 \ArrowLine(15,220)(15,260)  
 \Line(15,260)(15,320)  
 \ArrowLine(15,320)(15,350)  
 \ArrowLine(85,190)(85,220)  
 \Line(85,220)(85,280)  
 \ArrowLine(85,280)(85,320)  
 \ArrowLine(85,320)(85,350)  

 \DashArrowLine(85,280)(15,220){3}           
 \DashArrowLine(85,320)(15,260){3}           

 \PText( 5,195)(0)[b]{p}
 \PText( 5,345)(0)[b]{p'}
 \PText(95,195)(0)[b]{-p}
 \PText(95,345)(0)[b]{-p'}
 \PText( 5,240)(0)[b]{p''}
 \PText(95,300)(0)[b]{-p''}
 \PText(50,230)(0)[b]{k}
 \PText(50,300)(0)[b]{k'}
 \PText(50,180)(0)[b]{(a)}

 \Vertex(125,220){3}
 \Vertex(125,260){3}
 \Vertex(195,280){3}
 \Vertex(195,320){3}

 \ArrowLine(125,190)(125,220)  
 \ArrowLine(125,220)(125,260)  
 \Line(125,260)(125,320)  
 \ArrowLine(125,320)(125,350)  
 \ArrowLine(195,190)(195,220)  
 \Line(195,220)(195,280)  
 \ArrowLine(195,280)(195,320)  
 \ArrowLine(195,320)(195,350)  

 \DashArrowLine(195,320)(160,270){3}           
 \DashLine(160,270)(125,220){3}           
 \DashArrowLine(195,280)(160,270){3}           
 \DashLine(160,270)(125,260){3}           

 \PText(115,195)(0)[b]{p}
 \PText(115,345)(0)[b]{p'}
 \PText(205,195)(0)[b]{-p}
 \PText(205,345)(0)[b]{-p'}
 \PText(115,240)(0)[b]{p''}
 \PText(205,300)(0)[b]{-p''}
 \PText(140,220)(0)[b]{k}
 \PText(140,270)(0)[b]{k'}
 \PText(160,180)(0)[b]{(b)}

 \Vertex(15, 40){3}
 \Vertex(85, 60){3}
 \Vertex(15,100){3}
 \Vertex(85,140){3}

 \ArrowLine(15, 10)(15, 40)  
 \ArrowLine(15, 40)(15,100)  
 \Line(15,100)(15,140)  
 \ArrowLine(15,140)(15,170)  
 \ArrowLine(85, 10)(85, 40)  
 \Line(85, 40)(85, 60)  
 \ArrowLine(85, 60)(85,140)  
 \ArrowLine(85,140)(85,170)  

 \DashArrowLine(85,140)(15, 40){3}           
 \DashArrowLine(85, 60)(15,100){3}           

 \PText( 5, 15)(0)[b]{p}
 \PText( 5,160)(0)[b]{p'}
 \PText(95, 15)(0)[b]{-p}
 \PText(95,160)(0)[b]{-p'}
 \PText( 5, 70)(0)[b]{p''}
 \PText(95, 90)(0)[b]{-p''}
 \PText(40, 50)(0)[b]{k}
 \PText(40,100)(0)[b]{k'}
 \PText(50, 00)(0)[b]{(c)}

 \Vertex(125, 40){3}
 \Vertex(195, 80){3}
 \Vertex(195,120){3}
 \Vertex(125,140){3}

 \ArrowLine(125, 10)(125, 40)  
 \ArrowLine(125, 40)(125,140)  
 \ArrowLine(125,140)(125,170)  
 \ArrowLine(195, 10)(195, 40)  
 \Line(195, 40)(195, 80)  
 \ArrowLine(195, 80)(195,120)  
 \Line(195,120)(195,140)  
 \ArrowLine(195,140)(195,170)  

 \DashArrowLine(195,120)(125, 40){3}           
 \DashArrowLine(195, 80)(125,140){3}           

 \PText(115, 15)(0)[b]{p}
 \PText(115,160)(0)[b]{p'}
 \PText(205, 15)(0)[b]{-p}
 \PText(205,160)(0)[b]{-p'}
 \PText(115, 90)(0)[b]{p''}
 \PText(205,100)(0)[b]{-p''}
 \PText(150, 50)(0)[b]{k}
 \PText(150,130)(0)[b]{k'}
 \PText(160, 00)(0)[b]{(d)}

 \end{picture} 
  \end{center}
\caption{BW two-meson-exchange graphs: (a) planar and (b)--(d) crossed
        box. The dashed line with momentum ${\bf k}_{1}$ refers to the
        pion and the dashed line with momentum ${\bf k}_{2}$ refers
        to one of the other (vector, scalar, or pseudoscalar) mesons.
        To these we have to add the ``mirror'' graphs, and the
        graphs where we interchange the two meson lines.}
\label{bwfig}
   \end{figure}                     
  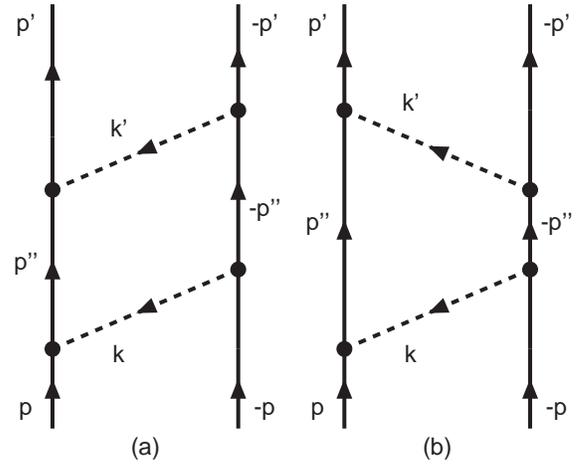
\begin{figure}[hbt]
 \begin{center} \begin{picture}(220,170)(0,0)
 \SetPFont{Helvetica}{9}
 \SetScale{1.0} \SetWidth{1.5}

 \Vertex(15, 40){3}
 \Vertex(15,100){3}
 \Vertex(85, 70){3}
 \Vertex(85,130){3}

 \ArrowLine(15, 10)(15, 40)  
 \ArrowLine(15, 40)(15,100)  
 \Line(15,100)(15,120)  
 \ArrowLine(15,120)(15,170)  
 \ArrowLine(85, 10)(85, 40)  
 \Line(85, 40)(85, 70)  
 \ArrowLine(85, 70)(85,130)  
 \ArrowLine(85,130)(85,170)  

 \DashArrowLine(85, 70)(15, 40){3}           
 \DashArrowLine(85,130)(15,100){3}           

 \PText( 5, 15)(0)[b]{p}
 \PText( 5,160)(0)[b]{p'}
 \PText(95, 15)(0)[b]{-p}
 \PText(95,160)(0)[b]{-p'}
 \PText( 5, 70)(0)[b]{p''}
 \PText(95, 90)(0)[b]{-p''}
 \PText(40, 35)(0)[b]{k}
 \PText(40,120)(0)[b]{k'}
 \PText(50, 00)(0)[b]{(a)}

 \Vertex(125, 40){3}
 \Vertex(125,130){3}
 \Vertex(195, 70){3}
 \Vertex(195,100){3}

 \ArrowLine(125, 10)(125, 40)  
 \ArrowLine(125, 40)(125,130)  
 \ArrowLine(125,130)(125,170)  
 \ArrowLine(195, 10)(195, 40)  
 \Line(195, 40)(195, 70)  
 \ArrowLine(195, 70)(195,100)  
 \Line(195,100)(195,130)  
 \ArrowLine(195,130)(195,170)  

 \DashArrowLine(195, 70)(125, 40){3}           
 \DashArrowLine(195,100)(125,130){3}           

 \PText(115, 15)(0)[b]{p}
 \PText(115,160)(0)[b]{p'}
 \PText(205, 15)(0)[b]{-p}
 \PText(205,160)(0)[b]{-p'}
 \PText(115, 85)(0)[b]{p''}
 \PText(205, 85)(0)[b]{-p''}
 \PText(150, 35)(0)[b]{k}
 \PText(150,130)(0)[b]{k'}
 \PText(160, 00)(0)[b]{(b)}

 \end{picture} 
  \end{center}
 \caption{Planar-box TMO two-meson-exchange graphs.
          Same notation as in Fig.~\protect\ref{bwfig}.
          To these we have to add the ``mirror'' graphs, and the
          graphs where we interchange the two meson lines.}
  \label{tmofig}
  \end{figure}                     

 \twocolumngrid

The transformation of (\ref{eq:30.22}) to the Lippmann-Schwinger 
equation can be effectuated by defining
\begin{eqnarray}
 T({\bf p}',{\bf p}) &=& N({\bf p}')\ M({\bf p}',{\bf p}|W)\ N({\bf p}), 
 \nonumber\\[0.2cm]
 V({\bf p}',{\bf p}) &=& N({\bf p}')\ K^{irr}({\bf p}',{\bf p}|W)\ N({\bf p}),       
\label{eq:30.24} \end{eqnarray}
where the transformation function is 
\begin{equation} 
 N({\bf p}) = \sqrt{\frac{{\bf p}_i^2-{\bf p}^2}{2M_N(E\left({\bf p}_i)-E({\bf p})\right)}}.
\label{eq:30.25} \end{equation}
Application of this transformation, yields the Lippmann-Schwinger equation
\begin{eqnarray}
    T({\bf p}',{\bf p}) &=& V({\bf p}',{\bf p}) +
    \int \frac{d^{3}p''}{(2\pi)^3}\
  \nonumber\\ && \times 
 V({\bf p}',{\bf p}'')\ g({\bf p}''; W)\; T({\bf p}'',{\bf p})
\label{eq:30.26} \end{eqnarray}
with the standard Green function     
\begin{equation}
    g({\bf p};W) = 
    \frac{M_N}{{\bf p}_i^{2}-{\bf p}^{2}+i\delta}.
\label{eq:30.27} \end{equation}
The corrections to the approximation $E_{2}^{(+)} \approx g({\bf p}; W)$ 
are of order $1/M^{2}$, which we neglect henceforth.
 
The transition from Dirac-spinors to
Pauli-spinors, is given in Appendix C of Ref.~\cite{Rij91}, where we write for the 
the Lippmann-Schwinger equation in the 4-dimensional Pauli-spinor space
\begin{eqnarray}
 {\cal T}({\bf p}',{\bf p})&=&{\cal V}({\bf p}',{\bf p}) + \int \frac{d^{3} p''}{(2\pi)^3}\
  \nonumber\\ && \times 
 {\cal V}({\bf p}',{\bf p}'')\  g({\bf p}''; W)\ {\cal T}({\bf p}'',{\bf p})\ .
 \label{eq:30.28} \end{eqnarray}

The ${\cal T}$-operator in Pauli spinor-space is defined by
\begin{eqnarray}
 && \chi^{(a)\dagger}_{\sigma'_{a}}\chi^{(b)\dagger}_{\sigma'_{b}}\; 
 {\cal T}({\bf p}',{\bf p})\;
 \chi^{(a)}_{\sigma_{a}}\chi^{(b)}_{\sigma_{b}}  =              
\nonumber\\ && 
 \bar{u}_{a}({\bf p}',\sigma'_{a})\bar{u}_{b}(-{\bf p}',\sigma'_{b})\
 \tilde{T}({\bf p}',{\bf p})\; u_{a}({\bf p},\sigma_{a}) u_{b}(-{\bf p},\sigma_{b}).
\nonumber\\
 \label{eq:30.29} \end{eqnarray}
and similarly for the ${\cal V}$-operator.
Like in the derivation of the OBE-potentials \cite{NRS78,NRS77}
we make off-shell and on-shell the approximation,
  $ E({\bf p})= M + {\bf p}^{2}/2M $
 and $ W = 2\sqrt{{\bf p}_i^{2}+M^{2}} = 2M + {\bf p}_i^{2}/M$ ,     
everywhere in the interaction kernels, which, of course,
is fully justified for low energies only. 
In contrast to these kinds of approximations, of course the full
${\bf k}^{2}$-dependence of the form factors is kept
throughout the derivation of the TME. 
Notice that the gaussian form factors suppress the high momentum
transfers strongly. This means that the contribution to the potentials
from intermediate states which are far off-energy-shell can not
be very large. 

Because of rotational invariance and parity conservation, the ${\cal T}$-matrix, which is
a $4\times 4$-matrix in Pauli-spinor space, can be expanded 
into the following set of in general 8 spinor invariants, see for example 
Ref.~\cite{SNRV71}. Introducing \cite{notation1}
\begin{equation}
  {\bf q}=\frac{1}{2}({\bf p}'+{\bf p})\ , \
  {\bf k}={\bf p}'-{\bf p}\ , \           
  {\bf n}={\bf p}\times {\bf p}',
\label{eq:30.30} \end{equation}
with, of course, ${\bf n}={\bf q}\times {\bf k}$,
we choose for the operators $P_{j}$ in spin-space
\begin{eqnarray}
&&  P_{1}=1,  \hspace{3mm} P_{2}= 
 \mbox{\boldmath $\sigma$}_1\cdot\mbox{\boldmath $\sigma$}_2,
 \nonumber\\[0.0cm]
&& P_{3}=(\mbox{\boldmath $\sigma$}_1\cdot{\bf k})(\mbox{\boldmath $\sigma$}_2\cdot{\bf k})
 -\frac{1}{3}(\mbox{\boldmath $\sigma$}_1\cdot\mbox{\boldmath $\sigma$}_2)
  {\bf k}^2,
 \nonumber\\[0.0cm]
&& P_{4}=\frac{i}{2}(\mbox{\boldmath $\sigma$}_1+
 \mbox{\boldmath $\sigma$}_2)\cdot{\bf n}, \hspace{3mm} 
 P_{5}=(\mbox{\boldmath $\sigma$}_1\cdot{\bf n})(\mbox{\boldmath $\sigma$}_2\cdot{\bf n}),
 \nonumber\\[0.0cm]
 && P_{6}=\frac{i}{2}(\mbox{\boldmath $\sigma$}_1-\mbox{\boldmath $\sigma$}_2)\cdot{\bf n}, 
 \nonumber\\[0.0cm]
 && P_{7}=(\mbox{\boldmath $\sigma$}_1\cdot{\bf q})(\mbox{\boldmath $\sigma$}_2\cdot{\bf k})
 +(\mbox{\boldmath $\sigma$}_1\cdot{\bf k})(\mbox{\boldmath $\sigma$}_2\cdot{\bf q}),
 \nonumber\\[0.0cm]
&& P_{8}=(\mbox{\boldmath $\sigma$}_1\cdot{\bf q})(\mbox{\boldmath $\sigma$}_2\cdot{\bf k})
 -(\mbox{\boldmath $\sigma$}_1\cdot{\bf k})(\mbox{\boldmath $\sigma$}_2\cdot{\bf q}).
\label{eq:30.31} \end{eqnarray}
Here we follow Ref.~\cite{MRS89}, where in contrast to Ref.~\cite{NRS78},
we have chosen $P_{3}$ to be a purely `tensor-force' operator.
The expansion in spinor-invariants reads
\begin{equation}
 {\cal T}({\bf p}',{\bf p}) = \sum_{j=1}^8\ \widetilde{T}_j({\bf p}^{\prime 2},{\bf p}^2,
 {\bf p}'\cdot{\bf p})\ P_j({\bf p}',{\bf p})\ .
\label{eq:30.32} \end{equation}
Similarly to (\ref{eq:30.32}) we expand the potentials $V$. 
In the case of the axial-vector meson exchange there will occur terms
proportional to
\begin{equation}
 P_5'=(\mbox{\boldmath $\sigma$}_1\cdot{\bf q})(\mbox{\boldmath $\sigma$}_2\cdot{\bf q})
 -\frac{1}{3}(\mbox{\boldmath $\sigma$}_1\cdot\mbox{\boldmath $\sigma$}_2){\bf q}^2.
\label{eq:30.33} \end{equation}
The treatment of such a Pauli-invariant using the Okubo-Marshak identity \cite{Ok58},
see also Ref.~\cite{SNRV71}, is not without problems because it involves the division with
${\bf k}^2$. Therefore, in the ESC04-models \cite{Rij04a,Rij04b} the replacement
$P_5' \rightarrow -P_3$ was chosen. For the ESC08-models a satisfactory treatment
has been developed, which is described in Appendix~\ref{app:B}.
For the treatment of the potentials with $P_8$ we use the identity \cite{BDI70}
\begin{equation}
 P_8 = -(1+\mbox{\boldmath $\sigma$}_1\cdot\mbox{\boldmath $\sigma$}_2) P_6.
\label{eq:30.34} \end{equation}
Under time-reversal $P_7 \rightarrow -P_7$ and $P_8 \rightarrow -P_8$.
Therefore for elastic scattering $V_7=V_8=0$. 
Anticipating the explicit results for the potentials in section~\ref{sec:3} we
notice the following:                     
 (i) For the general BB-reaction we will find no contribution to $V_7$. 
The operators $P_6$ and $P_8$ give spin singlet-triplet transitions. 
 (ii) In the case of non-strangeness-exchange ($\Delta S=0$), $V_6 \neq 0$  
and $V_8$=0. The latter follows from our approximation to neglect the
mass differences among the nucleons, between the $\Lambda$ and $\Sigma$'s, and
among the $\Xi$'s.
 (iii) In the case of strangeness-exchange ($\Delta S=\pm 1$), $V_6,V_8 \neq 0$.    
The contributions to $V_6$ come from graphs with both spin- and particle-exchange,
i.e. Majorana-type potentials having the $P_f P_\sigma P_6= -P_x P_6$-operator.
Here, $P_f P_\sigma$ reflect our convention for the two-particle wave functions,
see \cite{NRS77}. 
The contributions to $V_8$ come from graphs with particle-exchange and   
spin-exchange, because $P_8=-P_\sigma P_6$. Therefore, we only have to
apply $P_f$ in order to map the wave functions after such exchange onto
our two-particle wave-functions. So, we have the $P_f P_8= +P_x P_6$-operator.
Here, we used that for BB-systems the allowed physical states satisfy 
$P_f P_\sigma P_x=-1$.

  \begin{figure}[hbt]
 \begin{center} \begin{picture}(220,350)(0,0)
 \SetPFont{Helvetica}{9}
 \SetScale{1.0} \SetWidth{1.5}

 \Vertex(15,230){3}
 \Vertex(85,270){3}
 \Vertex(85,310){3}

 \ArrowLine(15,190)(15,230)  
 \Line(15,230)(15,310)  
 \ArrowLine(15,310)(15,350)  
 \ArrowLine(85,190)(85,220)  
 \Line(85,220)(85,270)  
 \ArrowLine(85,270)(85,310)  
 \ArrowLine(85,310)(85,350)  

 \DashArrowLine(85,270)(15,230){3}           
 \DashArrowLine(85,310)(15,230){3}           

 \PText( 5,195)(0)[b]{p}
 \PText( 5,345)(0)[b]{p'}
 \PText(95,195)(0)[b]{-p}
 \PText(95,345)(0)[b]{-p'}
 \PText(95,290)(0)[b]{-p''}
 \PText(50,230)(0)[b]{k}
 \PText(50,280)(0)[b]{k'}
 \PText(50,180)(0)[b]{(a)}

 \Vertex(125,310){3}
 \Vertex(195,230){3}
 \Vertex(195,270){3}

 \ArrowLine(125,190)(125,230)  
 \Line(125,230)(125,310)  
 \ArrowLine(125,310)(125,350)  
 \ArrowLine(195,190)(195,230)  
 \ArrowLine(195,230)(195,270)  
 \Line(195,270)(195,310)  
 \ArrowLine(195,310)(195,350)  

 \DashArrowLine(195,230)(125,310){3}           
 \DashArrowLine(195,270)(125,310){3}           

 \PText(115,195)(0)[b]{p}
 \PText(115,345)(0)[b]{p'}
 \PText(205,195)(0)[b]{-p}
 \PText(205,345)(0)[b]{-p'}
 \PText(205,250)(0)[b]{-p''}
 \PText(160,250)(0)[b]{k}
 \PText(160,300)(0)[b]{k'}
 \PText(160,180)(0)[b]{(b)}

 \Vertex(15, 85){3}
 \Vertex(85, 40){3}
 \Vertex(85,130){3}

 \ArrowLine(15, 10)(15, 40)  
 \Line(15, 40)(15, 85)  
 \Line(15, 85)(15,130)  
 \ArrowLine(15,130)(15,170)  
 \ArrowLine(85, 10)(85, 40)  
 \ArrowLine(85, 40)(85,130)  
 \ArrowLine(85,130)(85,170)  

 \DashArrowLine(85, 40)(15, 85){3}           
 \DashArrowLine(85,130)(15, 85){3}           

 \PText( 5, 15)(0)[b]{p}
 \PText( 5,160)(0)[b]{p'}
 \PText(95, 15)(0)[b]{-p}
 \PText(95,160)(0)[b]{-p'}
 \PText(95, 85)(0)[b]{-p''}
 \PText(40, 55)(0)[b]{k}
 \PText(40,110)(0)[b]{k'}
 \PText(50, 00)(0)[b]{(c)}

 \Vertex(125, 50){3}
 \Vertex(195,120){3}

 \ArrowLine(125, 10)(125, 50)  
 \Line(125, 50)(125,120)  
 \ArrowLine(125,120)(125,170)  
 \ArrowLine(195, 10)(195, 50)  
 \Line(195, 50)(195,120)  
 \ArrowLine(195,120)(195,170)  

 \DashArrowArc(195,50)(70,90,180){3}           
 \DashArrowArcn(125,120)(70,00,-90){3}           

 \PText(115, 15)(0)[b]{p}
 \PText(115,160)(0)[b]{p'}
 \PText(205, 15)(0)[b]{-p}
 \PText(205,160)(0)[b]{-p'}
 \PText(150, 40)(0)[b]{k}
 \PText(150,110)(0)[b]{k'}
 \PText(160, 00)(0)[b]{(d)}
 \end{picture} 
  \end{center}
 \caption{One- and Two-Pair exchange graphs.
          To these we have to add the ``mirror'' graphs, and the
          graphs where we interchange the two meson lines.}
  \label{pairfig}
  \end{figure}
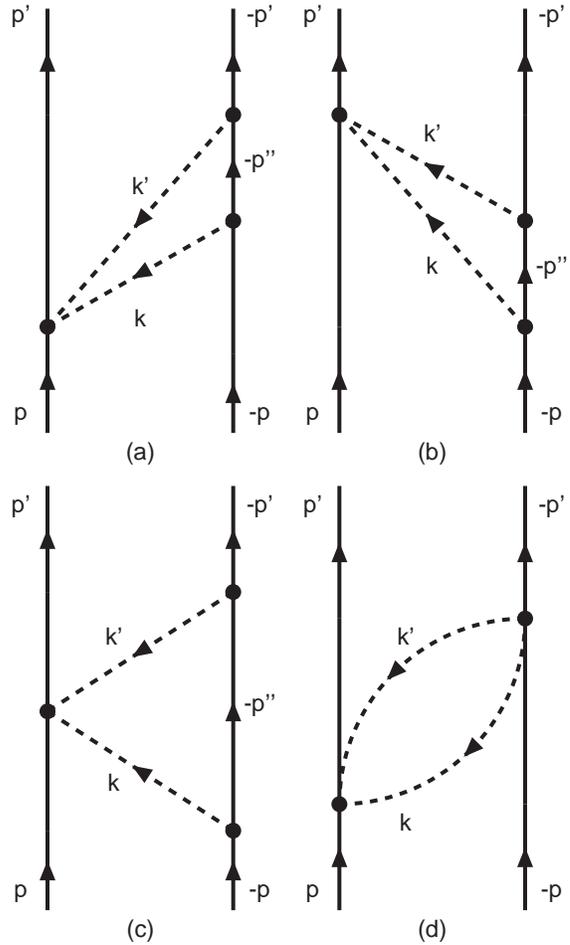                     

\section{Extended-Soft-Core Potentials in Momentum Space}
\label{sec:3}   
The potential of the ESC-model contains the contributions from 
(i) One-boson-exchanges, Fig.~\ref{obefig}, (ii) Uncorrelated 
Two-Pseudo-scalar exchange, Fig.~\ref{bwfig} and Fig.~\ref{tmofig}, 
and (iii) Meson-Pair-exchange, Fig~\ref{pairfig}. In this section we 
review the potentials and indicate the changes with respect to 
earlier papers on the OBE- and ESC-models.
The spin-1 meson-exchange is an important ingredient for the baryon-baryon force. 
In the ESC08-model we treat the vector-mesons and the axial-vector mesons 
according to the Proca- \cite{IZ80} and the B-field \cite{Nak72,NO90} formalism
respectively. For details, we refer to Appendix~\ref{app:C}.

\subsection{One-Boson-Exchange Interactions in Momentum Space}
\label{sect.3a}
The OBE-potentials are the 
same as given in \cite{NRS78,MRS89}, with the exception of 
 (i) the zero in the scalar form factor, and 
 (ii) the axial-vector-meson potentials.
Here, we review the OBE-potentials briefly, and give those potentials
which are not included in the above references.
The local interaction Hamilton densities for the different couplings
are \cite{BD65} \\ \\
        a) Pseudoscalar-meson exchange $(J^{PC}=0^{-+})$
         \begin{equation}
         {\cal H}_{PV}= \frac{f_{P}}{m_{\pi^{+}}}
         \bar{\psi}\gamma_{\mu}\gamma_{5}
                \psi\partial^{\mu}\phi_{P}. \label{eq:3.1}\end{equation}
This is the pseudovector coupling, and the
relation with the pseudoscalar coupling is 
$g_P = 2M_B/m_{\pi^+}$, where $M_B$ is the nucleon or hyperon mass.\\ \\     
        b) Vector-meson exchange $(J^{PC}=1^{--})$
       \begin{equation}
   {\cal H}_{V}=g_{V}\bar{\psi}\gamma_{\mu}\psi\phi_{V}^{\mu}
                +\frac{f_{V}}{4{\cal M}}\bar{\psi}\sigma_{\mu\nu}
                \psi (\partial^{\mu}\phi^{\nu}_{V}-\partial^{\nu}
                      \phi^{\mu}_{V}), \label{eq:3.2}\end{equation}
       where $\sigma_{\mu\nu}= i[\gamma_{\mu},\gamma_{\nu}]/2$,
       and the scaling mass ${\cal M}$, 
       will be taken to be the proton mass.\\ \\      
\noindent c)\ Axial-vector-meson exchange ( $J^{PC}=1^{++}$, 1$^{st}$ kind):
\begin{equation}
 {\cal H}_A = g_A[\bar{\psi}\gamma_\mu\gamma_5\psi] \phi^\mu_A + \frac{if_A}{{\cal M}}
 [\bar{\psi}\gamma_5\psi]\ \partial_\mu\phi_A^\mu.
\label{eq:OBE.1}\end{equation}
In ESC04 the $g_A$-coupling was included, but not the derivative $f_A$-coupling \cite{dercopax}.
Also, in ESC04  we used a local-tensor approximation (LTA) for the 
 $(\mbox{\boldmath $\sigma$}_1\cdot{\bf q})(\mbox{\boldmath $\sigma$}_2
  \cdot{\bf q})$ operator. Here, we improve on that considerably by avoiding such
 rather crude approximation. The details of our new treatment are given in 
Appendix~\ref{app:B}.
\\[0.2cm]
\noindent d)\ Axial-vector-meson exchange ( $J^{PC}=1^{+-}$, 2$^{nd}$ kind):
\begin{equation}
 {\cal H}_B = \frac{if_B}{m_B}
 [\bar{\psi}\sigma_{\mu\nu}\gamma_5\psi]\ \partial_\nu\phi_B^\mu\ .
\label{eq:OBE.2}\end{equation}
In ESC04 this coupling was not included. Like for the axial-vector mesons of the
1$^{st}$-kind we include an SU(3)-nonet with members $b_1(1235), h_1(1170), h_1(1380)$.
In the quark-model they are $Q\Bar{Q}(^1P_1)$-states.\\[0.2cm]
\noindent e)\ Scalar-meson exchange ($J^{PC}=0^{++}$):
\begin{equation}
 {\cal H}_S = g_S[\bar{\psi}\psi] \phi_S + \frac{f_S}{{\cal M}}
 [\bar{\psi}\gamma_\mu\psi]\ \partial^\mu\phi_S,
\label{eq:OBE.3}\end{equation}
which is the most general interaction up to the first derivative. 
However, charge conjugation gives 
${\cal C}[\bar{\psi}\gamma_\mu \psi] {\cal C}^{-1}=-[\bar{\psi}\gamma_\mu\psi]$, 
and therefore $f_S=0$.

\noindent f)\ Pomeron-exchange ($J^{PC}=0^{++}$):
       The vertices for this `diffractive'-exchange have the
       same Lorentz structure as those for scalar-meson-exchange.\\[0.2cm]
\noindent g)\ Odderon-exchange ($J^{PC}=1^{--}$):
\begin{equation}
 {\cal H}_O = g_O[\bar{\psi}\gamma_\mu\psi] \phi^\mu_O + \frac{f_O}{4{\cal M}}
 [\bar{\psi}\sigma_{\mu\nu}\psi] (\partial^\mu\phi^\nu_O-\partial^\nu\phi_O^\mu).
\label{eq:OBE.4}\end{equation}
Since the gluons are flavorless, Odderon-exchange is treated as an SU(3)-singlet.
Furthermore, since the Odderon represents a Regge-trajectory with an intercept
equal to that of the Pomeron, and is supposed not to contribute for small ${\bf k}^2$,
we include a factor ${\bf k}^2/{\cal M}^2$ in the coupling.\\[0.2cm]
 
Including form factors $f({\bf x}'-{\bf x})$ ,
the interaction hamiltonian densities are modified to
\begin{equation}
        H_{X}({\bf x})=\int\!d^{3}x'\,
  f({\bf x}'-{\bf x}){\cal H}_{X}({\bf x}'),
\end{equation}
 for $X= P,\ V,\ A$, and $S$ ($P =$ pseudo-scalar, $V =$ vector,
 $A=$ axial-vector, and $S =$ scalar). The         
potentials in momentum space are the same as for point interactions,
except that the coupling constants are multiplied by the Fourier
transform of the form factors.
 
In the derivation of the $V_{i}$ we employ the same approximations as in 
\cite{NRS78,MRS89}, i.e.
\begin{enumerate}
\item   We expand in $1/M$: 
    $E(p) = \left[ {\bf k}^{2}/4 +
    {\bf q}^{2}+M^{2}\right]^{\frac{1}{2}}$\\
    $\approx M+{\bf k}^{2}/8M + {\bf q}^{2}/2M$
 and keep only terms up to first order in ${\bf k}^{2}/M$ and
 ${\bf q}^{2}/M$. This except for the form factors where
 the full ${\bf k}^{2}$-dependence is kept throughout
 the calculations. Notice that the gaussian form factors
suppress the high ${\bf k}^{2}$-contributions strongly.
\item   In the meson propagators
$       (-(p_{1}-p_{3})^{2}+m^{2})  
        \approx({\bf k}^{2}+m^{2})$ .
\item   When two different baryons are involved at a $BBM$-vertex
        their average mass is used in the
        potentials and the non-zero component of the momentum transfer
        is accounted for by using an effective mass in
        the meson propagator (for details see \cite{MRS89}).     
\end{enumerate}
 
Due to the approximations we get only a linear dependence on
${\bf q}^{2}$ for $V_{1}$. In the following, separating the local and the
non-local parts, we write
\begin{equation}
  V_{i}({\bf k}^{2},{\bf q}^{2})=
  V_{i a}({\bf k}^{2})+V_{i b}({\bf k}^{2})({\bf q}^{2}+\frac{1}{4}{\bf k}^2),
\label{vcdec} \end{equation}
where in principle $i=1,8$. 
 
The OBE-potentials are now obtained in the standard way (see e.g.\
\cite{NRS78,MRS89}) by evaluating the $BB$-interaction in Born-approximation.
We write the potentials $V_{i}$ of Eqs.~(\ref{eq:30.33}) and
(\ref{vcdec}) in the form
\begin{equation}
  V_{i}({\bf k}\,^{2},{\bf q}\,^{2})=
   \sum_{X} \Omega^{(X)}_{i}({\bf k}\,^{2})
   \cdot \Delta^{(X)} ({\bf k}^{2},m^{2},\Lambda^{2}).
\label{nrexpv2} \end{equation}
Furthermore for $X=P,V$ 
\begin{equation}
   \Delta^{(X)}({\bf k}^{2},m^{2},\Lambda^{2})= e^{-{\bf k}^{2}/\Lambda^{2}}/  
                    \left({\bf k}^{2}+m^{2}\right),
\label{propm1} \end{equation}
and for $X=S,A$ a zero in the form factor
\begin{equation}
   \Delta^{(S)}({\bf k}^{2},m^{2},\Lambda^{2})= \left(1-{\bf k}^2/U^2\right)\
  e^{-{\bf k}^{2}/\Lambda^{2}}/  
  \left({\bf k}^{2}+m^{2}\right),
\label{propm2} \end{equation}
and for $X=D,O$
\begin{equation}
   \Delta^{(D)}({\bf k}^{2},m^{2},\Lambda^{2})=\frac{1}{{\cal M}^{2}}
   e^{-{\bf k}^{2}/(4m_{P,O}^{2})}.
\label{Eq:difdel}
\end{equation}
In the latter expression ${\cal M}$ is a universal
scaling mass, which is again taken to be the proton mass.
The mass parameter $m_{P}$ controls the ${\bf k}^{2}$-dependence of
the Pomeron-, $f$-, $f'$-, $A_{2}$-, and $K^{\star\star}$-potentials.
Similarly, $m_O$ controls the ${\bf k}^2$-dependence of the Odderon.\\

\noindent {\it In the following we give the OBE-potentials in momentum-space for the 
hyperon-nucleon systems. From these those for NN and YY can be deduced easily.
We assign the particles 1 and 3 to be hyperons, and particles 2 and 4 to be
nucleons. Mass differences among the hyperons and among the nucleons will be neglected.}\\

\onecolumngrid

For pseudo-scalar mesons, the graph's of Fig.~\ref{obefig} give for the 
 potential $ V({\bf k},{\bf q}) \approx K^{(2)}_{PS}({\bf p}',{\bf p}|W) $ 
\begin{eqnarray}
 V_{PS}({\bf k},{\bf q}) & = & -\frac{f_{13}f_{24}}{m_\pi^2}\
 \left(1-\frac{({\bf q}^2+{\bf k}^2/4)}{2M_YM_N}\right)\cdot 
 \left[
 \frac{1}{2\omega}\left\{\frac{1}{\omega + a}+\frac{1}{\omega -a}\right\}
 (\mbox{\boldmath $\sigma$}_1\cdot{\bf k})(\mbox{\boldmath $\sigma$}_2\cdot{\bf k})
 \right.\nonumber\\ && \hspace{-2.0cm} \left.
 +\frac{1}{M_Y+M_N}\left\{\frac{1}{\omega + a}-\frac{1}{\omega -a}\right\}
 (\mbox{\boldmath $\sigma$}_1\cdot{\bf q}\ \mbox{\boldmath $\sigma$}_2\cdot{\bf k} 
 -\mbox{\boldmath $\sigma$}_1\cdot{\bf k}\ \mbox{\boldmath $\sigma$}_2\cdot{\bf q})
 \right]
 \exp\left(-{\bf k}^2/\Lambda^2\right).
 \label{eq:psx1}\end{eqnarray}
Here, using the on-energy-shell approximation $E_1+E_2= E_3+E_4$, we have                      
\begin{eqnarray*}
 a &=& E_1+E_4 -W = \frac{1}{2}\left(\vphantom{\frac{A}{A}} E_1+E_4 -E_2-E_3\right) \\
  &\approx& \Delta M + \frac{1}{4}\Delta M\left(\frac{1}{M_1M_3}+\frac{1}{M_2M_4}\right)
 \left({\bf q}^2+{\bf k}^2/4\right), 
\end{eqnarray*}
where $\Delta M = (M_1+M_4-M_3-M_2)/2$, 
and we neglected the ${\bf q}\cdot{\bf k}$-term which is of order 
$(M_Y-M_N)/2M_YM_N$. Henceforth we neglect the 
non-adiabatic effects, i.e. $a \approx \Delta M$, in the OBE-potentials, 
except for the $P_8$-terms, where the leading term is proportional to $a$.
One notices that the $P_8$-term in (\ref{eq:psx1}) is only non-zero for K-exchange.

\subsection{Non-strange Meson-exchange}
For the non-strange mesons the mass differences at the vertices are neglected,
we take at the $YYM$- and the $NNM$-vertex the average hyperon and the average
nucleon mass respectively. This implies that we do not include contributions
to the Pauli-invariants $P_7$ and $P_8$.
 For vector-, and diffractive OBE-exchange we  
refer the reader to Ref.~\cite{MRS89}, where the contributions to the different
$\Omega^{(X)}_{i}$'s for baryon-baryon scattering are given in detail.
\begin{enumerate}
 \item[(a)]   Pseudoscalar-meson exchange:
      \begin{subequations}
      \begin{eqnarray}
       \Omega^{(P)}_{2a} & = & -g^P_{13}g^P_{24}\left( \frac{{\bf k}^{2}}
           {12M_{Y}M_{N}} \right) \ \ ,\ \ 
       \Omega^{(P)}_{3a}  =  -g^P_{13}g^P_{24}\left( \frac{1}
           {4M_{Y}M_{N}}  \right), \label{eq1a} \\
       \Omega^{(P)}_{2b} & = & +g^P_{13}g^P_{24}\left( \frac{{\bf k}^{2}}
           {24M_{Y}^2M_{N}^2} \right) \ \ ,\ \ 
       \Omega^{(P)}_{3b}  =  +g^P_{13}g^P_{24}\left( \frac{1}
           {8M_{Y}^2M_{N}^2}  \right). \label{eq1b}    
         \end{eqnarray}
      \end{subequations}
 \item[(b)]   Vector-meson exchange:
     \begin{eqnarray}  
       \Omega^{(V)}_{1a}&=&
   \left\{g^V_{13}g^V_{24}\left( 1-\frac{{\bf k}^{2}}{2M_{Y}M_{N}}\right)
           -g^V_{13}f^V_{24}\frac{{\bf k}^{2}}{4{\cal M}M_{N}} 
      -f^V_{13}g^V_{24}\frac{{\bf k}^{2}}{4{\cal M}M_{Y}}
 \vphantom{\frac{A}{A}}\right. \nonumber\\ && \left. \vphantom{\frac{A}{A}}
           +f^V_{13}f^V_{24}\frac{{\bf k}^{4}}
           {16{\cal M}^{2}M_{Y}M_{N}}\right\},\ \                 
    \Omega^{(V)}_{1b} =  g^V_{13}g^V_{24}\left(
    \frac{3}{2M_{Y}M_{N}}\right), \nonumber\\
  \Omega^{(V)}_{2a} &=& -\frac{2}{3} {\bf k}^{2}\,\Omega^{(V)}_{3a}, \ \ 
  \Omega^{(V)}_{2b}  =  -\frac{2}{3} {\bf k}^{2}\,\Omega^{(V)}_{3b}, 
 \nonumber\\
    \Omega^{(V)}_{3a}&=& \left\{
           (g^V_{13}+f^V_{13}\frac{M_{Y}}{{\cal M}})
           (g^V_{24}+f^V_{24}\frac{M_{N}}{{\cal M}}) 
          -f^V_{13}f^V_{24}\frac{{\bf k}^{2}}{8{\cal M}^{2}} \right\}
            /(4M_{Y}M_{N}), \nonumber\\                 
    \Omega^{(V)}_{3b}&=& -
           (g^V_{13}+f^V_{13}\frac{M_{Y}}{{\cal M}})
           (g^V_{24}+f^V_{24}\frac{M_{N}}{{\cal M}}) 
            /(8M_{Y}^2M_{N}^2), \nonumber\\                 
    \Omega^{(V)}_{4}&=&-\left\{12g^V_{13}g^V_{24}+8(g^V_{13}f^V_{24}+f^V_{13}g^V_{24})
           \frac{\sqrt{M_{Y}M_{N}}}{{\cal M}} 
     - f^V_{13}f^V_{24}\frac{3{\bf k}^{2}}{{\cal M}^{2}}\right\}
            /(8M_{Y}M_{N})              \nonumber\\
       \Omega^{(V)}_{5}&=&- \left\{
           g^V_{13}g^V_{24}+4(g^V_{13}f^V_{24}+f^V_{13}g^V_{24})
           \frac{\sqrt{M_{Y}M_{N}}}{{\cal M}}  
           +8f^V_{13}f^V_{24}\frac{M_{Y}M_{N}}{{\cal M}^{2}}\right\}
          /(16M_{Y}^{2}M_{N}^{2})        \nonumber\\
       \Omega^{(V)}_{6}&=&-\left\{(g^V_{13}g^V_{24}
           +f^V_{13}f^V_{24}\frac{{\bf k}^{2}}{4{\cal M}^{2}})
    \frac{(M_{N}^{2}-M_{Y}^{2})}{4M_{Y}^{2}M_{N}^{2}} 
      -(g^V_{13}f^V_{24}-f^V_{13}g^V_{24})
      \frac{1}{\sqrt{{\cal M}^{2}M_{Y}M_{N}}}\right\}.
 \nonumber\\
 \label{eq2}\end{eqnarray}
 \item[(c)]   Scalar-meson exchange:  \hspace{2em}
      \begin{eqnarray} 
      \Omega^{(S)}_{1} & = & 
      -g^S_{13} g^S_{24}
      \left( 1+\frac{{\bf k}^{2}}{4M_{Y}M_{N}} 
      -\frac{{\bf q}^2}{2M_YM_N}\right), \ \ 
      \Omega^{(S)}_{1b} =  
      +g^S_{13} g^S_{24} \frac{1}{2M_YM_N}
       \nonumber\\ &&\nonumber\\
      \Omega^{(S)}_{4}&=&
      -g^S_{13} g^S_{24} \frac{1}{2M_{Y}M_{N}},\ \
      \Omega^{(S)}_{5} = g^S_{13} g^S_{24}
        \frac{1}{16M_{Y}^{2}M_{N}^{2} } \nonumber\\ 
       \nonumber\\ &&\nonumber\\
      \Omega^{(S)}_{6}&=& -g^S_{13} g^S_{24}
        \frac{(M_{N}^{2}-M_{Y}^{2})}{4M_{Y}M_{N}}.
       \label{Eq:scal} \end{eqnarray}
\item[(d)] Axial-vector-exchange $J^{PC}=1^{++}$:
      \begin{eqnarray} 
      \Omega^{(A)}_{2a} & = & -g^A_{13}g^A_{24}\left[
         1-\frac{2{\bf k}^2}{3M_YM_N}\right]
         +\left[\left(g_{13}^A f_{24}^A\frac{M_N}{{\cal M}}
         +f_{13}^A g_{24}^A \frac{M_Y}{{\cal M}}\right)
         -f_{13}^A f_{24}^A \frac{{\bf k}^2}{2{\cal M}^2}\right]\
         \frac{{\bf k}^2}{6M_YM_N}
       \nonumber\\ && \nonumber\\
      \Omega^{(A)}_{2b} &=& 
        -g^A_{13}g^A_{24} \left(\frac{3}{2M_{Y}M_{N}}\right) 
        \nonumber\\ && \nonumber\\
      \Omega^{(A)}_{3}&=&
        -g^A_{13}g^A_{24} \left[\frac{1}{4M_{Y}M_{N}}\right]
         +\left[\left(g_{13}^A f_{24}^A\frac{M_N}{{\cal M}}
         +f_{13}^A g_{24}^A \frac{M_Y}{{\cal M}}\right)
         -f_{13}^A f_{24}^A \frac{{\bf k}^2}{2{\cal M}^2}\right]\
         \frac{1}{2M_YM_N}
       \nonumber\\ && \nonumber\\
	\Omega^{(A)}_{4}  &=&
     -g^A_{13}g^A_{24}   \left[\frac{1}{2M_{Y}M_{N}}\right] 
      \ \ ,\ \
      \Omega^{(A)}_{6} = 
     -g^A_{13}g^A_{24} \left[\frac{(M_{N}^{2}-M_{Y}^{2})}{4M_{Y}^2M_{N}^2}\right]
       \nonumber\\ && \nonumber\\
      \Omega^{(A)'}_{5} & = &
     -g^A_{13}g^A_{24}   \left[\frac{2}{M_{Y}M_{N}}\right] 
         \label{eq:axi1} \end{eqnarray}
Here, we used the B-field description with $\alpha_r=1$, see Appendix~\ref{app:C}.
 The detailed treatment of the potential proportional to $P_5'$, i.e. 
 with $\Omega_5^{(A)'}$, is given in Appendix~\ref{app:B}.
\item[(e)] Axial-vector mesons with $J^{PC}=1^{+-}$: 
      \begin{eqnarray} 
       \Omega^{(B)}_{2a} & = & +f^B_{13}f^B_{24}\frac{(M_N+M_Y)^2}{m_B^2}
       \left(1-\frac{{\bf k}^2}{4M_YM_N}\right)
       \left( \frac{{\bf k}^{2}}{12M_{Y}M_{N}} \right),\ \ 
       \Omega^{(B)}_{2b}   =   +f^B_{13}f^B_{24}\frac{(M_N+M_Y)^2}{m_B^2}
       \left( \frac{{\bf k}^{2}}{8M_{Y}^2M_{N}^2} \right)     
     \nonumber\\ 
       \Omega^{(B)}_{3a} & = & +f^B_{13}f^B_{24}\frac{(M_N+M_Y)^2}{m_B^2}
       \left(1-\frac{{\bf k}^2}{4M_YM_N}\right)
       \left( \frac{1}{4M_{Y}M_{N}} \right),\ \             
       \Omega^{(B)}_{3b}   =   +f^B_{13}f^B_{24}\frac{(M_N+M_Y)^2}{m_B^2}
       \left( \frac{3}{8M_{Y}^2M_{N}^2} \right). \nonumber\\     
     \label{eq:bxi1} \end{eqnarray}
 \item[(f)]   Diffractive-exchange (pomeron, $f, f', A_{2}$): \\
         The $\Omega^{D}_{i}$ are the same as for scalar-meson-exchange
         Eq.(\ref{Eq:scal}), but with
         $\pm g_{13}^{S}g_{24}^{S}$ replaced by
         $\mp g_{13}^{D}g_{24}^{D}$, and except for the zero in the form factor.
\item[(g)] Odderon-exchange:              
         The $\Omega^{O}_{i}$ are the same as for vector-meson-exchange
         Eq.(ref{eq2}), but with
         $ g_{13}^{V}\rightarrow g_{13}^{O}$, 
         $ f_{13}^{V}\rightarrow f_{13}^{O}$ and similarly for the couplings
         with the 24-subscript.

\end{enumerate}

As in Ref.~\cite{MRS89} in the derivation of the expressions for $\Omega_i^{(X)}$, 
given above, $M_{Y}$ and $M_{N}$ denote the mean hyperon and nucleon
mass, respectively \begin{math} M_{Y}=(M_{1}+M_{3})/2 \end{math}
and \begin{math} M_{N}=(M_{2}+M_{4})/2 \end{math},
 and $m$ denotes the mass of the exchanged meson.
Moreover, the approximation                            
        \begin{math}
              1/ M^{2}_{N}+1/ M^{2}_{Y}\approx
              2/ M_{N}M_{Y},
        \end{math}
is used, which is rather good since the mass differences
between the baryons are not large.\\

\noindent {\it The potentials for mesons with strangeness are given in 
paper II of this series.}

\subsection{One-Boson-Exchange Interactions in Configuration Space I}
\label{sect.IIIb}
In configuration space the BB-interactions are described by potentials
of the general form
\begin{subequations}
\begin{eqnarray}
 V &=& \left\{\vphantom{\frac{A}{A}} V_C(r) + V_\sigma(r)
\mbox{\boldmath $\sigma$}_1\cdot\mbox{\boldmath $\sigma$}_2
 + V_T(r) S_{12} + V_{SO}(r) {\bf L}\cdot{\bf S} + V_Q(r) Q_{12}
 \right.\nonumber\\ && \left.
 + V_{ASO}(r)\ \frac{1}{2}(\mbox{\boldmath $\sigma$}_1-
  \mbox{\boldmath $\sigma$}_2)\cdot{\bf L}
 -\frac{1}{2M_YM_N}\left(\vphantom{\frac{A}{A}} 
\mbox{\boldmath $\nabla$}^2 V^{n.l.}(r) + V^{n.l.}(r) 
 \mbox{\boldmath $\nabla$}^2\right)
\right\}\cdot P, \\
 V^{n.l.} &=& \left\{\vphantom{\frac{A}{A}} \varphi_C(r) + \varphi_\sigma(r)
\mbox{\boldmath $\sigma$}_1\cdot\mbox{\boldmath $\sigma$}_2
 + \varphi_T(r) S_{12}\right\}, 
 \label{eq:3b.a}\end{eqnarray}
\end{subequations}
where
\begin{subequations}
\begin{eqnarray}
 S_{12} &=& 3 (\mbox{\boldmath $\sigma$}_1\cdot\hat{r})
 (\mbox{\boldmath $\sigma$}_2\cdot\hat{r}) -
 (\mbox{\boldmath $\sigma$}_1\cdot\mbox{\boldmath $\sigma$}_2), \\
 Q_{12} &=& \frac{1}{2}\left[\vphantom{\frac{A}{A}} 
 (\mbox{\boldmath $\sigma$}_1\cdot{\bf L})(\mbox{\boldmath $\sigma$}_2\cdot{\bf L})
 +(\mbox{\boldmath $\sigma$}_2\cdot{\bf L})(\mbox{\boldmath $\sigma$}_1\cdot{\bf L})
 \right], \\
 \phi(r) &=& \phi_C(r) + \phi_\sigma(r) 
 \mbox{\boldmath $\sigma$}_1\cdot\mbox{\boldmath $\sigma$}_2, 
 \label{eq:3b.b}\end{eqnarray}
\end{subequations}
For the basic functions for the Fourier transforms with gaussian form factors,
we refer to Refs.~\cite{NRS78,MRS89}.                           
For the details of the Fourier transform for the potentials with $P_5'$, which 
occur in the case of the axial-vector mesons with $J^{PC}=1^{++}$, we refer 
to Appendix~\ref{app:B}. 

\noindent (a)\ Pseudoscalar-meson-exchange:
\begin{subequations}
\begin{eqnarray}
  V_{PS}(r) &=& \frac{m}{4\pi}\left[ g^P_{13}g^P_{24}\frac{m^2}{4M_YM_N}
 \left(\frac{1}{3}(\mbox{\boldmath $\sigma$}_1\cdot\mbox{\boldmath $\sigma$}_2)\
 \phi_C^1 + S_{12} \phi_T^0\right)\right] P, \\
  V_{PS}^{n.l.}(r) &=& \frac{m}{4\pi}\left[ g^P_{13}g^P_{24}\frac{m^2}{4M_YM_N}
 \left(\frac{1}{3}(\mbox{\boldmath $\sigma$}_1\cdot\mbox{\boldmath $\sigma$}_2)\
 \phi_C^1 + S_{12} \phi_T^0\right)\right] P. 
 \label{eq:3b.1}\end{eqnarray}
\end{subequations}
\noindent (b)\ Vector-meson-exchange:          
\begin{subequations}
\begin{eqnarray}
&& V_{V}(r) = \frac{m}{4\pi}\left[\left\{ g^V_{13}g^V_{24}\left[ \phi_C^0 +
 \frac{m^2}{2M_YM_N} \phi_C^1 
\right]
\right.\right.\nonumber\\ && \left.\left.  \hspace{0cm} 
 +\left[g^V_{13}f^V_{24}\frac{m^2}{4{\cal M}M_N}
 +f^V_{13}g^V_{24}\frac{m^2}{4{\cal M}M_Y}\right] \phi_C^1 + f^V_{13}f^V_{24}
\frac{m^4}{16{\cal M}^2 M_Y M_N} \phi_C^2\right\}
  \right.\nonumber\\ && \left.  \hspace{0cm} 
 +\frac{m^2}{6M_YM_N}\left\{\left[ \left(g^V_{13}+f^V_{13}\frac{M_Y}{{\cal M}}\right)\cdot
 \left(g^V_{24}+f^V_{24}\frac{M_N}{{\cal M}}\right)\right] \phi_C^1 
 +f^V_{13} f^V_{24}\frac{m^2}{8{\cal M}^2} \phi_C^2\right\}
 (\mbox{\boldmath $\sigma$}_1\cdot\mbox{\boldmath $\sigma$}_2)\
  \right.\nonumber\\ && \left.  \hspace{0cm} 
 -\frac{m^2}{4M_YM_N}\left\{\left[ \left(g^V_{13}+f^V_{13}\frac{M_Y}{{\cal M}}\right)\cdot
 \left(g^V_{24}+f^V_{24}\frac{M_N}{{\cal M}}\right)\right] \phi_T^0 
 +f^V_{13} f^V_{24}\frac{m^2}{8{\cal M}^2} \phi_T^1\right\} S_{12}
  \right.\nonumber\\ && \left.  \hspace{0cm} 
 -\frac{m^2}{M_YM_N}\left\{\left[ \frac{3}{2}g^V_{13}g^V_{24}
 +\left(g^V_{13}f^V_{24}+f^V_{13}g^V_{24}\right)
 \frac{\sqrt{M_YM_N}}{{\cal M}}\right] \phi_{SO}^0 
 +\frac{3}{8}f^V_{13} f^V_{24}\frac{m^2}{{\cal M}^2} \phi_{SO}^1\right\} {\bf L}\cdot{\bf S}
  \right.\nonumber\\ && \left.  \hspace{0cm} 
 +\frac{m^4}{16M_Y^2M_N^2}\left\{\left[ g^V_{13}g^V_{24}
 +4\left(g^V_{13}f^V_{24}+f^V_{13}g^V_{24}\right)
 \frac{\sqrt{M_YM_N}}{{\cal M}} 
 +8f^V_{13}f^V_{24}\frac{M_YM_N}{{\cal M}^2}\right]\right\} 
  \cdot\right.\nonumber\\ && \left.  \hspace{0cm} \times
\frac{3}{(mr)^2} \phi_T^0 Q_{12}
 -\frac{m^2}{M_YM_N}\left\{\left[ 
 \left(g^V_{13}g^V_{24}-f^V_{13}f^V_{24}\frac{m^2}{{\cal M}^2}\right)
  \frac{(M_N^2-M_Y^2)}{4M_YM_N}
  \right.\right.\right.\nonumber\\ && \left.\left.\left.  \hspace{0cm} 
  -\left(g^V_{13}f^V_{24}-f^V_{13}g^V_{24}\right)\frac{\sqrt{M_YM_N}}{{\cal M}}\right] \phi_{SO}^0
 \right\}\cdot\frac{1}{2}\left(
 \mbox{\boldmath $\sigma$}_1-\mbox{\boldmath $\sigma$}_2\right)\cdot{\bf L}\right] P,
\\
&& V_{V}^{n.l.}(r) = \frac{m}{4\pi}\left[ \frac{3}{2} g^V_{13}g^V_{24}\ \phi_C^0 
  \right.\nonumber\\ && \left.  \hspace{0cm} 
 +\frac{m^2}{6M_YM_N}\left\{\left[ \left(g^V_{13}+f^V_{13}\frac{M_Y}{{\cal M}}\right)\cdot
 \left(g^V_{24}+f^V_{24}\frac{M_N}{{\cal M}}\right)\right] \phi_C^1 \right\}
 (\mbox{\boldmath $\sigma$}_1\cdot\mbox{\boldmath $\sigma$}_2)\
  \right.\nonumber\\ && \left.  \hspace{0cm} 
 -\frac{m^2}{4M_YM_N}\left\{\left[ \left(g^V_{13}+f^V_{13}\frac{M_Y}{{\cal M}}\right)\cdot
 \left(g^V_{24}+f^V_{24}\frac{M_N}{{\cal M}}\right)\right] \phi_T^0 \right\} S_{12}
\right]. 
 \label{eq:3b.2}\end{eqnarray}
\end{subequations}
Note: the spin-spin and tensor non-local terms are not included in ESC08c.\\

\noindent (c)\ Scalar-meson-exchange:          
\begin{eqnarray}
 V_{S}(r) &=& -\frac{m}{4\pi}\left[ \hat{g}^S_{13}\hat{g}^S_{24}\left\{\left[ \phi_C^0 
 -\frac{m^2}{4M_YM_N} \phi_C^1\right] + \frac{m^2}{2M_YM_N} \phi_{SO}^0\ {\bf L}\cdot{\bf S}
 +\frac{m^4}{16M_Y^2M_N^2}
 \cdot\right.\right.\nonumber\\ && \left.\left. \times
\frac{3}{(mr)^2} \phi_T^0 Q_{12} 
 +\frac{m^2}{M_YM_N} \left[\frac{(M_N^2-M_Y^2)}{4M_YM_N}\right] \phi_{SO}^0\cdot
 \frac{1}{2}\left(\mbox{\boldmath $\sigma$}_1-\mbox{\boldmath $\sigma$}_2\right)\cdot{\bf L}
  \right.\right.\nonumber\\ && \left.\left.  \hspace{0.0cm}
  +\frac{1}{4M_YM_N}\left(\mbox{\boldmath $\nabla$}^2 \phi_C^0 
 + \phi_C^0 \mbox{\boldmath $\nabla$}^2\right) \right\}\right] P,
 \label{eq:3b.3}\end{eqnarray}
where
\begin{equation}
 \hat{g}^S_{13}= g^S_{13}-i\frac{M_3-M_1}{{\cal M}} f^S_{13}\ \ ,\ \
 \hat{g}^S_{24}= g^S_{24}-i\frac{M_4-M_2}{{\cal M}} f^S_{24}.        
 \label{eq:3b.3a}\end{equation}
\noindent (d)\ Axial-vector-meson exchange $J^{PC}=1^{++}$:
\begin{eqnarray}
&& V_{A}(r) = -\frac{m}{4\pi}\left[ 
 \left\{ g^A_{13}g^A_{24}\left(\phi_C^0 +\frac{2m^2}{3M_YM_N} \phi_C^1\right)
 +\frac{m^2}{6M_YM_N}\left(g^A_{13}f^A_{24}\frac{M_N}{{\cal M}}
 +f^A_{13}g^A_{24}\frac{M_Y}{{\cal M}}\right)\phi_C^1
\right.\right.\nonumber\\ && \left.\left.
 +f^A_{13}f^A_{24}\frac{m^4}{12M_YM_N{\cal M}^2}\phi_C^2\right\}
 (\mbox{\boldmath $\sigma$}_1\cdot\mbox{\boldmath $\sigma$}_2)
  -\frac{3}{4M_YM_N} g^A_{13}g^A_{24}\left(\mbox{\boldmath $\nabla$}^2 \phi_C^0 
 + \phi_C^0 \mbox{\boldmath $\nabla$}^2\right) 
 (\mbox{\boldmath $\sigma$}_1\cdot\mbox{\boldmath $\sigma$}_2)
 \right.\nonumber\\ && \left. 
 - \frac{m^2}{4M_YM_N}\left\{\left[g^A_{13}g^A_{24}-2\left(g^A_{13}f^A_{24}
 \frac{M_N}{{\cal M}}+f^A_{13}g^A_{24}\frac{M_Y}{{\cal M}}\right)\right] \phi_T^0
 -f^A_{13}f^A_{24}\frac{m^2}{{\cal M}^2} \phi_T^1\right\} S_{12}
 \right.\nonumber\\ && \left. 
 +\frac{m^2}{2M_YM_N}g^A_{13}g^A_{24} \left\{\phi_{SO}^0\ {\bf L}\cdot{\bf S}
 +\frac{m^2}{M_YM_N} \left[\frac{(M_N^2-M_Y^2)}{4M_YM_N}\right] \phi_{SO}^0\cdot
 \frac{1}{2}\left(\mbox{\boldmath $\sigma$}_1-\mbox{\boldmath $\sigma$}_2\right)\cdot{\bf L}
 \right\}\right] P.     
 \label{eq:3b.4}\end{eqnarray}
\noindent (e)\ Axial-vector-meson exchange $J^{PC}=1^{+-}$:
\begin{subequations}
\begin{eqnarray}
 V_{B}(r) &=& -\frac{m}{4\pi}\frac{(M_N+M_Y)^2}{m^2}\left[ 
 f^B_{13}f^B_{24}\left\{\frac{m^2}{12M_YM_N}\left(\phi_C^1+
 \frac{m^2}{4M_YM_N} \phi_C^2\right)
 (\mbox{\boldmath $\sigma$}_1\cdot\mbox{\boldmath $\sigma$}_2)
 \right.\right.\nonumber\\ && \left.\left. 
  -\frac{m^2}{8M_YM_N}\left(\mbox{\boldmath $\nabla$}^2 \phi_C^1 
 + \phi_C^1 \mbox{\boldmath $\nabla$}^2\right) 
 (\mbox{\boldmath $\sigma$}_1\cdot\mbox{\boldmath $\sigma$}_2)
 +\left[\frac{m^2}{4M_YM_N}\right] \phi^0_T\ S_{12}\right\}\right] P, \\
 V_{B}^{n.l.}(r) &=& -\frac{m}{4\pi}\frac{(M_N+M_Y)^2}{m^2}\left[ 
 f^B_{13}f^B_{24}\left\{
  \frac{3m^2}{4M_YM_N} \left(\frac{1}{3}  
  \mbox{\boldmath $\sigma$}_1\cdot\mbox{\boldmath $\sigma$}_2\ \phi_C^1
  + S_{12}\ \phi_T^0\right)\right\}\right] P. 
 \label{eq:3b.5}\end{eqnarray}
\end{subequations}
\noindent (f)\ Diffractive exchange:           
\begin{eqnarray}
&& V_{D}(r) = \frac{m_P}{4\pi}\left[ g^D_{13}g^D_{24} 
 \frac{4}{\sqrt{\pi}}\frac{m_P^2}{{\cal M}^2}\cdot\left[\left\{
 1+\frac{m_P^2}{2M_YM_N}(3-2 m_P^2r^2) + \frac{m_P^2}{M_YM_N} {\bf L}\cdot{\bf S}
  \right.\right.\right.\nonumber\\ && \left.\left.\left.  \hspace{0.75cm} 
 +\left(\frac{m_P^2}{2M_YM_N}\right)^2 Q_{12}
 +\frac{m_P^2}{M_YM_N} \left[\frac{(M_N^2-M_Y^2)}{4M_YM_N}\right]\cdot
 \frac{1}{2}\left(\mbox{\boldmath $\sigma$}_1-\mbox{\boldmath $\sigma$}_2\right)\cdot{\bf L}
 \right\}\ e^{-m_P^2r^2} 
  \right.\right.\nonumber\\ && \left.\left.  \hspace{1.5cm} 
  +\frac{1}{4M_YM_N}\left(\mbox{\boldmath $\nabla$}^2 e^{-m_P^2r^2} 
 + e^{-m_P^2r^2}\mbox{\boldmath $\nabla$}^2\right) \right]\right] P.
 \label{eq:3b.6}\end{eqnarray}
\noindent (g)\ Odderon-exchange:                      
\begin{subequations}
\begin{eqnarray}
 V_{O,C}(r) &=& +\frac{g^O_{13}g^O_{24}}{4\pi}\frac{8}{\sqrt{\pi}}\frac{m_O^5}{{\cal M}^4}
 \left[\left(3-2m_O^2r^2\right) \right.\nonumber\\ && \left. 
 -\frac{m_O^2}{M'M}\left( 15 - 20 m_O^2r^2+4 m_O^4r^4\right)
 \right]\exp(-m_O^2 r^2)\ , \\
 V_{O,n.l.}(r) &=& -\frac{g^O_{13}g^O_{24}}{4\pi}\frac{8}{\sqrt{\pi}}\frac{m_O^5}{{\cal M}^4}
 \frac{3}{4M'M}\left\{\mbox{\boldmath $\nabla$}^2
 \left[(3-2m_O^2r^2)\exp(-m_O^2 r^2)\right]+ \right.\nonumber\\
 && \left. + \left[(3-2m_O^2r^2)\exp(-m_O^2 r^2)\right] 
 \mbox{\boldmath $\nabla$}^2 \right\}\ , \\
 V_{O,\sigma}(r) &=& 
-\frac{g^O_{13}g^O_{24}}{4\pi}\frac{8}{3\sqrt{\pi}}\frac{m_O^5}{{\cal M}^4}
 \frac{m_O^2}{M_YM_N}
 \left[15-20 m_O^2r^2+4 m_O^4 r^4\right]\exp(-m_O^2 r^2)\cdot \nonumber\\
 && \times\left(1+\kappa^O_{13}\frac{M_Y}{\cal M}\right) 
 \left(1+\kappa^O_{24}\frac{M_N}{\cal M}\right) 
 , \\
 V_{O,T}(r) &=& -\frac{g^O_{13}g^O_{24}}{4\pi}\frac{8}{3\sqrt{\pi}}\frac{m_O^5}{{\cal M}^4}
 \frac{m_O^2}{M_YM_N}\cdot m_O^2 r^2
 \left[7-2 m_O^2r^2\right]\exp(-m_O^2 r^2)\cdot \nonumber\\
 && \times\left(1+\kappa^O_{13}\frac{M_Y}{\cal M}\right) 
 \left(1+\kappa^O_{24}\frac{M_N}{\cal M}\right) 
 , \\
 V_{O,SO}(r) &=& -\frac{g^O_{13}g^O_{24}}{4\pi}\frac{8}{\sqrt{\pi}}\frac{m_O^5}{{\cal M}^4}
 \frac{m_O^2}{M_YM_N} \left[5-2 m_O^2r^2\right]\exp(-m_O^2 r^2)\cdot \nonumber\\
 && \times\left\{3+\left(\kappa^O_{13}+\kappa^O_{24}\right)\frac{\sqrt{M_YM_N}}{\cal M}\right\}
 , \\
 V_{O,Q}(r) &=& +\frac{g^O_{13}g^O_{24}}{4\pi}\frac{2}{\sqrt{\pi}}\frac{m_O^5}{{\cal M}^4}
 \frac{m_O^4}{M_Y^2M_N^2} 
 \left[7-2 m_O^2r^2\right]\exp(-m_O^2 r^2)\cdot \nonumber\\
 && \times\left\{1+4\left(\kappa^O_{13}+\kappa^O_{24}\right)\frac{\sqrt{M_YM_N}}{\cal M}
 +8\kappa_{13}\kappa_{24}\frac{M_YM_N}{{\cal M}^2}\right\}
 , \\
 V_{O,ASO}(r) &=& -\frac{g^O_{13}g^O_{24}}{4\pi}\frac{4}{\sqrt{\pi}}\frac{m_O^5}{{\cal M}^4}
 \frac{m_O^2}{M_YM_N} \left[5-2 m_O^2r^2\right]\exp(-m_O^2 r^2)\cdot \nonumber\\
 &&  \times\left\{ \frac{M_N^2-M_Y^2}{M_YM_N}
-4\left(\kappa^O_{24}-\kappa^O_{13}\right)
 \frac{\sqrt{M_YM_N}}{\cal M} \right\}\ .
 \label{eq:3b.7}\end{eqnarray}
\end{subequations}

\subsection{One-Boson-Exchange Interactions in Configuration Space II}
\label{sect.IIIc}
Here we give the extra potentials due to the 
zero's in the scalar and axial-vector form factors.
\begin{enumerate}
\item[a)] Again, for $X=V,D$ we refer to the configuration space potentials 
in Ref.~\cite{MRS89}. For $X=S$ we give here the additional terms w.r.t. those 
in \cite{MRS89}, which are due to the zero in the scalar form factor. They are

\begin{eqnarray}
 &&  \Delta V_{S}(r) = - \frac{m}{4\pi}\ \frac{m^2}{U^2}\
 \left[ g^S_{13} g^S_{24}\left\{
 \left[\phi^1_C - \frac{m^2}{4M_YM_N} \phi^2_C\right]
+\frac{m^2}{2M_YM_N}\phi^1_{SO}\ {\bf L}\cdot{\bf S}
 \right.\right. \nonumber\\ && \left.\left.
+\frac{m^4}{16M_Y^2M_N^2}\phi^1_T\ Q_{12}  
 + \frac{m^2}{4M_YM_N}\frac{M_N^2-M_Y^2}{M_YM_N}\ \phi^{(1)}_{SO}\cdot
 \frac{1}{2}(\mbox{\boldmath $\sigma$}_1-\mbox{\boldmath $\sigma$}_2)\cdot{\bf L}
  \right\}\right]\ . \nonumber\\
 \label{eq:3.15}\end{eqnarray}


\item[b)] For the axial-vector mesons, the configuration space potential 
 corresponding to (\ref{eq:axi1}) is             
\begin{eqnarray}
 &&  V_{A}^{(1)}(r) = - \frac{g_{A}^{2}}{4\pi}\ m  \left[
 \phi^{0}_{C}\ (\mbox{\boldmath $\sigma$}_1\cdot\mbox{\boldmath $\sigma$}_2) 
  -\frac{1}{12M_YM_N}
  \left( \nabla^{2} \phi^{0}_{C}+\phi^{0}_{C}\nabla^{2}\right)
 (\mbox{\boldmath $\sigma$}_1\cdot\mbox{\boldmath $\sigma$}_2) 
 \right. \nonumber \\ && \nonumber \\ & & \left. \hspace*{1.4cm}
   + \frac{3m^{2}}{4M_YM_N}\ \phi^{0}_{T}\ S_{12}
 +\frac{m^{2}}{2M_YM_N}\ \phi^{0}_{SO}\ {\bf L}\cdot{\bf S}
 \right. \nonumber \\ && \nonumber \\ & & \left. \hspace*{1.4cm}
 + \frac{m^2}{4M_YM_N}\frac{M_N^2-M_Y^2}{M_YM_N}\ \phi^{(0)}_{SO}\cdot
 \frac{1}{2}(\mbox{\boldmath $\sigma$}_1-\mbox{\boldmath $\sigma$}_2)\cdot{\bf L}
  \right]\ .
 \label{eq:3.16}\end{eqnarray}
The extra contribution to the potentials coming from the zero in the axial-vector
meson form factor are obtained from the expression (\ref{eq:3.16}) by making 
substitutions as follows
\begin{eqnarray}
   \Delta V_{A}^{(1)}(r) &=&  V_{A}^{(1)}\left(\phi_C^0 \rightarrow \phi_C^1,
 \phi_T^0 \rightarrow \phi_T^1, \phi_{SO}^0 \rightarrow \phi_{SO}^1\right)
 \cdot\frac{m^2}{U^2}\ .
\label{eq:3.17}\end{eqnarray}
Note that we do not include the similar $\Delta V_A^{(2)}(r)$ since they involve
${\bf k}^4$-terms in momentum-space. 
\end{enumerate}


\subsection{PS-PS-exchange Interactions in Configuration Space}                       
\label{sect.d}
In Fig.~\ref{bwfig} and Fig.~\ref{tmofig} the included two-meson exchange 
graphs are shown schematically. Explicit expressions for  
$K^{irr}(BW)$ and $K^{irr}(TMO)$ were derived \cite{Rij91}, where also the 
terminology BW and TMO is explained.
The TPS-potentials for nucleon-nucleon have been given in detail in \cite{RS96a,RS96b}
The generalization to baryon-baryon is similar to that for the OBE-potentials.
So, we substitute $M \rightarrow \sqrt{M_YM_N}$, and include all PS-PS 
possibilities with coupling constants as in the OBE-potentials. 
As compared to nucleon-nucleon in \cite{RS96a,RS96b} here we have in addition 
the potentials with double K-exchange.  The masses
are the physical pseudo-scalar meson masses. For the intermediate two-baryon
states we take into account of the different thresholds.
We have not included uncorrelated PS-vector, PS-scalar, or PS-diffractive 
exchange. This because the range of these potentials is similar to 
that of the vector-,scalar-,and axial-vector-potentials. Moreover, for 
potentially large potentials, in particularly those with scalar mesons involved,
there will be very strong cancellations between the planar- and crossed-box
contributions.

\subsection{MPE-exchange Interactions}
In Fig.~\ref{pairfig} both the one-pair graphs and the two-pair graphs are shown.
In this work we include only the one-pair graphs. The argument for neglecting 
the two-pair graph is to avoid some 'double-counting'. Viewing the pair-vertex 
as containing heavy-meson exchange means that the contributions from $\rho(750)$
and $\epsilon=f_0(760)$ to the two-pair graphs is already accounted for by 
our treatment of the broad $\rho$ and $\epsilon$ OBE-potential.
For a more complete discussion of the physics behind MPE we refer to our 
previous papers \cite{Rij93,RS96a,RS96b}.
The MPE-potentials for nucleon-nucleon have been given in Ref.~\cite{RS96a,RS96b}.
The generalization to baryon-baryon is similar to that for the TPS-potentials.
For the intermediate two-baryon
states we neglect the different two-baryon thresholds. This because,
although in principle possible, it complicates the computation of the 
potentials considerably. 
For a proper appreciation of the physics it is useful to scale the 
phenomenological meson-pair baryon-baryon interaction Hamiltonians different from
the originally used scalings \cite{RS96a,RS96b}. Below we give these Hamiltonians:
\begin{subequations}
\begin{eqnarray}
&& {\cal H}_S = \bar{\psi}\psi \left[g_{(\pi\pi)_0}
\mbox{\boldmath $\pi$}\cdot\mbox{\boldmath $\pi$} + 
 g_{(\sigma\sigma)}\sigma^2\right]/
{\cal M}, \label{eq:3.17a}\\
&& {\cal H}_V = g_{(\pi\pi)_1}\left[\bar{\psi}\gamma_\mu\mbox{\boldmath $\tau$}\psi\right]\cdot
\left(\mbox{\boldmath $\pi$}\times\partial^\mu 
\mbox{\boldmath $\pi$}/m_\pi\right)/{\cal M}
 \nonumber\\ && \hspace{1cm} 
 -\frac{f_{(\pi\pi)_1}}{2M}
\left[\bar{\psi}\sigma_{\mu\nu}\mbox{\boldmath $\tau$}\psi\right]\partial^\nu\cdot\left(
\mbox{\boldmath $\pi$}\times\partial^\mu \mbox{\boldmath $\pi$}/m_\pi\right)/{\cal M},
\label{eq:3.17b}\\
&& {\cal H}_A = g_{(\pi\rho)_1}\left[\bar{\psi}\gamma_5\gamma_\mu
 \mbox{\boldmath $\tau$}\psi\right]\cdot
\mbox{\boldmath $\pi$}\times\mbox{\boldmath $\rho$}/{\cal M},\label{eq:3.17c}\\
&& {\cal H}_B = i g_{(\pi\omega)}\left[\bar{\psi}\gamma_5\sigma_{\mu\nu}
 \mbox{\boldmath $\tau$}\psi\right]\cdot\partial^\nu\left(
\mbox{\boldmath $\pi$}\phi_\omega^\mu\right)/(m_\pi {\cal M}),\label{eq:3.17d}\\
&& {\cal H}_P = g_{(\pi\sigma)}\left[\bar{\psi}\gamma_5\gamma_\mu
 \mbox{\boldmath $\tau$}\psi\right]\cdot
\left(\mbox{\boldmath $\pi$}\partial^\mu\sigma - 
 \sigma\partial^\mu\mbox{\boldmath $\pi$}\right)/(m_\pi{\cal M}).
\label{eq:3.17e}\end{eqnarray}
\end{subequations}
Here, we systematically scaled the partial derivatives with $m_\pi$.\\
The generalization of the pair-couplings to baryon-baryon is described in 
Ref.~\cite{Rij04b}, section III.
Also here in $N\!N$, we have in addition to  \cite{RS96a,RS96b}  
included the pair-potentials with KK-, KK*-, and K$\kappa$-exchange.
The convention for the MPE coupling constants is the same as in Ref.~\cite{RS96a,RS96b}.

\subsection{The Schr\"{o}dinger equation with Non-local potential}
\label{sect.5}
The non-local potentials are of the central-, spin-spin, and tensor type. The method 
of solution of the Schr\"{o}dinger equation for nucleon-nucleon is described in 
Ref.~\cite{NRS78}. Here, the non-local tensor is in momentum space 
of the form ${\bf q}^2\ \tilde{v}_T({\bf k})$. 

\section{ ESC-couplings and the QPC-model}                     
\label{sec:4} 
In the ESC-model for baryon-baryon the meson-baryon couplings are in principle
only restricted by the requirements of relativistic covariance, time-reversal and
parity. However, dynamical input based on e.g. QCD, the QM, chiral-symmetry, and
flavor SU(3), is essential in order to be able to link the NN-, YN-, and YY-systems.
It appeared that in the ESC-model the $^3P_0$ quark-antiquark pair-creation
model \cite{Mic69,LeY73} leads to a scheme for the meson-baryon-baryon couplings which
is very similar to that found in the fits of the ESC-model \cite{Rij04a,Rij04b}.
The couplings found in the ESC08-model fit very well in the $(^3P_0+^3S_1)$-scheme
with a ratio $^3P_0/^3S_1 = 2:1$.

\subsection{ QPC-model Coupling Non-strange Mesons}                     
\label{sec:4a} 
 According to the Quark-Pair-Creation (QPC) model, in the $^3P_0$-version
 \cite{Mic69,LeY73}, the
baryon-baryon-meson couplings are given in terms of the quark-pair creation 
constant $\gamma_M$, and the radii of the (constituent) gaussian quark wave
functions, by \cite{LeY73,LeY75}
\begin{equation}
  g_{BBM}(\pm) = \gamma_{q\bar{q}}\ \frac{3}{\sqrt{2}}\ \pi^{-3/4}\ 
  X_M\left(I_M,L_M,S_M,J_M\right)\ F^{(\pm)}_M\ ,
\label{eq:qpc.1}\end{equation}
where $\pm = -(-)^{L_f}$ with $L_f$ is the orbital angular momentum of the final
BM-state, $X_M(\ldots)$ is a isospin, spin etc. recoupling coefficient, and 
\begin{eqnarray}
 F^{(+)} &=& \frac{3}{2}\ \left(m_MR_M\right)^{+1/2}\ (\Lambda_{QPC}R_M)^{-2}, 
 \nonumber\\ 
 F^{(-)} &=& \frac{3}{2}\ \left(m_MR_M\right)^{-1/2}\ (\Lambda_{QPC}R_M)^{-2}\cdot
 3\sqrt{2} (M_M/M_B). 
\label{eq:qpc.2}\end{eqnarray}
are coming from the overlap integrals, see Appendix~\ref{app:D}.
Here, the superscripts $\mp$ refer to the parity of the mesons $M$: $(-)$ for 
$J^{PC}=0^{+-}, 1^{--}$, and $(+)$ for $J^{PC}= 0^{++}, 1^{++}$. The radii of the 
baryons, in this case nucleons, and the mesons are respectively denoted by $R_B$ and $R_M$.
 
The QPC($^3P_0$)-model gives several interesting relations, such as  
$g_\omega = 3 g_\rho, g_\epsilon = 3 g_{a_0}$, and              
 $g_{a_0} \approx g_\rho,  g_\epsilon \approx g_\omega$.              
These relations can be seen most easily by applying the Fierz-transformation
to the $³P_0$-pair-creation Hamiltonian, see Appendixref{app:D}.

From $\rho \rightarrow e^+ e^-$, employing the current-field-identities (C.F.I's)
one can derive, see for example \cite{Roy67}, the following relation with the QPC-model      
\begin{equation}
 f_\rho = \frac{m_\rho^{3/2}}{\sqrt{2}|\psi_\rho(0)|} \Leftrightarrow
 \gamma\left(\frac{2}{3\pi}\right)^{1/2}
 \frac{m_\rho^{3/2}}{|'\psi_\rho(0)'|}\ ,           
\label{eq:gam0}\end{equation}
which, neglecting the difference between the wave functions on the left and
right hand side, gives for the pair creation constant 
$\gamma \rightarrow \gamma_0=\frac{1}{2}\sqrt{3\pi} =1.535$. However, 
since in the QPC-model gaussian wave functions are used, the $q\bar{q}$-potential
is a harmonic-oscillator one. This does not account for the $1/r$-behavior,
due to one-gluon-exchange (OGE), at short distance. This implies a OG-correction
\cite{LP96} to the wave function, which gives for $\gamma$ \cite{Chai80}
\begin{equation}
 \gamma = \gamma_0 \left(1-\frac{16}{3}\frac{\alpha(m_M)}{\pi}\right)^{-1/2}\ .
\label{eq:gam}\end{equation}
In Table~\ref{tab.gam} $\gamma(\mu)$ is shown,
Using from \cite{PDG02} the parameterization
\begin{equation}
 \alpha_s(\mu) = 4\pi/\left(\beta_0\ln(\mu^2/\Lambda_{QCD}^2)\right)\ ,
\label{eq:alphas}\end{equation}
with $\Lambda_{QCD} = 100$ MeV and $\beta_0 = 11-\frac{2}{3} n_f$ for $n_f=3$, 
\begin{table}[hbt]
 \caption{Pair-creation constant $\gamma$ as a function of $\mu$. }       
\begin{center}
\begin{tabular}{c|c|c} \hline & \\
  $\mu$ [GeV] & $\alpha_s(\mu)$   & $\gamma(\mu)$     \\
        &        &               \\
\hline
        &        &               \\
  $\infty$ &  0.00         &  1.535        \\
  80.0  &  0.10         &  1.685        \\
  35.0  &  0.20         &  1.889        \\
  1.05  &  0.30         &  2.191        \\
  0.55  &  0.40         &  2.710        \\
  0.40  &  0.50         &  3.94         \\
  0.35  &  0.55         &  5.96         \\
        &        &               \\
\hline
\end{tabular}
\end{center}
\label{tab.gam}
\end{table}
and taking the typical scale $m_M \approx 1$ GeV, the above formula gives 
$\gamma = 2.19$. This value we will use later
when comparing the QPC-model predictions and the ESC08c-model
coupling constants.\\
The formulas (\ref{eq:qpc.2}) are valid for the most simple QPC-model.
For a realistic description of the coupling constants
of the ESC08-model we include two sophistocations: (i) inclusion of both
the $^3P_0$- and the $^3S_1$-mechanism, (ii) inclusion of SU(6)-breaking.
For details, see \cite{THAR11}.
For the latter we use the (\underline{56}) and (\underline 70) SU(6)-irrep
mixing \cite{LeY75}, and a short-distance quark-gluon form factor.
In Table~\ref{tab:cc7} we show the $^3P_0-^3S_1$-model results and 
the values obtained in the ESC08c-fit. 
In this table we fixed $\gamma_M = 2.19$ for the vector-, scalar-, and 
axial-vector-mesons. 
From Table~\ref{tab.gam} 
one sees that at the scale of $m_M \approx 1$ GeV such a value is reasonable.
Here, one has to realize that the QPC-predictions are kind of "bare" couplings,
which allows vertex corrections from meson-exchange.
For the pseudo-scalar, a different value has to be used, showing indeed 
some 'running'-behavior as expected from QCD. 
In \cite{Chai80}, for the decays $\rho, \epsilon \rightarrow 2\pi$ etc. it was found
$\gamma =3.33$, which is close to our $\gamma_\pi=4.19$. 
For the mesonic decays of the charmonium states $\gamma=1.12$.
One notices the similarity between the QPC($^3P_0$)-model predictions 
and the fitted couplings. Here, for $f_1(1420)$
we have to take a larger radius $r_M=1.10$ fm in order to reduce the couplings
in the QPC-model. 
Of course, these results are sensitive to the $r_M$ values. We found that for all
solutions with a very good $\chi^2_{NN}$ the $r_M$ values varied by $\pm 0.2$ fm.

\noindent {\it The ESC08c-couplings and the QPC-couplings agree very well. 
In particularly,
the SU(6)-breaking is improving the agreement significantly. All this strengthens the
claim that the ESC08c-couplings are realistic ones.}\\
\begin{table}
 \caption{SU(6)-breaking in coupling constants, using (\underline{56}) and
(\underline{70})-irrep mixing with angle $\varphi = -22^o$ for the
$^3P_0$- and $^3S_1$-model. Gaussian Quark-gluon cut-off $\Lambda_{QQG}=986.6$ MeV.
Ideal mixing for vector and scalar meson nonets.
For pseudoscalar- and axial-nonets the mixing angles are $-13^0$ and $+50.0^o$ respectively,
imposing the OZI-rule.
Here, $\Lambda_{QPC} = 255.0$ MeV, $\gamma(\alpha_s=0.30)=2.19$ etc.       
The weights are A=0.697 and B=0.303 for the $^3P_0$ and $^3S_1$ respectively.
The values in parentheses in the column QPC denote the results for $\varphi=0^o$.
}
\begin{center}
\begin{ruledtabular}
\begin{tabular}{l|cc|c|c|c|c} \hline & & & & & & \\
  Meson          & $r_M[fm]$ & $\gamma_M$  
  & $^3S_1$ & $^3P_0$ & QPC & ESC08c \\
                 &       &       &          &         &   &   \\
\hline
                 &       &       &          &         &   &   \\
 $\pi(140)$      & 0.30  & 5.51 & $g=-2.74$ & $g=+6.31$ & 3.57 (3.77) & 3.65 \\
                 &       &       &          &         &   &   \\
 $\eta'(957)$    & 0.70  & 2.22 & $g=-2.49$ & $g=+5.72$ & 3.23 (3.92)  & 3.14 \\
                 &       &       &          &         &   &   \\
 $\rho(770)$     & 0.80  & 2.37 & $g=-0.17$ & $g=+0.80$ & 0.63 (0.77) & 0.65 \\
                 &       &       &          &         &   &   \\
 $\omega(783)$   & 0.70  & 2.35 & $g=-0.96$ & $g=+4.43$ & 3.47 (3.43) & 3.46 \\
                 &       &       &          &         &   &   \\
 $a_0(962)$      & 0.90  & 2.22 & $g=+0.19$ & $g=+0.43$ & 0.62 (0.64) & 0.59 \\
                 &       &       &          &         &   &   \\
 $\epsilon(760)$ & 0.70  & 2.37 & $g=+1.26$ & $g=+2.89$ & 4.15 (4.15) & 4.15 \\
                 &       &       &          &         &   &   \\
 $a_1(1270)$     & 0.70  & 2.09 & $g=-0.13$ & $g=-0.58$ & -0.71 (-0.71) & -0.79 \\
                 &       &       &          &         &   &   \\
 $f_1(1420)$     & 1.10  & 2.09 & $g=-0.14$ & $g=-0.66$ & -0.80 (-0.81)& -0.76\\
                 &       &       &          &         &   &   \\
\hline
\end{tabular}
\end{ruledtabular}
\end{center}
\label{tab:cc7}
\end{table}
 

\section{ ESC08-model: Fitting $NN\oplus YN \oplus YY$-data} 
\label{sec:5} 
In the simultaneous $\chi^2$-fit of the $NN$-, $YN$-, and YY-data a 
{\it single set of parameters} was used, which means the same parameters for all 
BB-channels.
The input $NN$-data are the same as in Ref.~\cite{Rij04a}, and we refer the reader 
to this paper for a description of the employed phase shift analysis \cite{Sto93,Klo93}. 
Note that in addition to the NN-phases, including their correlations, 
in the ESC08-models also the $NN$-low energy parameters 
and the deuteron binding energy are fitted. 
The YN-data are those used in Ref.~\cite{Rij04b} with the addition of higher
energy data, see paper II.
Of course, it is to be expected that the accurate and 
very numerous $NN$-data essentially fix most of the parameters. Only some of 
the parameters, for example certain $F/(F+D)$-ratios, are quite    
influenced by the $YN$-data. 
In the fitting procedure the following constraints are applied:
 (i) A strong restriction imposed on YN-models is the absence of S=-1 bound states.
 (ii) During the fitting process sometimes constraints are imposed 
in the form of 
 'pseudo-data' for some YN scattering lengths. These constraints are based 
 on experiences with Nijmegen YN-models in the past or to impose constraints
from the G-matrix results. In some cases it is necessary 
to add some extra weight of the YN-scattering data w.r.t. the NN-data in the
fitting process.
 (iii) After obtaining a solution for the
scattering data the corresponding model is tested by checking the corresponding 
G-matrix results for the well-depths for $U_\Sigma >0$ and $U_\Xi <0$, and
sufficient s-wave spin splitting in the $U_\Lambda$.
If not satisfactory we refit the scattering data etc. This iterative process 
implements the constraints from the G-matrix well-depth's results, 
and plays a vital role in obtaining the final results of the combined fit. 
(For the G-matrix approach to hyperon-nucleus systems, see e.g. Ref.~\cite{Yam10}.)
The fitting process is discussed more elaborately in paper II.

The $\chi^2$ is a very shallow function of the quark-core parameter.
Accordingly solutions have been obtained using different assumptions about the
quark-core-effects, all with a strength of about 20\% of the 
diffractive contribution.
In previous work \cite{PTP185.a}, models ESC08a and ESC08a'', the solutions were obtained 
by assuming quark-core effects only for the channels where the [51]-component is dominant: 
$\Sigma^+p(^3S_1,I=3/2), \Sigma N(^1S_0,I=1/2)$, and $\Xi N(^1S_0,I=1)$.
The solution ESC08c is obtained by application of the quark-core effects
according to equation~(8.4) in \cite{PTP185.a}, see paper II              
for a full description of the Pauli-blocking scheme.

Like in the $NN$-fit, described in Ref.~\cite{Rij04a}, also in the 
simultaneous $\chi^2$-fit of the $NN$- and $YN$-data,
it appeared again that the OBE-couplings could be constrained successfully
by the 'naive' predictions of the QPC-model \cite{Mic69,LeY73}. Although these 
predictions, see section \ref{sec:4}, are 'bare' ones, we tried to keep during the 
searches many OBE-couplings in the neighborhood of the QPC-values.
Also, it appeared that we could either fix the $F/(F+D)$-ratios 
to those as suggested by the QPC-model, 
or apply the same restraining strategy as for the OBE-couplings.                      
\subsection{ Fitted BB-parameters}                       
\label{sec:5a} 

The treatment of the broad mesons $\rho$ and $\epsilon$ is similar to that in the 
OBE-models \cite{NRS78,MRS89}. For the $\rho$-meson the same parameters are used 
as in these references. However, for the $\epsilon=f_0(760)$ assuming 
$m_\epsilon=760$ MeV and $\Gamma_\epsilon = 640$ MeV the Bryan-Gersten parameters      
\cite{Bry72} are used. For the chosen mass and width they are: 
$ m_1=496.39796$ MeV, $m_2=1365.59411$ MeV, and $\beta_1=0.21781, \beta_2=0.78219$.
Other meson masses are given in Table~\ref{table4}.
The sensitivity for the values of the cut-off masses of the $\eta$ and $\eta'$ 
is very weak. 
Therefore we have set the \{1\}-cut-off imass for the pseudoscalar nonet equal to 
that for the \{8\}.  Likewise, for the two nonets of the axial-vector mesons, 
see table~\ref{table5}.

Summarizing the parameters we have for baryon-baryon (BB):\\
 (i) NN Meson-couplings: $f_{NN\pi},f_{NN\eta'}$,     
 $ g_{NN\rho}, g_{NN\omega}$, 
  $f_{NN\rho},f_{NN\omega}$, $g_{NNa_0},g_{NN\epsilon}$,
   $g_{NNa_1}$, $f_{NNa_1}$, $g_{NNf'_1}$, $f_{NNf'_1}$, 
  $f_{NNb_1}$, $f_{NNh'_1}$\\
  (ii) $F/(F+D)$-ratios: $\alpha^{m}_{V}$, $\alpha_{A}$ \\
  (iii) NN Pair couplings: $g_{NN(\pi\pi)_1},f_{NN(\pi\pi)_1}$, $g_{NN(\pi\rho)_1}$,
  $g_{NN\pi\omega}, g_{NN\pi\eta}, g_{NN\pi\epsilon}$ \\
  (iv) Diffractive couplings and masslike parameters $g_{NNP}$, $g_{NNO}$, $f_{NNO}$, $m_P$, $m_O$ \\
  (v) Cut-off masses: $\Lambda_{8}^P = \Lambda_{1}^P$, $\Lambda_{8}^V$, $\Lambda_{1}^V$,
  $\Lambda_{8}^S$, $\Lambda_{1}^S$, and  $\Lambda_{8}^A$ = $\Lambda_{1}^A$.
 
The pair coupling $g_{NN(\pi\pi)_0}$ was kept fixed at zero.    
Note that in the interaction Hamiltonians of the pair-couplings 
(\ref{eq:3.17b})-(\ref{eq:3.17e}) 
the partial derivatives are scaled by $m_\pi$, and there is a scaling mass $M_N$.

The ESC-model described here, is fully consistent with $SU(3)$-symmetry  
using a straightforward extension of the NN-model to YN and YY. 
This the case for the OBE- and
TPS-potentials, as well as for the Pair-potentials.
For example $g_{(\pi\rho)_1} = g_{A_8VP}$, and
besides $(\pi\rho)$-pairs one sees also that $K K^*(I=1)$- and 
$K K^*(I=0)$-pairs contribute to the $NN$ potentials.
All $F/(F+D)$-ratio's are taken as fixed with heavy-meson saturation in mind.
The approximation we have made in this paper is to neglect the baryon mass
differencesi in the TPS-potentials, i.e. we put $m_\Lambda = m_\Sigma = m_N$. This because we
have not yet worked out the formulas for the inclusion of these mass 
differences, which is straightforward in principle.

\subsection{ Coupling Constants, $F/(F+D)$ Ratios, and Mixing Angles}             
\label{sec:5b} 
In Table~\ref{table5} we give the ESC08c meson masses, and the fitted 
couplings and cut-off parameters. Note that the axial-vector couplings for the
B-mesons are scaled with $m_{B_1}$.
The mixing for the pseudo-scalar, vector, and scalar mesons, as well as the 
handling of the diffractive potentials, has been described elsewhere, see
e.g. Refs.~\cite{MRS89,RSY99}. The mixing scheme of the axial-vector mesons is completely
similar as for the vector etc. mesons, except for the mixing angle.        
In the paper II \cite{RNY10b}
the $SU(3)$ singlet and octet couplings are 
listed, and also the $F/(F+D)$-ratios and mixing angles.
Also the Pauli-blocking effect parameter $a_{PB}$, 
described in \cite{PTP185.a}, section 8, for ESC08c is given.
As mentioned above, we searched for solutions where 
all OBE-couplings are compatible with the QPC-predictions. This time the QPC-model
contains a mixture of the $^3P_0$ and $^3S_1$ mechanism, whereas in 
Ref.~\cite{Rij04a} only the $^3P_0$-mechanism was considered.
For the pair-couplings all $F/(F+D)$-ratios were fixed to the predictions of
the QPC-model. 

\begin{table}
\caption{Meson couplings and parameters employed in the ESC08c-potentials.
         Coupling constants are at ${\bf k}^{2}=0$.
         An asterisk denotes that the coupling constant is constrained via SU(3).
         The masses and $\Lambda$'s are given in MeV.}
\label{table4}
\begin{center}
\begin{ruledtabular}
\begin{tabular}{crccr} \hline\hline
meson & mass & $g/\sqrt{4\pi}$ & $f/\sqrt{4\pi}$ & \multicolumn{1}{c}{$\Lambda$}\\
\hline
 $\pi$         &  138.04 &           &   0.2687   &   1056.13\    \\
 $\eta$        &  547.45 &           & \hspace{2mm}0.1265$^\ast$   & ,, \hspace{5mm} \\
 $\eta'$       &  957.75 &           &   0.2309   &  ,, \hspace{5mm} \\ \hline
 $\rho$        &  768.10 &  0.6446   &   3.7743   &    695.67\    \\
 $\phi$        & 1019.41 &--1.3390$^\ast$ & \hspace{2mm}3.1678$^\ast$ & 
 ,, \hspace{5mm}  \\
 $\omega$      &  781.95 &  3.4570   & --0.8575   &    758.58\\ \hline
 $a_1 $        & 1270.00 &--0.7895   & --0.8192   &   1051.80\    \\
 $f_1 $        & 1420.00 &  \hspace{3mm}0.7311$^\ast$ &\hspace{2mm}  0.3495$^\ast$  &      ,,  \hspace{5mm} \\
 $f_1'$        & 1285.00 &--0.7613   & \hspace{2mm}--0.4467 &      ,, \hspace{5mm}  \\ \hline
 $b_1 $        & 1235.00 &           & --0.2022   &   1056.13    \\
 $h_1 $        & 1380.00 &           & \hspace{2mm}--0.0621$^\ast$   &     
 ,,  \hspace{5mm} \\
 $h_1'$        & 1170.00 &           & --0.0335   &      ,, \hspace{5mm}  \\ \hline
 $a_{0}$       &  962.00 &  0.5853   &            &    994.89\    \\
 $f_{0}$       &  993.00 &\hspace{0mm}--1.6898$^\ast$   &            &  
 ,, \hspace{5mm}  \\
 $\varepsilon$ &  760.00 &  4.1461   &            &   1113.57 \\ \hline
 Pomeron       &  220.50 &  3.5815   &            &              \\
 Odderon       &  273.35 &  4.6362   & --4.7602   &              \\
\hline
\end{tabular}
\end{ruledtabular}
\end{center}
\label{table5}
\end{table}

One notices that all the BBM $\alpha$'s have values rather close to that 
which are expected from the QPC-model. In the ESC08c solution $\alpha_A \approx 0.31$,
which is not too far from $\alpha_A \sim 0.4$. 
As in previous works, e.g. Ref.~\cite{NRS78}, $\alpha_V^e=1$ is 
kept fixed.
Above, we remarked that the axial-nonet parameters may be sensitive to whether
or not the heavy pseudoscalar nonet with the $\pi$(1300) are included.

\begin{table}[hbt]
\caption{Pair-meson coupling constants employed in the ESC08c MPE-potentials.     
         Coupling constants are at ${\bf k}^{2}=0$.
         The F/(F+D)-ratio are QPC-predictions, except that 
 $\alpha_{(\pi\omega)}=\alpha_{pv}$, which is very close to QPC.}
\label{tab.gpair}
\begin{center}
\begin{ruledtabular}
\begin{tabular}{cclrc} \hline\hline
 $J^{PC}$ & $SU(3)$-irrep & $(\alpha\beta)$  &\multicolumn{1}{c}{$g/4\pi$} & $F/(F+D)$ \\
 \hline \\
 $0^{++}$ & $\{1\}$  & $g(\pi\pi)_{0}$   &  ---    &  ---    \\
 $0^{++}$ & ,,       & $g(\sigma\sigma)$ &  ---    &  ---    \\
 $0^{++}$ &$\{8\}_s$ & $g(\pi\eta)$      & -1.2371 &  1.000  \\ \hline
 $1^{--}$ &$\{8\}_a$ & $g(\pi\pi)_{1}$   &  0.2703 &  1.000  \\
          &          & $f(\pi\pi)_{1}$   &--1.6592 &  0.400  \\ \hline
 $1^{++}$ & ,,       & $g(\pi\rho)_{1}$  &  5.1287 &  0.400  \\
 $1^{++}$ & ,,       & $g(\pi\sigma)$    &--0.2989 &  0.400  \\
 $1^{++}$ & ,,       & $g(\pi P)$        &  ---    &  ---    \\ \hline
 $1^{+-}$ &$\{8\}_s$ & $g(\pi\omega)$    &--0.2059 &  0.365  \\
\hline
\end{tabular}
\end{ruledtabular}
\end{center}
\end{table}

In Table~\ref{tab.gpair} we listed the fitted Pair-couplings for the MPE-potentials.
We recall that only One-pair graphs are included, in order to avoid double
counting, see Ref.~\cite{Rij04a}. The $F/(F+D)$-ratios are all fixed, assuming heavy-boson 
domination of the pair-vertices. The ratios are taken from the QPC-model for 
$Q\bar{Q}$-systems with the same quantum numbers as the dominating boson.
For example, the $\alpha$-parameter for the axial $(\pi\rho)_1$-pair could fixed
at the quark-model prediction 0.40, see Table~\ref{tab.gpair}.
The $BB$-Pair couplings are calculated, assuming unbroken $SU(3)$-symmetry, 
from the $NN$-Pair coupling and the $F/(F+D)$-ratio using $SU(3)$.
Unlike in Ref.~\cite{RS96a,RS96b}, we did not fix pair couplings using
a theoretical model, e.g. based on heavy-meson saturation and chiral-symmetry.
So, in addition to the 14 parameters used in Ref.~\cite{RS96a,RS96b} we now have
6 pair-coupling fit parameters. 
In Table~\ref{tab.gpair} the fitted pair-couplings are given.
 Note that the $(\pi\pi)_0$-pair coupling gets contributions from the $\{1\}$ and
 the $\{8_s\}$ pairs as well, giving in total $g_{(\pi\pi)} \approx 0.10$, 
which has the same sign as in \cite{RS96a,RS96b}. 
 The $f_{(\pi\pi)_1}$-pair coupling has opposite
 sign as compared to Ref.~\cite{RS96a,RS96b}.
In a model with a more complex and realistic
meson-dynamics \cite{SR97} this coupling is predicted as found in the present 
ESC-fit. The $(\pi\rho)_1$-coupling is large as expected from $A_1$-saturation, see 
Ref.~\cite{RS96a,RS96b}. We conclude that the pair-couplings are in general 
not well understood quantitatively, and deserve more study.

In Table~\ref{table4} we show the OBE-coupling constants and the 
gaussian cut-off's $\Lambda$. The used  $\alpha =: F/(F+D)$-ratio's 
for the OBE-couplings are:
pseudo-scalar mesons $\alpha_{pv}=0.365$, 
vector mesons $\alpha_V^e=1.0, \alpha_V^m=0.472$, 
and scalar-mesons $\alpha_S=1.00$, which is calculated using the physical 
$S^* =: f_0(993)$ coupling etc..
In Table~\ref{tab.gpair} we show the MPE-coupling constants.        
The used  $\alpha =: F/(F+D)$-ratio's for the MPE-couplings are:
$(\pi\eta)$ pairs $\alpha(\{8_s\})=1.0$, 
$(\pi\pi)_1$ pairs $\alpha_V^e(\{8\}_a)=1.0, \alpha_V^m(\{8\}_a)=0.400$, 
and the $(\pi\rho)_1$ pairs $\alpha_A(\{8\}_a)=0.400$. 
The $(\pi\omega)$ pairs $\alpha(\{8_s\})$ has been set equal to
$\alpha_{pv}=0.365$.

 \begin{figure}   
\resizebox{8.cm}{11.43cm}        
 {\includegraphics[50,50][554,770]{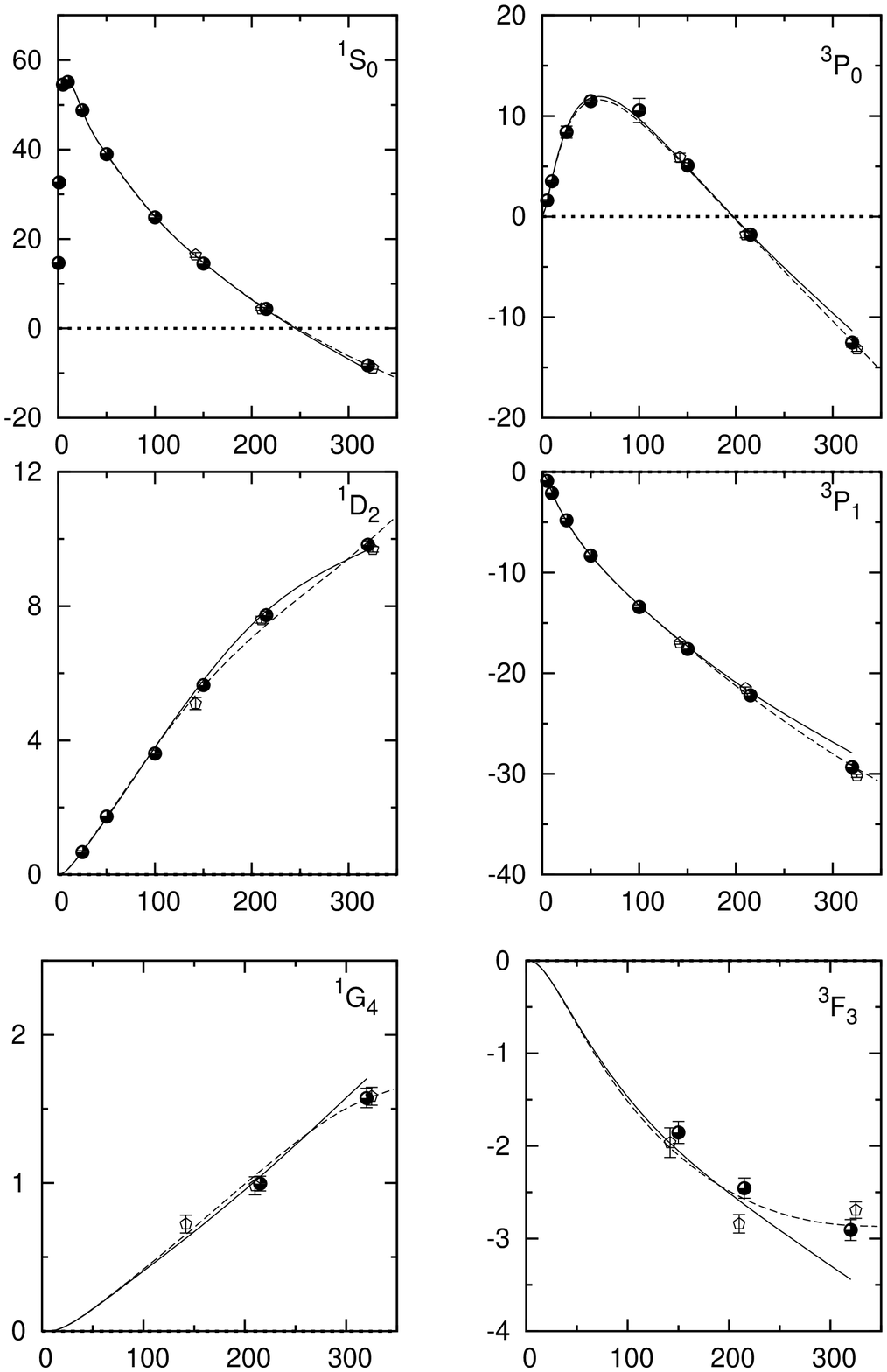}}
\caption{Solid line: proton-proton $I=1$ phase shifts for the ESC08c-model. 
 The dashed line: the m.e. phases of the Nijmegen93 PW-analysis \cite{Sto93}. 
 The black dots:  the s.e. phases of the Nijmegen93 PW-analysis.
 The diamonds:  Bugg s.e. \cite{Bugg92}.}
\label{ppi1.fig}
 \end{figure}

 \begin{figure}   
\resizebox{8.cm}{11.43cm}
 {\includegraphics[50,50][554,770]{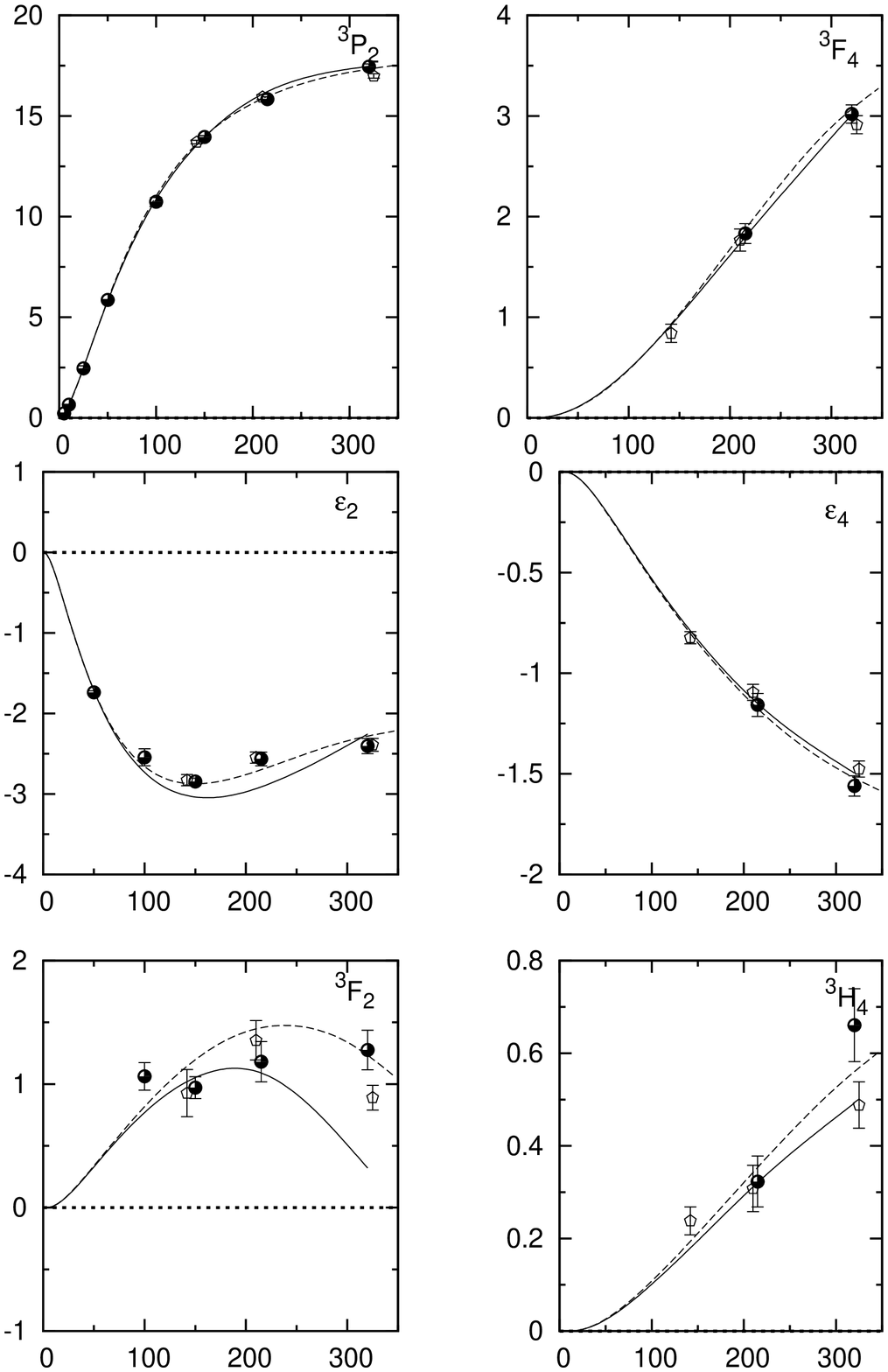}}
\caption{Solid line: proton-proton $I=1$ phase shifts for the ESC08c-model. 
 The dashed line: the m.e. phases of the Nijmegen93 PW-analysis \cite{Sto93}. 
 The black dots:  the s.e. phases of the Nijmegen93 PW-analysis.
 The diamonds:  Bugg s.e. \cite{Bugg92}.}
\label{ppi1c.fig}
 \end{figure}

 \begin{figure}   
\resizebox{8.cm}{11.43cm}
 {\includegraphics[50,50][554,770]{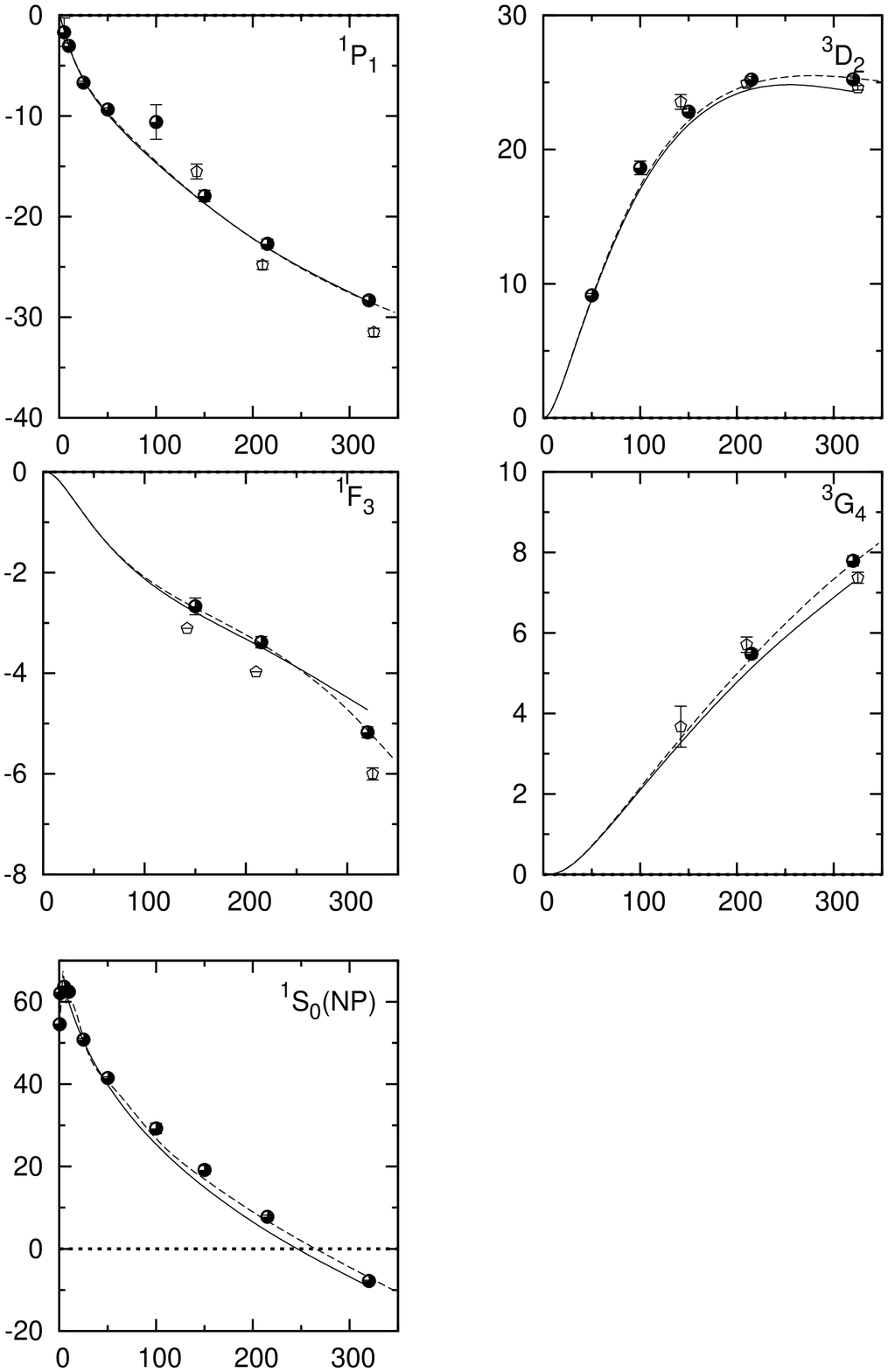}}
\caption{Solid line: neutron-proton $I=0$, and the I=1 $^1S_0(NP)$
 phase shifts for the ESC08c-model. 
 The dashed line: the m.e. phases of the Nijmegen93 PW-analysis \cite{Sto93}. 
 The black dots:  the s.e. phases of the Nijmegen93 PW-analysis.
 The diamonds:  Bugg s.e. \cite{Bugg92}.}
\label{npi0.fig}
 \end{figure}

 \begin{figure}   
\resizebox{8.cm}{11.43cm}
 {\includegraphics[50,50][554,770]{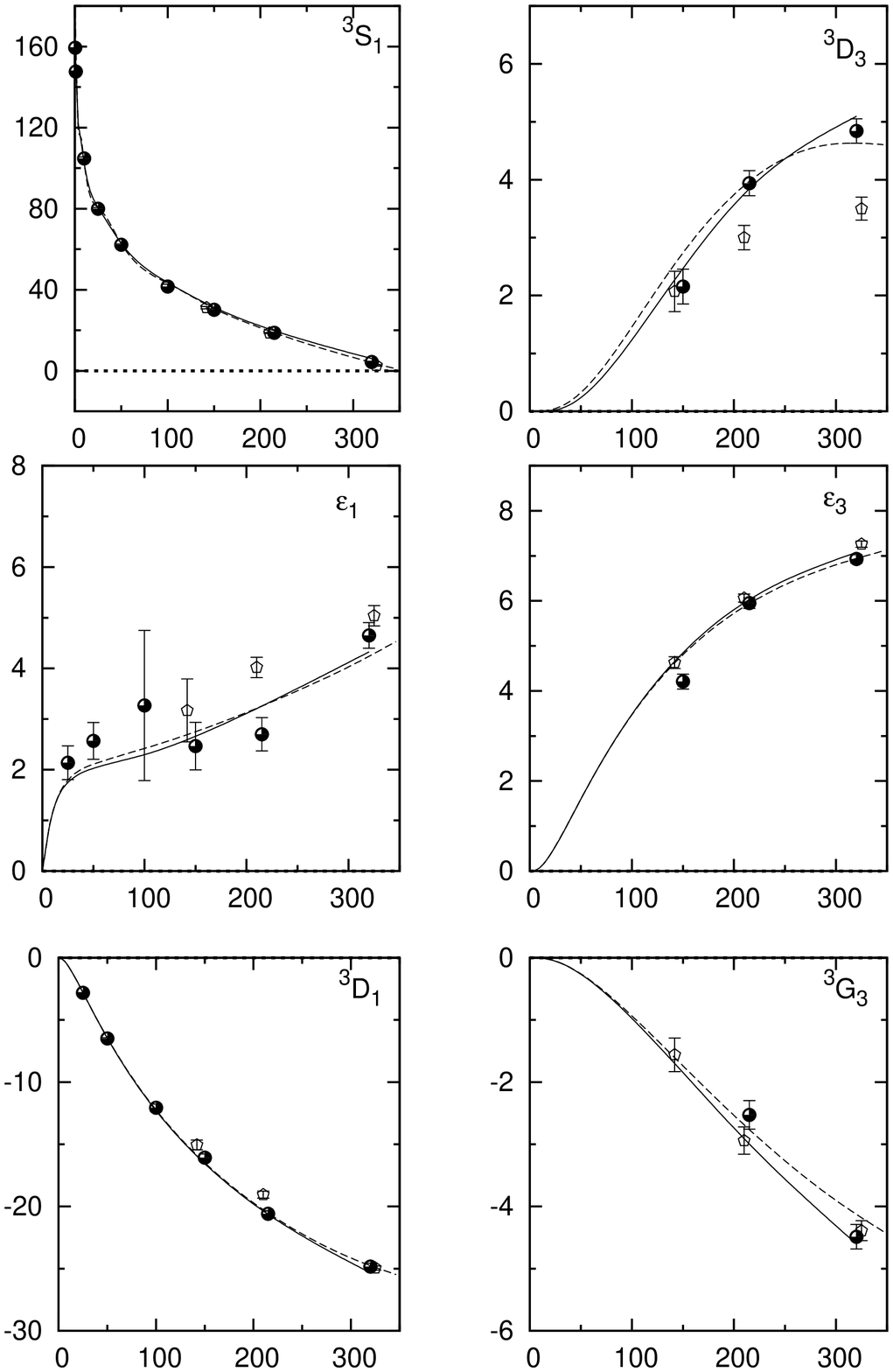}}
\caption{Solid line: neutron-proton $I=0$ phase shifts for the ESC08c-model. 
 The dashed line: the m.e. phases of the Nijmegen93 PW-analysis \cite{Sto93}. 
 The black dots:  the s.e. phases of the Nijmegen93 PW-analysis.
 The diamonds:  Bugg s.e. \cite{Bugg92}.}
\label{npi0c.fig}
 \end{figure}

\section{ ESC08-model , $N\!N$-Results}                                  
\label{sec:6}
\subsection{ Nucleon-nucleon Fit, Low-energy and Phase Parameters}                                  
\label{sec:6a} 
For a more detailed discussion on the NN-fitting we refer to Ref.~\cite{Rij04a}.
Here, we fit to the 1993 Nijmegen representation of the $\chi^2$-hypersurface of the 
$NN$ scattering data below $T_{lab}=350$ MeV \cite{Sto93,Klo93}, 
and also the low-energy parameters are fitted for $pp, np$ and $nn$. 
In this simultaneous fit of $NN$ and $YN$, we obtained for ESC08c  
for the phase shifts $\chi^2/Ndata =1.081$. 
For a comparison with Ref.~\cite{Rij04a}, and for use of this model for the description
of $NN$, we give in Table~\ref{tab.nnphas3} the nuclear-bar
phases for $pp$ in case $I=1$, and for $np$ in the case of $^1S_0(I=1)$ 
and the $I=0$-phases.
Here, $\Delta\chi^2$ denotes the access in $\chi^2$ of the ESC-model w.r.t. the
phase shift analysis \cite{Sto93,Klo93}.

The deuteron has been included in the fitting procedure, as well as the low-energy parameters.
The fitted binding energy $E_B= 2.224593$ MeV, which
is very close to $E_B(experiment)=2.224644$. 
The charge-symmetry breaking is described phenomenologically by having 
 next to $g_{\rho nn}$ free couplings for $g_{\rho np}$, 
and $g_{\rho pp}$. This phenomenological treatment is successful for the 
various NN-channels, 
especially for the $np(^1S_0,I=1)$-phases, which were included in the NN-fit.
\begin{table}[hbt]
\caption{ ESC08c nuclear-bar $pp$ and $np$ phases in degrees.}
\begin{tabular}{crrrrrrrrrr} \hline\hline & & & & & &&&&&\\
 $T_{\rm lab}$ & 0.38& 1 & 5  & 10 & 25 & 50 & 100 & 150 & 215 & 320 \\ \hline
     &    &     &     &     &    &&&&& \\
 $\sharp$ data &144  & 68  & 103 & 290& 352 & 572 & 399 & 676 & 756 & 954 \\
     &    &     &     &     &   &&&&&\\
$\Delta \chi^{2}$& 11  & 52  & 11  & 28  & 28 &  75 & 21 &  96 & 140 & 124 \\
     &    &     &     &     &    &&&&&\\ \hline
     &    &     &     &     &   &&&&& \\
 $^{1}S_{0}(np)$ & 54.57  & 62.02 & 63.48 & 59.73& 50.49    
                 & 39.82  & 25.40 & 14.99 & 4.37 & --9.02  \\
 $^{1}S_{0}$ & 14.61  & 32.62 & 54.75 & 55.17& 48.68    
             & 38.98  & 25.04 & 14.77 & 4.21 &--9.14  \\
 $^{3}S_{1}$ & 159.39 & 147.77& 118.25& 102.73& 80.83  
             & 63.07  & 43.68 & 31.34 & 19.60 & 5.66 \\
 $\epsilon_{1}$ & 0.03  & 0.11 & 0.68 & 1.17 & 1.81   
                & 2.13  & 2.44 & 2.82 & 3.43 & 4.56 \\
 $^{3}P_{0}$ & 0.02   &  0.14 & 1.61 & 3.81 & 8.78    
             & 11.75  &  9.64 & 4.85 &--1.70 &--11.20 \\
 $^{3}P_{1}$ & --0.01  &--0.08  &--0.89  & --2.04  & --4.87     
             & --8.25  &--13.22 &--17.32 & --21.94 & --28.16 \\
 $^{1}P_{1}$ & --0.05  &--0.19  &--1.50  & --3.07  & --6.40     
             & --9.82  &--14.68 &--18.82 & --23.51 & --29.57 \\
 $^{3}P_{2}$ &  0.00  & 0.01  & 0.22  &  0.67  &  2.51     
             &  5.80  & 10.88 & 14.03  &  16.30 &  17.28 \\
 $\epsilon_{2}$ &--0.00  &--0.00 &--0.05 &--0.20 &--0.81    
                &--1.71  &--2.69 &--2.95 &--2.79 &--2.15 \\
 $^{3}D_{1}$ & --0.00  &--0.01  &--0.18  & --0.68  & --2.83    
             &--6.52  &--12.41 &--16.70 & --20.74 & --25.12 \\
 $^{3}D_{2}$ & 0.00  & 0.01  & 0.22  &  0.85  &  3.70     
             & 8.95  & 17.27 & 22.17 &  24.93 &  24.83 \\
 $^{1}D_{2}$ & 0.00  & 0.00  & 0.04  &  0.17  &  0.69     
             & 1.69  & 3.78  & 5.71  &  7.66  &   9.30 \\
 $^{3}D_{3}$ & 0.00  & 0.00  & 0.00  &  0.00  &  0.03    
             & 0.24  & 1.20  & 2.37  &  3.71  &  5.07  \\
 $\epsilon_{3}$ & 0.00  & 0.00 & 0.01 & 0.08 & 0.55   
                & 1.60  & 3.47 & 4.83 & 6.01 & 7.05 \\
 $^{3}F_{2}$ & 0.00  & 0.00  & 0.00  &  0.01  &  0.11     
             & 0.34  & 0.80  & 1.11  &  1.17  &  0.45  \\
 $^{3}F_{3}$ & --0.00  & --0.00  &--0.00  & --0.03  & --0.23     
             &--0.67  &--1.46  &--2.04  & --2.62  & --3.44  \\
 $^{1}F_{3}$ & --0.00  & --0.00  &--0.01  & --0.06  & --0.41     
             &--1.10  &--2.12  &--2.77  & --3.45  & --4.65  \\
 $^{3}F_{4}$ & 0.00  & 0.00  & 0.00  &  0.00  &  0.02     
             & 0.12  & 0.51  & 1.05  &  1.82  &  3.00  \\
 $\epsilon_{4}$ & --0.00  & --0.00 & --0.00 &--0.00 &--0.05    
                &--0.19  &--0.53 &--0.83 &--1.13 &--1.45 \\
 $^{3}G_{3}$ &--0.00 &--0.00  &--0.00  &--0.00  & --0.05    
             &--0.26 &--0.93  &--1.73  &--2.77  & --4.17  \\
 $^{3}G_{4}$ & 0.00 & 0.00  & 0.00  & 0.01  &  0.17     
             & 0.71  & 2.12  & 3.53  &  5.18  &  7.33  \\
 $^{1}G_{4}$ & 0.00 & 0.00  & 0.00  & 0.00  &  0.04     
             & 0.15  & 0.41  & 0.69  &  1.06  &  1.71  \\
 $^{3}G_{5}$ &--0.00 &--0.00  &--0.00  &--0.00  & --0.01      
             &--0.05  &--0.17  &--0.25  & --0.28  & --0.17  \\
 $\epsilon_{5}$ & 0.00 & 0.00  & 0.00 & 0.00 & 0.04    
                & 0.20  & 0.71 & 1.22 & 1.83 & 2.62 \\
     &    &     &     &     &  &&&&&  \\
\hline
\end{tabular}
\label{tab.nnphas3}   
\end{table}

 \begin{table}[hbt]
\caption{ESC08c Low energy parameters: S-wave scattering lengths and 
effective ranges, deuteron binding energy $E_B$, and electric 
quadrupole $Q_e$.
The asterisk denotes that the low-energy parameters were not searched.}  
\begin{center}
\begin{tabular}{ccccc} \hline\hline & & & & \\
     & \multicolumn{3}{c}{experimental data}& ESC08c
     \\ &&&& \\ \hline
 $a_{pp}(^1S_0)$ & --7.823 & $\pm$ & 0.010 & --7.7710\\
 $r_{pp}(^1S_0)$ &  2.794 & $\pm$ & 0.015 &  2.7601$^\ast$ \\ \hline
 $a_{np}(^1S_0)$ & --23.715 & $\pm$ & 0.015 & --23.7316\\
 $r_{np}(^1S_0)$ &  2.760 & $\pm$ & 0.030 &  2.6983$^\ast$ \\ \hline
 $a_{nn}(^1S_0)$ & --18.63     & $\pm$ & 0.48  & --17.177\\
 $r_{nn}(^1S_0)$ &  2.860   & $\pm$ & 0.15  &  2.8417$^\ast$ \\ \hline
 $a_{np}(^3S_1)$ &  5.423 & $\pm$ & 0.005 &  5.4384$^\ast$ \\
 $r_{np}(^3S_1)$ &  1.761 & $\pm$ & 0.005 &  1.7481$^\ast$\\ \hline
  $E_B$         &  --2.224644 & $\pm$ & 0.000046 & --2.224593 \\
  $Q_e$         &  0.286 & $\pm$ & 0.002 &  0.2742 \\ \hline\hline
\end{tabular}
\end{center}
 \label{tab.lowenergy}
 \end{table}

\begin{table}
\caption{ ESC08c $\chi^2$ and $\chi^2$ per datum at the ten energy bins for the    
 Nijmegen93 Partial-Wave-Analysis. $N_{data}$ lists the number of data
 within each energy bin. The bottom line gives the results for the 
 total $0-350$ MeV interval.
 The $\chi^{2}$-access for the ESC model is denoted    
 by  $\Delta\chi^{2}$ and $\Delta\hat{\chi}^{2}$, respectively.}  
\begin{ruledtabular}
\begin{tabular}{crrrrrr} & & & & & \\
 $T_{\rm lab}$ & $\sharp$ data & $\chi_{0}^{2}$\hspace*{5.5mm}&
 $\Delta\chi^{2}$&$\hat{\chi}_{0}^{2}$\hspace*{3mm}&
 $\Delta\hat{\chi}^{2}_0$ \\ &&&&& \\ \hline
0.383 & 144 & 137.555 & 14.7 & 0.960 & 0.102  \\
  1   &  68 &  38.019 & 60.5 & 0.560 & 0.890  \\
  5   & 103 &  82.226 &  8.1 & 0.800 & 0.078  \\
  10  & 290 & 257.995 & 28.4 & 1.234 & 0.098  \\
  25  & 352 & 272.197 & 33.1 & 0.773 & 0.094  \\
  50  & 571 & 538.522 & 37.2 & 0.957 & 0.065  \\
  100 & 399 & 382.499 & 19.6 & 0.959 & 0.049  \\
  150 & 676 & 673.055 & 72.5 & 0.996 & 0.107  \\
  215 & 756 & 754.525 &118.5 & 0.998 & 0.157  \\
  320 & 954 & 945.379 &189.5 & 0.991 & 0.199  \\ \hline
      &    &     &     &     &    \\
Total &4313&4081.971& 582.0 &0.948 &0.133  \\
      &    &     &     &     &     \\
\end{tabular}
\end{ruledtabular}
\label{tab.chidistr} 
\end{table}

We emphasize that we use the single-energy (s.e.) phases and $\chi^2$-surface 
\cite{Klo93} as a means to fit the NN-data. 
The multi-energy (m.e.) phases of the PW-analysis \cite{Sto93}
in Fig.~\ref{ppi1.fig}-Fig.~\ref{npi0.fig} are the dashed lines in these figures.
One notices that the central value of the s.e. phases do not correspond
to the m.e. phases in general,
illustrating that there has been a certain amount
of noise fitting in the s.e. PW-analysis, see e.g. $\epsilon_1$ and $^1P_1$ 
at $T_{lab}=100$ MeV.
The m.e. PW-analysis reaches $\chi^2/N_{data}=0.99$, using 
 39 phenomenological parameters plus normalization parameters.
The related phenomenological PW-potentials NijmI,II and Reid93 \cite{SKTS94},
with respectively 41, 47, and 50 parameters, turn out all with $\chi^2/Ndata=1.03$.
This should be compared to the ESC-model, which has $\chi^2/N_{data}=1.08$
using for NN 32 parameters. These are 14 QPC-constrained meson-nucleon-nucleon couplings,
6 meson-pair-nucleon-nucleon couplings, 6 gaussian cut-off parameters, 3 diffractive 
couplings, and 2 diffractive mass parameters. The 3 remaining fitting parameters
 (2 F/(F+D) ratios and the Pauli blocking fraction) are mainly or totally 
determined by the YN-fit.
From the figures it is obvious that the ESC-model deviates from the m.e.
PW-analysis in particular at the highest energy. 

In Table~\ref{tab.lowenergy} the results for the low energy parameters are given.
In order to discriminate between the $^1S_0$-wave for pp, np, and nn, we introduced 
some charge independence breaking by taking $g_{pp\rho} \neq g_{np\rho} \neq g_{nn\rho}$.
With this device we fitted the difference between the $^1S_0(pp)$ and $^1S_0(np)$ 
phases, and the different scattering lengths and effective ranges as well. We found
$g_{np\rho} = 0.5889,\ g_{pp\rho} = 0.6389$, which are not far from
$g_{nn\rho} = 0.6446$, see Table~\ref{table4}.  
The NN low-energy parameters are fitted very well, see Table~\ref{tab.lowenergy}.
For a discussion of the theoretical and experimental situation w.r.t. these low 
energy parameters, see \cite{Mil90}.\\
The binding energy of the deuteron is fitted excellently. The electric
quadrupole moment result is typical for models without meson-exchange
current effects. Further properties of the deuteron in this model are:
$P_D=6.07 \%, D/S=0.0257, N_G^2=0.7721$, and $\rho_{-\epsilon,-\epsilon}=1.7273$.

  \begin{figure}[hbt]
  \resizebox{3.5cm}{!}       
  {\includegraphics[200,000][400,850]{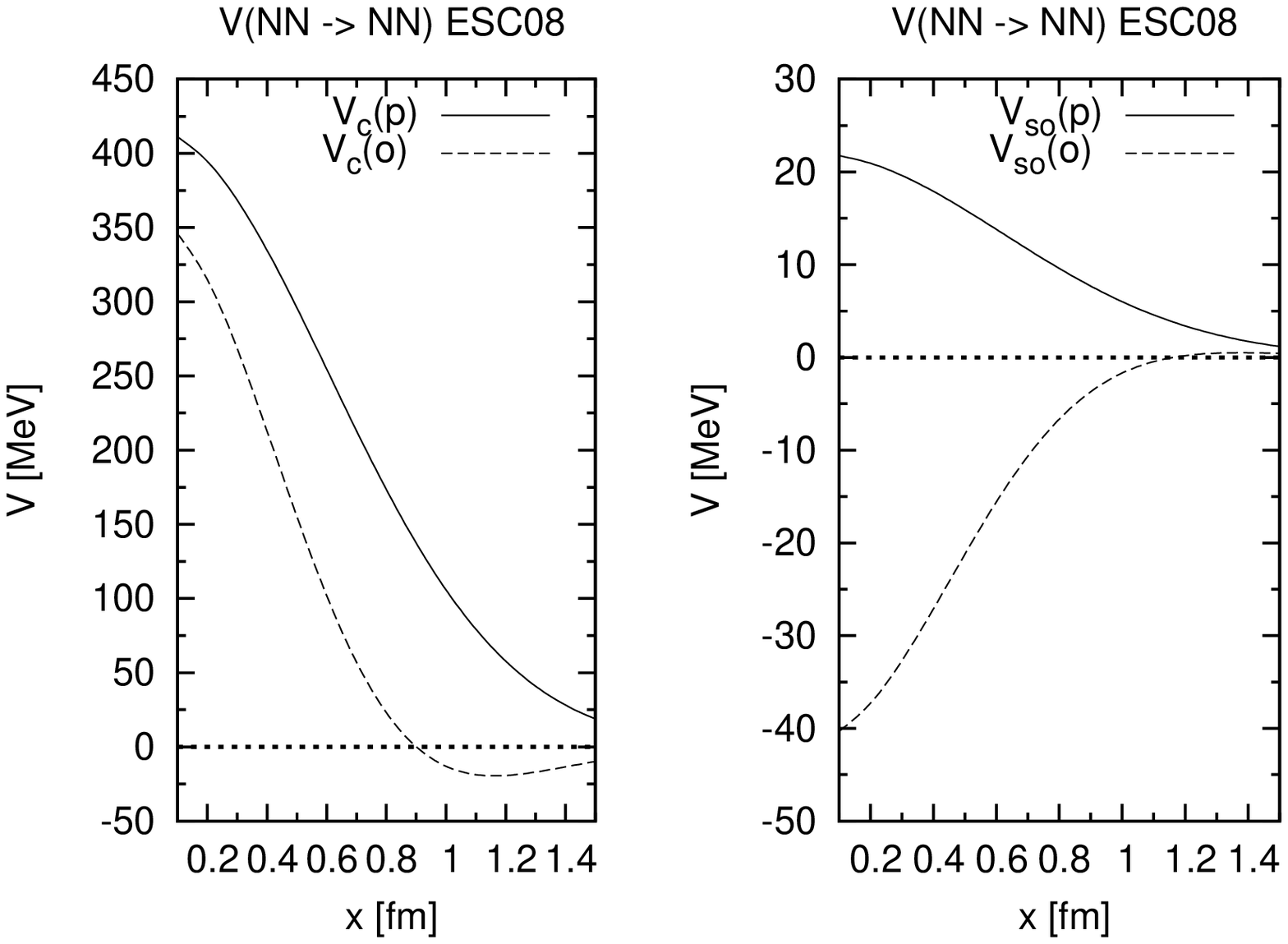}}
 \caption{Pomeron and Odderon central- and spin-orbit potentials.}     
 \label{fig:pom-odd2}
  \end{figure}

\subsection{ Nucleon-nucleon Potentials}                           
\label{sec:6c} 
The hyperon-nucleon OBE-, TPS-, and Pair-potentials for ESC04 model are shown in 
Ref.~\cite{Rij04b}. These potentials are rather similar to those of ESC08c, and 
therefore we refer the reader the cited YN-paper. Also, these NN-potentials 
are qualitatively rather similar in character.

The odderon potential is a novel feature of ESC08-model. In Fig.~\ref{fig:pom-odd2} 
the central and spin-orbit potentials are shown. The spin-spin, tensor, and
quadratic spin-orbit potentials are very small. One notices from this figure that
the pomeron potential is like an 'anti-scalar' potential whereas the odderon
is a normal vector-exchange potential. Note the strong cancellation in the
spin-orbit giving a negligible summed contribution. The upshot is a universal
central repulsion from the pomeron+odderon.\\

\section{Nuclear Saturation and Three-body repulsion}
\label{sec:7}
The lowest-order Brueckner G-matrix calculations 
with the continuous (CON) choice for intermediate single particle potentials
were shown to simulate well the results including higher hole-line contributions
up to $3\sim 4$ $\rho_0$~\cite{Baldo98,Baldo02}.
Here, the Brueckner G-matrix theory is considered a good starting point 
for studies of many-body systems on the basis of
free-space baryon-baryon interaction models. We study 
the properties of high-density nuclear matter 
on the basis of the lowest-order G-matrix theory with the CON choice.

As is well known, the experimental nuclear saturation properties, the density $\rho_N$, 
the binding energy per nucleon E/A, the compression modulus K, cannot be reproduced
quantitatively with nuclear two-body interactions only, see e.g.~\cite{Lag81}.
Essential for giving the correct energy curve $E(\rho_{N})$ is the inclusion of many-nucleon
interactions. Here, ithe most important seems to be the three-nucleon interaction (TNI), 
composed of an attractive (TNA) and a repulsive (TNR) part. Integrating over the third
particle results in a dependence on the nuclear-matter density $\rho_{N}$ of the 
'effective' two-nucleon potential (see below).
Since TNA contributes only moderately as a function of $\rho_{N}$, 
the saturation curve is not so remarkably changed by the TNA \cite{Lag81}.
Its inclusion is nevertheless important for obtaining the right
nuclear saturation point.
On the other hand, it turns out that the TNR contribution increases rapidly in 
the high-density region, giving high values for the incompressibility.
Maximum masses of neutron stars can be reproduced with use of
the stiff equation of state (EoS) realized by the TBR contributions.
The soft-core two-baryon potentials give a too soft EoS.
In particular, ESC08 gives for the mass of the neutron star $1.35 M_{solar}$
\cite{Sch11}, implying for this model the necessity for a TNR contribution.
Therefore, we incorporate the TNR contribution in the ESC-model
together with an additional TNA one, 
giving it a key role in stiffening the EoS for symmetric and neutron-star matter. 
As will be shown below, this enables to satisfy both the nuclear saturation point 
and the observed maximum mass of neutron stars.

At high densities hyperon-mixing occurs in neutron-star matter, which 
brings about a significant softening of the EoS canceling the TNR effect
for the maximum mass \cite{BBS00,VPREH00,Nis02}.
To compensate this adverse effect Nishizaki, Takatsuka 
and one of the authors (Y.Y.) \cite{Nis02} made the conjecture that 
the TNR-type repulsion works universally for $Y\!N\!N$ and $Y\!Y\!N$ 
as well as for $N\!N\!N$.  
They demonstrated that the TNR-stiffening of the EoS
can be recovered clearly by this assumption.
Universal repulsions among three baryons were called 
the three-baryon repulsion (TBR).
It is our aim to realize the TBR assumption consistently with the ESC modeling
of the baryon-baryon systems.  The presence of three-body forces (3BF) 
is a natural possibility in nuclei and hypernuclei, 
generating effective two-body forces, which very likely improve the binding 
energies and well-depth's. The latter will appear indeed the case for the ESC-model
as shown in the YN-paper \cite{RNY10b} of this series.

 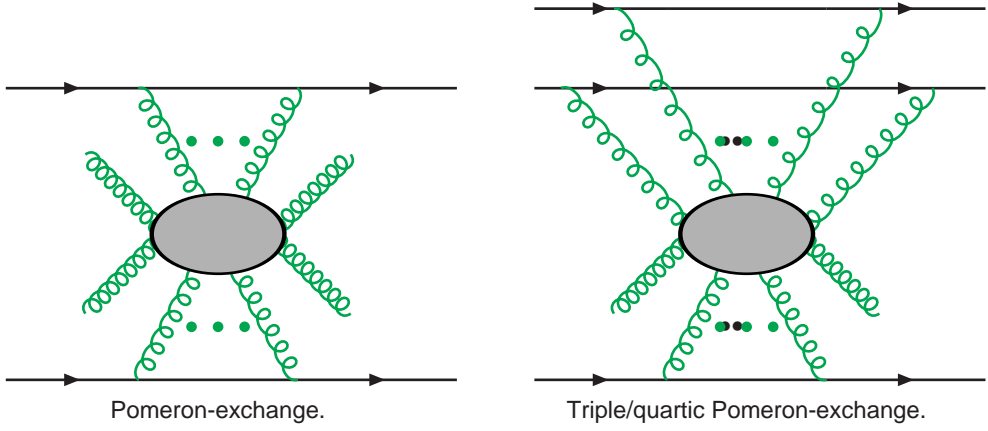
\begin{figure}[hbt]
 \vspace*{15mm}
 \begin{center} \begin{picture}(180,130)(100,0)
 \SetPFont{Helvetica}{9}
 \SetScale{1.0} \SetWidth{1.0}

 \ArrowLine(10,120)(60,120)   
 \Line(60,120)(120,120)   
 \ArrowLine(120,120)(180,120)   
 \ArrowLine(10,10)(60,10)   
 \Line(60,10)(120,10)   
 \ArrowLine(120,10)(180,10)   

 \SetColor{Green}
 \Gluon(60,120)(90,65){3}{8}
 \Gluon(120,120)(90,65){3}{8}
 \Gluon(60,10)(90,65){3}{8}
 \Gluon(120,10)(90,65){3}{8}
 \Gluon(40,95)(70,65){3}{8}
 \Gluon(40,35)(70,65){3}{8}
 \Gluon(140,95)(110,65){3}{8}
 \Gluon(140,35)(110,65){3}{8}
 \Vertex(80,100){2}
 \Vertex(90,100){2}
 \Vertex(100,100){2}
 \Vertex(80,30){2}
 \Vertex(90,30){2}
 \Vertex(100,30){2}
 \SetColor{Black}
 \GOval(90,65)(15,25)(0.0){0.7}
 \PText(90,0)(0)[c]{Pomeron-exchange.}        
 \SetScale{1.0} \SetWidth{1.0}
 \SetOffset(200,0)

 \ArrowLine(10,150)(60,150)   
 \Line(60,150)(120,150)   
 \ArrowLine(120,150)(180,150)   

 \ArrowLine(10,120)(60,120)   
 \Line(60,120)(120,120)   
 \ArrowLine(120,120)(180,120)   

 \ArrowLine(10,10)(60,10)   
 \Line(60,10)(120,10)   
 \ArrowLine(120,10)(180,10)   

 \SetColor{Green}
 \Gluon(40,150)(90,65){3}{8}
 \Gluon(140,150)(90,65){3}{8}

 \Gluon(60,10)(90,65){3}{8}
 \Gluon(120,10)(90,65){3}{8}
 \Gluon(20,120)(70,65){3}{8}
 \Gluon(40,35)(70,65){3}{8}
 \Gluon(160,120)(110,65){3}{8}
 \Gluon(140,35)(110,65){3}{8}
\Text(80,100)[l]{$\bullet \bullet$ }        
\Text(80,30)[l]{$\bullet \bullet$ }        
 \Vertex(80,100){2}
 \Vertex(90,100){2}
 \Vertex(100,100){2}
 \Vertex(80,30){2}
 \Vertex(90,30){2}
 \Vertex(100,30){2}
 \SetColor{Black}
 \GOval(90,65)(15,25)(0.0){0.7}
 \PText(90,0)(0)[c]{Triple/quartic Pomeron-exchange.}  
 \end{picture}   
\caption{Multi-gluon exchange processes.}                         
 \label{fig.mglue}
  \end{center}
  \end{figure}                     

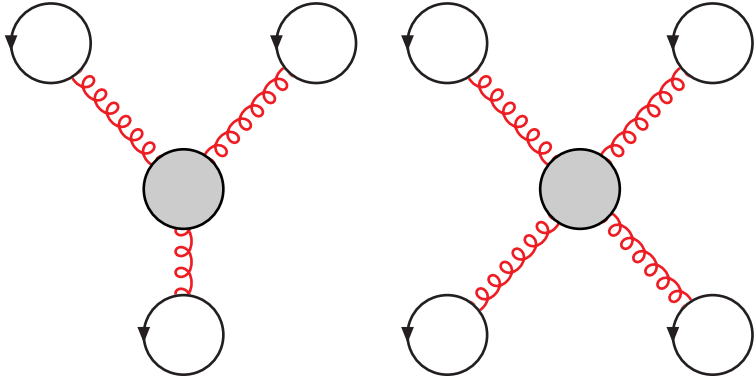
\begin{figure}[hbt]
 \begin{center} \begin{picture}(180,150)(100,0)
 \SetPFont{Helvetica}{9}
 \SetScale{1.0} \SetWidth{1.0}
 \SetOffset( 30,0)

\SetColor{Red}
\Gluon( 75, 70.0)(25,125.0){3}{10}
\Gluon( 75, 70.0)(125,125.0){3}{10}
\Gluon( 75, 70.0)( 75,15.0 ){3}{6}
\SetColor{Black}

\CCirc(25,125){15}{White}{White}
\CCirc(125,125){15}{White}{White}
\CCirc( 75,15.0){15}{White}{White}

\ArrowArc(25,125)(15,0,360)
\ArrowArc(125,125)(15,0,360)
\ArrowArc( 75,15)(15,0,360)

\GCirc(75, 70.0){15}{0.8}
 \SetOffset(180,0)

\SetColor{Red}
\Gluon( 75, 70.0)(25,125.0){3}{10}
\Gluon( 75, 70.0)(125,125.0){3}{10}
\Gluon( 75, 70.0)(25, 15.0){3}{10}
\Gluon( 75, 70.0)(125, 15.0){3}{10}

\SetColor{Black}

\CCirc(25,125){15}{White}{White}
\CCirc(125,125){15}{White}{White}
\CCirc(25, 15){15}{White}{White}
\CCirc(125, 15){15}{White}{White}

\ArrowArc(25,125)(15,0,360)
\ArrowArc(125,125)(15,0,360)
\ArrowArc(25, 15)(15,0,360)
\ArrowArc(125, 15)(15,0,360)

\GCirc( 75, 70.0){15}{0.8}

\end{picture} \end{center}
\caption{Triple- and quartic-pomeron 3- and 4-body interaction.}
 \label{fig.mpp2} 
\end{figure}

Since in QCD the gluons are flavor blind it is natural to relate the universality 
of the TBR repulsion to multi-gluon exchange, see Fig.~\ref{fig.mglue}. 
In the Nijmegen soft-core OBE and
ESC models pomeron-exchange can be viewed as due to an even number of
gluon-exchanges contributing a universal repulsion in BB-systems. Like for the 
two-baryon systems, in ESC we introduce the multi-gluon three-body forces 
with the multi-pomeron exchange potential (MPP) \cite{MPP05,RMPP08,PTP185.a}.
In Fig.~\ref{fig.mpp2} the triple- and quartic-pomeron vertices are illustrated.
We convert the three-body potential into an effective two-body potential by
integrating out the third nucleon. 
As demonstrated in \cite{YFYR13}, the MPP gives the stiff 
EoS of neutron matter enough to assure the large observed values
of two massive neutron stars with mass $1.97\pm 0.04 M_{solar}$ 
for PSR J1614-2230 \cite{Demorest10} and $2.01\pm 0.04 M_{solar}$ for 
PSR J0348-0432 \cite{Antoniadis13}.

In \cite{Rij04b} the medium effect on the vector masses was assumed as 
the dominant mechanism for generating extra repulsion at higher densities.
However, the mass shift of the vector meson masses due to the nuclear medium
has been put in doubt \cite{Mos10}. Therefore, in the ESC08-model, in contrast
to \cite{Rij04b}, we assume that the dominant mechanism is triple and
quartic pomeron exchange \cite{Kai74,Bro77}. 

For the triple pomeron vertex we take the Lagrangian \cite{MPP05,RMPP08} 
\begin{eqnarray}
 {\cal L}_3 = g_P^{(3)} {\cal M} \sigma_P^3(x)/3! \ .
\label{eq:tbf.1}\end{eqnarray}
Then, the three-body local potential by pomeron exchange is given by
\begin{eqnarray}
&&V({\bf x}_1,{\bf x}_2,{\bf x}_3) = g_P^{(3)} (g_P)^3 \Pi_{i=1}^3
\int \  \frac{d^3k_i}{(2\pi)^3} 
\Pi_{i=1}^3 
e^{-i{\bf p}_i\cdot{\bf x}}\cdot (2\pi)^3
\delta({\bf k}_1+{\bf k}_2+{\bf k}_3)  
\nonumber
\\
&& \times \exp(-{\bf k}_1^2/4m_P^2) \exp(-{\bf k}_2^2/4m_P^2)
\exp(-{\bf k}_3^2/4m_P^2) \cdot {\cal M}^{-5} .
\label{eq:tbf.2}
\end{eqnarray}
Here, the (low-energy) pomeron propagator is the same as used in the
two-body pomeron potential used in all Nijmegen soft-core 
OBE and ESC models.                            

The effective two-body potential in a baryonic medium is obtained
by integrating over the coordinate ${\bf x}_3$.
\begin{eqnarray}
V_{eff}({\bf x}_1,{\bf x}_2)&=& \rho_{NM} \int d^3x_3 
V({\bf x}_1,{\bf x}_3,{\bf x}_3) 
\nonumber
\\
&=& g_P^{(3)} (g_P)^3 \frac{\rho_{NM}}{{\cal M}} \cdot
\frac{1}{4\pi} \frac{4}{\sqrt{\pi}}
\left(\frac{m_P}{\sqrt{2}}\right)^3
\exp\left(-\frac12 m_P^2 r_{12}^2 \right) \ .
\label{eq:tbf.3}
\end{eqnarray}

In a similar way, one can obtain a four-body interaction
$V({\bf x}_1,{\bf x}_2,{\bf x}_3,{\bf x}_4)$ and a corresponding
effective two-body potential with a quartic pomeron coupling
$g_P^{(4)}$ \cite{RMPP08}. The expressions for the $N$-body interaction and
the effective two-body potential by multiple-pomeron exchange are given 
in Ref.~\cite{MPP05,RMPP08,YFYR13}. 
Here, we restrict ourselves to the triple and quartic pomeron 
couplings where there is information from the ISR pp-data \cite{Kai74}.
Since the pomeron is an SU(3)-singlet, the MPP in nuclear medium
leads to the density-dependent universal repulsion, which can be associated with the
proposal in \cite{Nis02}.
Estimates for $g_P^{(3)}$ and $g_P^{(4)}$ can be obtained from 
\cite{Kai74,Bro77}, which shows that $g_P^{(4)} >> g_P^{(3)}$.
The pomeron coupling $g_P$ is fitted to the NN-data etc., 
see Table~\ref{table4}.
In \cite{YFYR13}, the MPP strengths ($g_P^{(3)}$ and $g_P^{(4)}$) were 
determined by analyzing the $^{16}$O$+^{16}$O elastic scattering at 
$E/A=70$ MeV with use of G-matrix folding potentials, where the TNR 
effect appears clearly in the angular distribution. As shown in \cite{FSY}, 
in such a high scattering energy the frozen-density approximation gives 
a good prescription, where G-matrices including TNR at about two times of 
normal density contribute to nucleus-nucleus folding potentials.

In addition to MPP, in order to assure the nuclear saturation property 
precisely, we introduce also a TNA part phenomenologically as
a density-dependent two-body interaction
\begin{eqnarray*}
V_{TNA}(r;\rho_N)= V^0_{TNA}\, \exp(-(r/2.0)^2)\, \rho_N\, 
\exp(-\eta \rho_N)\, (1+P_r)/2 \ ,
\end{eqnarray*}
whose form is similar to the TNA part given in \cite{Lag81}.
$P_r$ is a space-exchange operator. By a $(1+P_r)$ factor, the TNA part
works only in even states, which is needed to reproduce nucleus-nucleus 
angular distributions precisely. Then, 
$V^0_{TNA}$ and $\eta$ are treated as adjustable parameters.
Strengths of the MPP part ($g_P^{(3)}$ and $g_P^{(4)}$) 
and the TNA part ($V_{TNA}$ and $\eta$) are determined so as to reproduce 
the $^{16}$O$+^{16}$O angular distribution $E/A=70$ MeV, and 
the energy and density at the saturation point of symmetric matter.
The ratio of $g_P^{(3)}$ and $g_P^{(4)}$ unsettled in our analysis was taken 
adequately in reference to the results in Ref.~\cite{Kai74,Bro77}.
Quantitatively, however, it is quite uncertain.

In Table~\ref{tabMPP}, we give the three parameter sets of 
($g_P^{(3)}$, $g_P^{(4)}$, $V^0_{TNA}$, $\eta$) 
named as MPa, MPb and MPc, respectively. 
MPa and MPb are very similar to MP1a and MP2a, respectively,
given in Ref.~\cite{YFYR13}. The $g_P^{(3)}$ value of MPc
is the same as that of MPa, but with $g_P^{(4)}=0$.
These sets reproduce equally well the saturation property.
In the case of $^{16}$O$+^{16}$O elastic scattering at $E/A=70$ MeV,
the G-matrix folding potentials derived from the three sets
give the angular distributions similar to the solid and dashed curves
in Fig.1 in Ref.~\cite{YFYR13}, reproducing nicely the experimental data.

\begin{table}
\caption{Parameters of MPP+TNA parts.
$V^0_{TNA}$ and $\eta$ are in MeV and fm$^{-3}$, respectively.}
\label{tabMPP}
 \begin{tabular}{|c|cccc|}
 \hline
 & $g_P^{(3)}$ & $g_P^{(4)}$ & $V^0_{TNA}$ & $\eta$ \\
 \hline
 MPa &  2.34 & 30.0   &  $-32.8$  & 3.5  \\
 MPb &  2.94 &  0.0   &  $-45.0$  & 5.4  \\
 MPc &  2.34 &  0.0   &  $-43.0$  & 7.3  \\
 \hline
 \end{tabular}
\end{table}

\begin{figure}[ht]
\begin{center}
\includegraphics*[width=10cm,height=10cm]{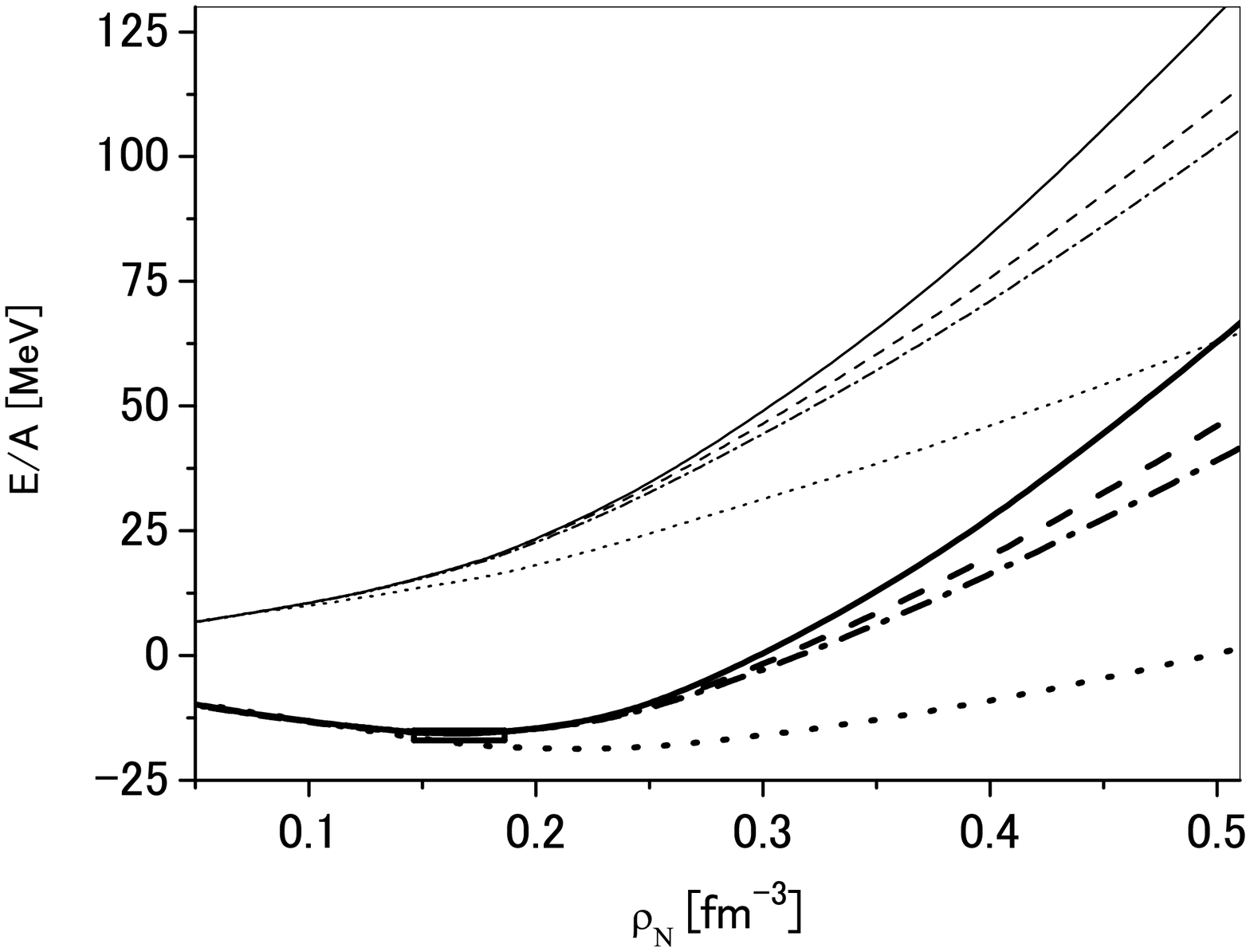}
\caption{Energies per nucleon drawn as a function of $\rho_N$
in symmetric nuclear matter (thick curves) and neutron
matter (thin curves). Dotted, solid, dashed and dot-dashed curves are
in cases of ESC08c only, MPa, MPb and MPc,respectively. 
The box shows the empirical value.
}
\label{saturation1}
\end{center}
\end{figure}

In Fig.~\ref{saturation1}, 
we show the energy curves of symmetric nuclear matter (thick curves) 
and neutron matter (thin curves), namely binding energy per nucleon 
($E/A$) as a function of $\rho_N$. They are obtained from G-matrix 
calculations with ESC08c only, and including the MPP+TNA parts. 
The box in the figure shows the area where nuclear 
saturation is expected to occur empirically.
The dotted curves are obtained only with the two-body interaction ESC08c. 
The saturation point in symmetric nuclear matter 
is found to deviate substantially from the box. On the other hand,
the solid, dashed and dot-dashed curves are obtained with including 
MPa, MPb and MPc contributions, respectively.
As is clearly seen, saturation densities and minimum values of $E/A$ in 
these cases are nicely close to the empirical value shown by the box:
For MPa and MPb (MPc), we obtain the value of $\sim -15.8$ ($-15.5$) MeV for 
the binding energy per nucleon at the saturation density $\sim 0.16$ fm$^{-3}$.
The incompressibilities $K$ for MPa, MPb and MPc are obtained as 
310 MeV, 280 MeV and 260 MeV, respectively, at the saturation densities.

\begin{figure}[ht]
\begin{center}
\includegraphics*[width=10cm,height=10cm]{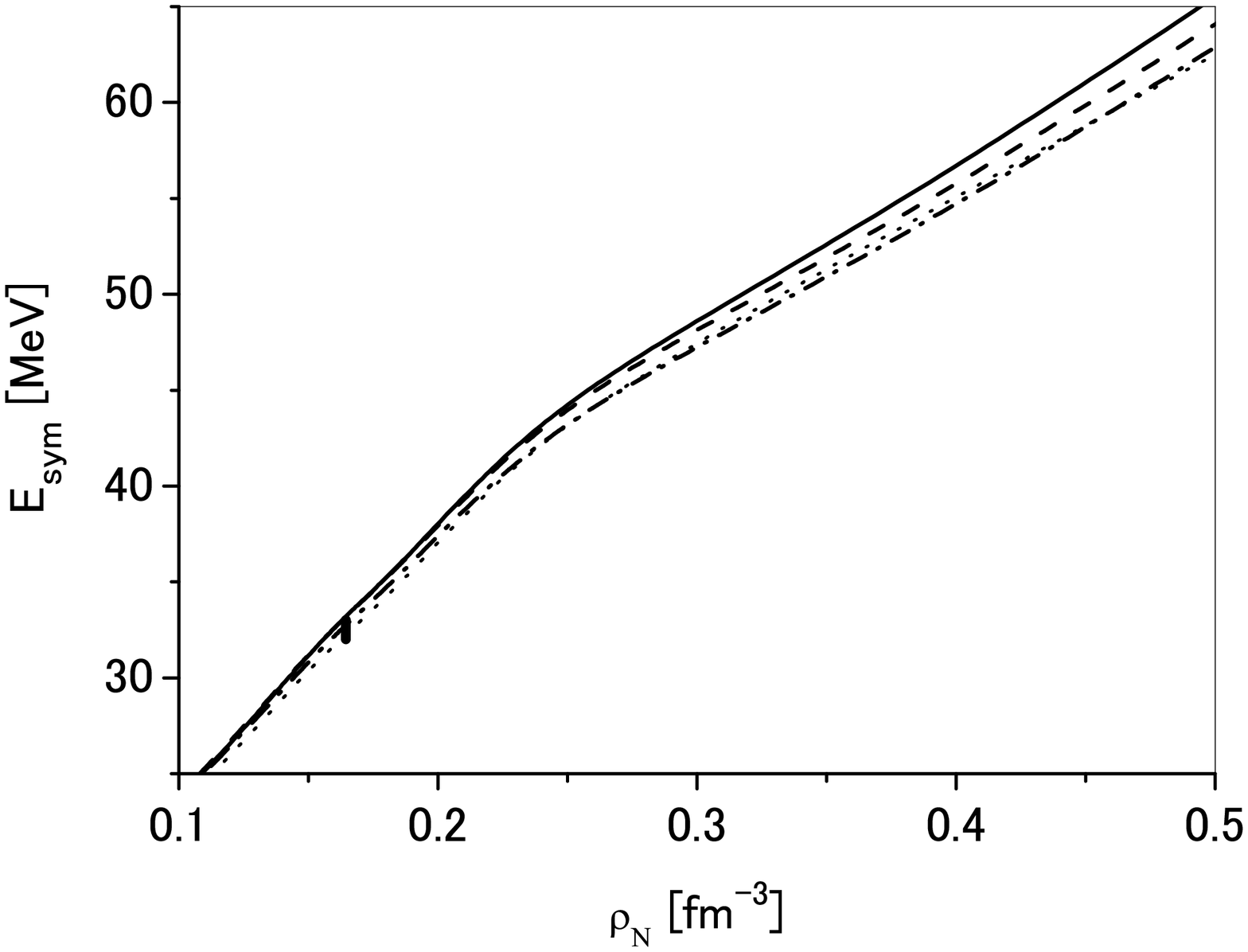}
\caption{Symmetric energies as a function of $\rho_N$
in cases of ESC08c only and including MPa, MPb and MPc by dotted, 
solid, dashed and dot-dashed curves, respectively. The short bar 
denotes the experimental value $32.5\pm 0.5$ MeV at normal density.
}
\label{saturation2}
\end{center}
\end{figure}

The difference between the $E/A$ curves for neutron matter and
symmetric matter gives the symmetry energy $E_{sym}(\rho)$.
In Fig.~\ref{saturation2}, obtained values of $E_{sym}$ are drawn 
as a function of $\rho_N$ in the cases of ESC08c only (dotted) and including
MPa (solid), MPb (dashed) and MPc (dot-dashed). 
The values of $E_{sym}$ at the saturation density 0.16 fm$^{-3}$
are 32.2 MeV (ESC08c only), 33.1 MeV (MPa), 
33.1 MeV (MPb) and 32.7 MeV (MPc).
The slope parameter is defined by 
$L=3\rho_0 \left[\frac{\partial E_{sym}(\rho)}{\partial \rho}\right]_{\rho_0}$.
The values of $L$ at the same density are 69.0 MeV (ESC08c only), 
70.4 MeV (MPa), 69.2 MeV (MPb) and 67.1 MeV (MPc). 
The above values of $E_{sym}$ and $L$ are in nice agreement to the values 
$E_{sym}=32.5\pm0.5$ MeV and $L=70\pm15$ MeV determined 
recently on the basis of experimental data~\cite{Yoshida}.
It should be noted the values of $E_{sym}$ and $L$ for the three sets
are similar to the values for ESC08c only, owing to the 
isospin-independent nature of the present three-body interaction.
Then, our three sets are specified mainly by the incompressibilities.

In the case of pure neutron-matter EOS, the mass-radius relations 
of neutron stars can be derived from the present ESC08c+MPP+TNA models 
in the same way as those in \cite{YFYR13}. 
Calculated values of maximum masses of neutron stars are
$2.5 M_{solar}$ for MPa, $2.2 M_{solar}$ for MPb and 
$2.1 M_{solar}$ for MPc,
being larger than the observed value $1.97 M_{solar}$.
The difference between the values for MPa and MPc comes from
the four-body repulsive part included in the former.
When the TNA parts are switched off in the three sets,
we have almost the same values of maximum masses:
The TNA parts contribute very slightly to maximum masses.

Thus, the inclusion of MPP and additional TNA provide a solution 
for both the nuclear saturation and the neutron-star mass problem.
It should be noted here that our MPP contributions exist universally 
in every baryonic system. It is very interesting to investigate
the relation between the universal MPP repulsions and
the softening effect induced by hyperon mixing to neutron-star matter.
The result will be published in the near future.

Another mechanism for generating extra repulsion at higher densities is
suggested by the relativistic mean-field theory (RMFT), see e.g.\cite{RMFT88}.
Here, at higher densities the scalar field becomes suppressed and
the vector field becomes dominating.
The effect would be similar to that employed in \cite{Rij04b}.
For the Dirac-Brueckner approach to the EoS see \cite{TerHaar87}.

\section{Discussion and Conclusions }
\label{sec:8}
The presentation in this paper reports on the present stage of the ESC-model.
Compared to ESC04 \cite{Rij04a,Rij04b,Rij04c} the model has been developed
further. The new version ESC08 has in addition to meson-exchange also
incorporated quark-core effects. Furthermore, the multi-gluon sector has been
completed by the inclusion of the Odderon. Moreover, the treatment of the
axial-vector mesons is now in a very satisfactory shape by employing the
B-field formalism. 
The ESC-approach to the nuclear force is a promising one. 
It opens the possibility to make a connection between 
the at present available baryon-baryon experimental data on the one hand, 
and with the underlying quark structure of the baryons and mesons on the other hand.
Namely, a successful description
of both the $N\!N$- and $Y\!N$-scattering data is obtained with meson-baryon
coupling parameters which all comply with the QPC-model.
We note that by studying the relation between the QPC-processes and the BBM-couplings, we 
determined the ratio $\gamma(^3P_0)/\gamma(^3S_1)=2:1$. 
In the literature, the $^3P_0$-QPC and the $^3S_1$-QPC in the SCQCD \cite{Mil89} 
has been 
studied by \cite{KI87} and \cite{KP88} respectively. In this paper we give therefore
an estimation of the relative importance of the QPC processes.
At the same time we comply with the strong constraint of no bound states 
in the $S=-1$-systems.  Therefore, the ESC-models, ESC04 and ESC08, 
are an important step in the determination of    
the baryon-baryon interactions for low energy scattering and the description
of hypernuclei in the context of broken SU$(3)$-symmetry.
The values for many parameters, which in previous Nijmegen work were considered to be free to 
a large extent, follow now rather well the pattern shown in quark-model predictions.          
This is particularly the case for the $F/(F+D)$-ratios of the OBE- and MPE-interactions.

For the nuclear matter description we introduced the multi-pomeron 
(multi-gluon) exchange
three-body force potential, achieving three things (i) right nuclear saturation,
 (ii) correct neutron star maas, and (iii) better hyperonic well depth's 
$U_Y$ for $Y= \Lambda, \Sigma, \Xi$ (see the companion papers II,III).
The combined fit for NN and YN is extremely good in ESC08. It is for the first time
that the quality of the NN-fit does not suffer from the inclusion of the YN-data.
The $\Lambda N$ p-waves seem to be better, which is the result of the truly 
simultaneous $NN+YN$-fitting. This is also reflected in the better
 Scheerbaum $K_\Lambda$-value \cite{Sch76},
making the well-known small spin-orbit splitting smaller, see Ref.~\cite{Hiya00}. 

The G-matrix results showed for ESC04 that basic features of hypernuclear data are reproduced
nicely, improving on the soft-core OBE-models NSC89 \cite{MRS89}
and NSC97 \cite{RSY99}. In spite of this superiority of ESC04 for hypernuclear data,
some problems remained. In particular the well depth $U_\Sigma$ was attractive, 
which is very unlikely in view of several other studies e.g. 
Ref.'s~\cite{Bat94,Dab99,Nou02,Koh04}
Furthermore, it has been shown \cite{Nis02} that the EOS for nuclear matter is too soft for
the soft-core models.
From this we learn that a good fit to the present scattering data not necessarily
means success in the G-matrix results. To explain this one can think of two reasons:     
(i) the G-matrix results are sensitive to the two-body interactions below 1 fm, 
whereas the present YN-scattering data are not, (ii) other than two-body forces play an
important role. 
Since the problem with $U_\Sigma$ hints at a special feature in 
the $\Sigma^+p(^3S_1)$-channel, 
it is likely that it is a two-body problem. As we have shown in ESC08 it can be solved by the
inclusion of the quark-core effects. For the softness of the EOS 
a natural possibility is the presence of three-body forces (3BF) in nuclear and hyperonic
matter, see Ref.~\cite{Nis02}. This also solves the nuclear saturation problem \cite{Rij04b}.

It is important to stress the role of the information on hypernuclei in our analysis.
We imposed for the ESC08-solution that $U_\Sigma >0$ and $U_\Xi < 0$.
This induced the occurrence of strong tensor-forces with the consequence of a
bound state in the S=-2 systems. Namely, deuteron-like bound states in the 
$\Xi N(^3S_1-^3D_1,I=0,1))$-system. 

Summarizing the results of the ESC-approach to baryon-baryon interactions, it can be 
stated that this is a very successful one. It has been shown that ESC-models are able
to give with single parameter-set's extremely satisfactory descriptions of the    
NN$\oplus$YN-data, and at the same time lead to successful G-matrix results.
For the coupling constants (i) flavor SU(3)-symmetry can be maintained, and 
(ii) they show rather well the pattern as predicted by the QPC-model.
The tensor-couplings play an important role, especially in the prediction of 
a deuteron-like S=-2 bound states. We conclude that these ESC-model predictions, 
as well as the applications to the S=-3,-4 systems and hyperonic matter, 
have a rather sound physical basis. 

We close by remarking that the determination of the MPE-couplings opens the possibility
to compute the 3BF-potentials for baryon-systems where all meson-pair vertices are fixed
by the ESC-model.

\section*{Acknowledgements}
 We wish to thank E. Hiyama, T. Motoba and H.-J. Schulze for many 
stimulating discussions. 


 \appendix 

\section{B-field formalism for vector- and axial-vector mesons}
\label{app:C} 
 As an alternative to the usual Proca-formalism for vector mesons, 
 Nakanishi and collaborators \cite{Nak72,NO90} introduced the
B-field formalism. 
In the non-abelian theory, e.g. isospin SU$_I$(2),
one introduces the B-field through the Lagrangian
\begin{eqnarray}
 {\cal L}_A &=& -\frac{1}{4}{\cal F}^i_{\mu\nu} {\cal F}^{\mu\nu i} 
  +\frac{1}{2} m^2 A_\mu^i A^{\mu i} 
 +B^i\partial_\mu A^{\mu i} +\frac{\alpha}{2} B^i B^i\ ,
\label{eq:C.1}\end{eqnarray}
where the field tensor and the covariant derivative $D_\mu$ are given by
\begin{subequations}
\begin{eqnarray}
 {\cal F}_{\mu\nu}^i &=& \partial_\mu A_\nu^i-\partial_\nu A_\mu^i 
 +g_A\epsilon^{ijk} A_\mu^jA_\nu^k, \\
  D_\mu &=& \partial_\mu -ig_A t_i\ A_\mu^i\ .
\label{eq:C.2}\end{eqnarray}
\end{subequations}
We assume that the $A^i_\mu$-field is coupled to the conserved, or almost conserved,
hadronic 'strong' current $J_{H,\mu}$.
The field equations, neglecting the non-abelian term in the axial
field tensor, become
\begin{subequations}
\begin{eqnarray}
  A^i_\mu &:& \partial^\mu{\cal F}_{\mu\nu}^i + m^2 A_\mu^i =
 -J^i_{H,\mu} + \partial_\mu B^i\ , \\
  B^i &:& \partial^\mu A_\mu^i + \alpha B^i = 0\ .
\label{eq:C.4}\end{eqnarray}
\end{subequations}
Exploiting now that approximately $\partial^\mu J_\mu^i=0$, one derives from
the field equation for $A^i_\mu$, upon taking the derivative $\partial_\mu$ etc.,
that $B^i$ is a free field, i.e.
\begin{eqnarray}
   \left(\Box + \alpha m^2\right) B^i = 0\ .                                    
\label{eq:C.5}\end{eqnarray}
This theory can be quantized in a satisfactory way, giving an axial-vector-meson
propagator which is covariant, see Nakanishi \& Ojima \cite{NO90}
It implies that in the propagator one has for the spectral function of the
propagator projection operator 
\begin{equation}
   \Pi^{\mu\nu}(k)= \left[-\eta^{\mu\nu}+ \frac{k^{\mu}k^{\nu}}{m^2}\right]
 \delta(k^2-m^2) - \frac{k^{\mu}k^{\nu}}{m^2}\
 \delta(k^2-\alpha_r m^2)\ ,
\label{eq:C.6}\end{equation}
where $\alpha_r> 0 $ is the renormalized B-field parameter $\alpha$ 
giving it a mass $\sqrt{\alpha_r}m$ \cite{NO90}. The propagator becomes
\begin{eqnarray}
   P^{\mu\nu}(k) &=& -\frac{\eta^{\mu\nu}}{k^2-m^2+i\epsilon} 
+ (1-\alpha_r)\frac{k^{\mu}k^{\nu}}{(k^2-m^2+i\epsilon)
(k^2-\alpha_rm^2+i\epsilon)} \nonumber\\
 & \Rightarrow & -\frac{\eta^{\mu\nu}}{k^2-m^2+i\epsilon}\ \ ,\ \ {\rm for} \ \ 
 \alpha_r=1\ .
\label{eq:C.7}\end{eqnarray}
The case $\alpha_r=1$ reminds one of the Feynman-gauge in the massless case.
Now, in the case of coupling to a conserved current, the potential will be
independent of $\alpha_r$. Therefore, we will use the "Feynman-gauge" in 
this paper. 
It implies that the $k^\mu k^\nu$-terms in the vector-meson propagators will not
contribute to the potentials in the B-field formalism. This in contrast to the 
Proca-formalism, see e.g. Ref.~\cite{IZ80}.
For the axial-vector mesons we will use the B-field formalism, whereas for the vector mesons
we continue to use the Proca formalism, like in Refs.~\cite{MRS89,RSY99,Rij04b}.

\section{Exact treatment non-local-tensor (NLT) Operator}
\label{app:B}
From results given in Ref.~\cite{RKS91}, we derive a new method for the 
treatment of the non-local-tensor (NLT)
$\boldsymbol{\sigma}_1\cdot{\bf q}\boldsymbol{\sigma}_2\cdot{\bf q}$-operator.
Starting from  
\begin{equation}
 \widetilde{V}({\bf k},{\bf q}) = \int d^3r'\ \int d^3r\ 
 e^{i{\bf p'}\cdot{\bf r'}} V({\bf r'},{\bf r}) e^{-i{\bf p}\cdot{\bf r}}\ ,
\label{eq:6.1}\end{equation}
where 
\begin{equation}
 V({\bf r'},{\bf r})  =  \delta^3({\bf r'}-{\bf r})\ f(r)\ Q_{12}\ , \\
\label{eq:6.2}\end{equation}
with the quadratic-spin-orbit operator
 $ Q_{12} = \left(
\boldsymbol{\sigma}_1\cdot{\bf L}\boldsymbol{\sigma}_2\cdot{\bf L} +
\boldsymbol{\sigma}_2\cdot{\bf L}\boldsymbol{\sigma}_1\cdot{\bf L}\right)/2$ .
Introducing the functions $g(r)$ and $h(r)$ by
\begin{equation}
 r_i f(r) = -\nabla_i g(r)\ \ ,\ \ 
 r_i r_j f(r) = \left[-\nabla_i\nabla_j+ \delta_{ij}\left(\frac{1}{r}\frac{d}{dr}
 \right)\right]\ h(r)\ , 
\label{eq:6.3}\end{equation}
executing the Fourier transformation in (\ref{eq:6.1}) leads to the identity
\begin{eqnarray}
 \widetilde{V}({\bf k},{\bf q}) &=&                         
\left[\boldsymbol{\sigma}_1\cdot{\bf q}\times{\bf k}\right]
\left[\boldsymbol{\sigma}_2\cdot{\bf q}\times{\bf k}\right]\ \widetilde{h}({\bf k}^2)
\nonumber\\ 
 && -\left[\vphantom{\frac{A}{A}} 
\boldsymbol{\sigma}_1\cdot{\bf q}\boldsymbol{\sigma}_2\cdot{\bf q} -
 {\bf q}^2 \boldsymbol{\sigma}_1\cdot\boldsymbol{\sigma}_2\right]\ \widetilde{g}({\bf k}^2)
\nonumber\\ 
 && +\frac{1}{4}\left[\vphantom{\frac{A}{A}} 
\boldsymbol{\sigma}_1\cdot{\bf k}\boldsymbol{\sigma}_2\cdot{\bf k} -
 {\bf k}^2 \boldsymbol{\sigma}_1\cdot\boldsymbol{\sigma}_2\right]\ \widetilde{g}({\bf k}^2)\ ,
\label{eq:6.4}\end{eqnarray}
where $\widetilde{h}({\bf k}^2)$ and $\widetilde{g}({\bf k}^2)$ are the 
Fourier transforms of respectively $h(r)$ and $g(r)$. 
 
The strategy is now to derive the configuration potentials with the 
 $\boldsymbol{\sigma}_1\cdot{\bf q}\boldsymbol{\sigma}_2\cdot{\bf q}$-operator
by utilizing (\ref{eq:6.4}), which we rewrite as 
\begin{eqnarray}
&& \left[\vphantom{\frac{A}{A}} 
\boldsymbol{\sigma}_1\cdot{\bf q}\boldsymbol{\sigma}_2\cdot{\bf q} -
 {\bf q}^2 \boldsymbol{\sigma}_1\cdot\boldsymbol{\sigma}_2\right]\ \widetilde{g}({\bf k}^2)
 = \nonumber\\ && \left\{
\left[\boldsymbol{\sigma}_1\cdot{\bf q}\times{\bf k}\right]
\left[\boldsymbol{\sigma}_2\cdot{\bf q}\times{\bf k}\right]\ \widetilde{h}({\bf k}^2)
 -\widetilde{V}({\bf k},{\bf q})\right\}                  
\nonumber\\ 
 && +\frac{1}{4}\left[\vphantom{\frac{A}{A}} 
\boldsymbol{\sigma}_1\cdot{\bf k}\boldsymbol{\sigma}_2\cdot{\bf k} -
 {\bf k}^2 \boldsymbol{\sigma}_1\cdot\boldsymbol{\sigma}_2\right]\ \widetilde{g}({\bf k}^2)\ ,
\label{eq:6.5}\end{eqnarray}

In our application
\begin{equation}
 \widetilde{g}({\bf k}^2) = \frac{\exp(-{\bf k}^2/\Lambda^2)}{{\bf k}^2+m^2}\ 
\ ,\ \ g(r) = \frac{m}{4\pi}\phi^0_C(r,m,\Lambda)\ .
\label{eq:6.6}\end{equation}
Then, from (\ref{eq:6.3}) one derives that
\begin{equation}
 f(r) = -\frac{1}{r}\frac{d}{dr} g(r) = -\frac{m}{4\pi} 
 \frac{1}{r}\frac{d}{dr}\phi^0_C(r,m,\Lambda) = 
 \frac{m^3}{4\pi} \phi^0_{SO}(r,m,\Lambda)\ .
\label{eq:6.7}\end{equation}
In momentum space, one easily derives the relation
 $d\widetilde{f}({\bf k}^2)/d{\bf k}^2 = 
 -\widetilde{g}({\bf k}^2)/2$ ,
which leads to 
\begin{equation}
\widetilde{f}({\bf k}^2) = \frac{1}{2} \exp\left(m^2/\Lambda^2\right)\
 E_1\left[({\bf k}^2+m^2)/\Lambda^2\right]\ ,
\label{eq:6.8b}\end{equation}
where $E_1(x)$ is the standard exponential integral function.\\

Next, we turn to the determination of $h(r)$. From (\ref{eq:6.3}) one readily
derives the momentum space differential equation
\begin{equation}
 \boldsymbol{\nabla}^2_k\ \widetilde{g}({\bf k}^2) = \left({\bf k}\cdot\boldsymbol{\nabla}_k
 + 3\right)\ \widetilde{h}({\bf k}^2)\ .
\label{eq:6.9}\end{equation}
Trying the form
\begin{equation}
 \widetilde{h}({\bf k}^2) = \left(A + \frac{B}{{\bf k}^2+m^2}\right)
 \widetilde{g}({\bf k}^2)\ ,
\label{eq:6.10}\end{equation}
one obtains from (\ref{eq:6.9}) the solution $A=-2/\Lambda^2$ and $B=-2$. So,
\begin{eqnarray}
&& \hspace*{-5mm} \widetilde{h}({\bf k}^2) = 
 -2\left(\frac{1}{\Lambda^2}+\frac{1}{{\bf k}^2+m^2}\right)
 \widetilde{g}({\bf k}^2) =              
 -2\left(\frac{1}{\Lambda^2}-\frac{d}{dm^2}\right)
 \widetilde{g}({\bf k}^2) 
 = 2 \frac{d\widetilde{g}({\bf k}^2)}{d{\bf k}^2}  
\label{eq:6.11}\end{eqnarray}
Using the (approximate) axial-current conservation, and the "Feynman gauge" in the
B-field formalism, we have from the $\Omega_i^{(A)}$ in (\ref{eq:axi1}) the following
expression for ${\cal V}_A^{(1)}$ 
\begin{eqnarray}
   \tilde{\cal V}_{A}^{(1)} &=& -g_{A}^{2}\left[
  \left(1-\frac{1}{MM'}{\bf k}^2
  +\frac{3({\bf q}^{2}+{\bf k}^2/4)}{2M'M}\right)
   \boldsymbol{\sigma}_{1}\cdot\boldsymbol{\sigma}_{2}
   + \frac{2}{M M'} \left(\vphantom{\frac{A}{A}}
  (\boldsymbol{\sigma}_{1}\cdot{\bf q})(\boldsymbol{\sigma}_{2}\cdot{\bf q})
 \vphantom{\frac{A}{A}} \right.\right.\nonumber\\ && \left.\left.\vphantom{\frac{A}{A}}
  -{\bf q}^2\boldsymbol{\sigma}_{1}\cdot\boldsymbol{\sigma}_{2}\right)
   - \frac{1}{4M' M}\left(
  (\boldsymbol{\sigma}_{1}\cdot{\bf k})(\boldsymbol{\sigma}_{2}\cdot{\bf k})   
  -\frac{1}{3}{\bf k}^2 \boldsymbol{\sigma}_{1}\cdot\boldsymbol{\sigma}_{2}\right)
 \right.\nonumber\\ && \left. 
   + \frac{i}{4M' M} (\boldsymbol{\sigma}_{1}+\boldsymbol{\sigma}_{2})
  \cdot{\bf q}\times{\bf k} \right]\cdot
  \widetilde{g}({\bf k}^2)\ ,
 \label{eq:6.12}\end{eqnarray}
Here, the superscript $(1)$ refers to the circumstance that this comes from the 
$g_{\mu\nu}$-term in the axial-vector-meson propagator.
Then, using the identity (\ref{eq:6.5}) we get from (\ref{eq:6.12})
\begin{eqnarray}
   \tilde{\cal V}_{A}^{(1)} &=& -g_{A}^{2}\left[
  \left(1-\frac{2{\bf k}^2}{3MM'}
  +\frac{3({\bf q}^{2}+{\bf k}^2/4)}{2M'M}\right)
   \boldsymbol{\sigma}_{1}\cdot\boldsymbol{\sigma}_{2}
 \right.\nonumber\\ && \left. \hspace*{5mm}
   + \frac{1}{4M' M}\left(
  (\boldsymbol{\sigma}_{1}\cdot{\bf k})(\boldsymbol{\sigma}_{2}\cdot{\bf k})   
  -\frac{1}{3}{\bf k}^2 \boldsymbol{\sigma}_{1}\cdot\boldsymbol{\sigma}_{2}\right)
   \right. \nonumber \\ & & \left. \hspace*{5mm}
   + \frac{i}{4M' M} (\boldsymbol{\sigma}_{1}+\boldsymbol{\sigma}_{2})
  \cdot{\bf q}\times{\bf k} \right]\cdot
  \widetilde{g}({\bf k}^2) \nonumber\\ && 
 -g_A^2\left[\frac{2}{MM'} \left\{
\left[\boldsymbol{\sigma}_1\cdot{\bf q}\times{\bf k}\right]
\left[\boldsymbol{\sigma}_2\cdot{\bf q}\times{\bf k}\right]\ \widetilde{h}({\bf k}^2)
 -\widetilde{V}({\bf k},{\bf q})\right\}\right]\ .        
 \label{eq:6.13}\end{eqnarray}
Making now our standard approximation of the Fourier transformation of the 
$\left[\boldsymbol{\sigma}_1\cdot{\bf q}\times{\bf k}\right]
\left[\boldsymbol{\sigma}_2\cdot{\bf q}\times{\bf k}\right]$-operator, 
cfrm. Ref.~\cite{NRS78}, the
configuration space potentials corresponding with (\ref{eq:6.13}) read
\begin{eqnarray}
   {\cal V}_{A}^{(1)} &=& - \frac{g_{A}^{2}}{4\pi}\ m  \left[
 \left(\phi^{0}_{C} + \frac{2m^{2}}{3M'M}\ \phi^{1}_{C}\right)
  (\boldsymbol{\sigma}_{1}\cdot\boldsymbol{\sigma}_{2})  -\frac{3}{4M'M}
  \left( \nabla^{2} \phi^{0}_{C}+\phi^{0}_{C}\nabla^{2}\right)
 (\boldsymbol{\sigma}_{1}\cdot\boldsymbol{\sigma}_{2})
 \right. \nonumber \\ & & \left. \hspace*{1.4cm}
   - \frac{m^{2}}{4M' M}\ \phi^{0}_{T}\ S_{12}
 +\frac{m^{2}}{2M'M}\ \phi^{0}_{SO}(m,r)\ {\bf L}\cdot{\bf S}
  \right] \nonumber\\ && 
  +\frac{g_{A}^{2}}{4\pi}\ \frac{2m^2}{M'M}\left[\phi_{SO}^0(r) 
  +\frac{3}{(mr)^2}\left\{ 3-\frac{2m^2}{\Lambda^2}  
  +\left(m\frac{d}{dm}\right)\right\}\ \phi_T^0(r)\right]\ Q_{12}\ .
 \label{eq:6.14}\end{eqnarray}

\noindent Now it happens that the second term in the coefficient of
$Q_{12}$ in (\ref{eq:6.14}) becomes 
by virtue of the properties of the Gaussian Yukawa-functions, 
see Appendix~\ref{app:A},
\begin{eqnarray}
&& \frac{3}{(mr)^2}\left\{\vphantom{\frac{A}{A}} \ldots \right\} =
 -\frac{3}{(mr)^2}\ \psi_T^0(r) = -\phi_{SO}^0(r)\ ,
 \label{eq:6.15}\end{eqnarray}
and so the coefficient of $Q_{12}$ in (\ref{eq:6.14}) vanishes!

 
\section{Axial-derivative Coupling and CAC} 
\label{app:Q}
In the B-field theory the conservation of the axial-current conservation (CAC) 
is an important ingredient. Therefore, an analysis of the realization of CAC 
in the ESC-model is opportune.
Isolating the derivative coupling terms in the axial-vector meson-exchange
potential we have
\begin{subequations}
\begin{eqnarray}
&& V_{A,a}(r) = -\frac{m}{4\pi} 
 \frac{m^2}{2M_YM_N}\left(g^A_{13}f^A_{24}\frac{M_N}{{\cal M}}
 +f^A_{13}g^A_{24}\frac{M_Y}{{\cal M}}\right)\left[ \frac{1}{3}
 (\mbox{\boldmath $\sigma$}_1\cdot\mbox{\boldmath $\sigma$}_2)\ \phi_C^1
 + S_{12}\ \phi_T^0\right] P, \\
&& V_{A,b}(r) = -\frac{m}{4\pi}\ 
 f^A_{13}f^A_{24}\frac{m^2}{{\cal M}^2}\frac{m^2}{4M_YM_N}\left[
 \frac{1}{3}
 (\mbox{\boldmath $\sigma$}_1\cdot\mbox{\boldmath $\sigma$}_2)\ \phi_C^2
 + S_{12}\ \phi_T^1\right] P.    
 \label{app:Q.1}\end{eqnarray}
\end{subequations}
Depending on the sign of $g_A f_A$ the first potential $V_{A,a}(r)$ is a
B-type ($g_A f_A>0$) or a P-type ($g_A f_A < 0$) potential, and the 
second potential $V_{A,b}(r)$ is a B-type potential.

\noindent Axial-vector current conservation at the meson-pole requires
\begin{equation}
 \partial_\mu J_A^\mu =0:\ \frac{f^A}{g^A} = -2 \frac{M_N{\cal M}}{m_A^2},
 \label{app:Q.5}\end{equation}
For NN the response of the axial potentials upon the change 
$f^A \rightarrow f_0^A + \Delta f^A$ from (\ref{app:Q.1}) is 
\begin{eqnarray}
\Delta V_{A}(r) &=& \Delta V_{A,a}(r)+\Delta V_{A,b}(r) = 
-\frac{m}{4\pi} \frac{m^2}{2M_N^2}\left[ 2\left(g_A+\frac{m^2}{2M_N^2} f_A\right)
 \Delta f^A \vphantom{\frac{A}{A}} \right.\nonumber\\ && \left. 
 + \frac{m^2}{2M_N^2} (\Delta f^A)^2\right]\cdot
 \left[ \frac{1}{3}(\mbox{\boldmath $\sigma$}_1\cdot\mbox{\boldmath $\sigma$}_2)\ \phi_C^1
 + S_{12}\ \phi_T^0\right].    
\label{app:Q.6}\end{eqnarray}
Now, it turns out that for ESC08c, with the parameters presented in this paper, the 
expression $\left[ \ldots \right] > 0$ for the axial mesons $a_1(1270), f_1(1420), f_1(1285)$. 
The coupling constant for the compensating B-meson potential is
\begin{equation}
 f_B^2(A) = \frac{3m^2}{2M_N^2}\left[ 2\left(g_A+\frac{m^2}{2M_N^2} f_A\right)
 \Delta f^A + \frac{m^2}{2M_N^2} (\Delta f^A)^2\right] 
\label{app:Q.7}\end{equation}
From the results for the couplings it appears that changes in the derivative 
couplings can be made in order to satisfy 
(\ref{app:Q.5}), which can be compensated by changing the B-meson couplings.

\section{Non-local tensor-correction}                                   
\label{app:NTC}
In this appendix we repeat the treatment of the non-local correction 
correction to the tensor-potential similar to that for the central
non-local potential
\begin{equation}
 \Delta \widetilde{V}_{T} = \left({\bf q}^2+\frac{1}{4}{\bf k}^2\right)
 \widetilde{v}_T\ S_{12}.
\label{app:NTC1} \end{equation} 
This incorporation of this kind of potential in the solution of the
Schr\"{o}dinger equation is given in \cite{Graz78}. For completeness we
repeat here the treatment of this type of potential, which is exact when
there is no non-local spin-orbit potential.
For definiteness we consider the contribution to the $\pi$-exchange potential
\begin{equation}
 \widetilde{v}_{T} = \frac{f_P^2}{2MM'\ m_\pi^2}
 \left({\bf q}^2+\frac{1}{4}{\bf k}^2\right)/({\bf k}^2+m^2). 
\label{app:NTC2} \end{equation} 
In configuration space this leads to the potential
\begin{eqnarray}
 V_T(r) &=& \frac{f_P^2}{4\pi} \frac{m}{4MM'}\left[
 \frac{1}{3}(\bm{\sigma}_1\cdot\bm{\sigma}_2)
 \left(\nabla^2 \phi^1_C+\phi_C^1 \nabla^2\right)
 +\left(\nabla^2 \phi_T^0 S_{12}+\phi_T^0 S_{12}\nabla^2\right)\right] 
 \nonumber\\ &\equiv & 
 -\left[\left(\nabla^2 \phi(r)+\phi(r) \nabla^2\right) +
  \left(\nabla^2 \chi(r) S_{12} +\chi(r) S_{12}  \nabla^2\right)\right].
\label{app:NTC3} \end{eqnarray} 
Here we put $\bm{\sigma}_1\cdot\bm{\sigma}_2=1$, 
because this potential contributes for spin-triplet states only.
The radial Schr\"{o}dinger equation reads
\begin{eqnarray}
&& \left\{\vphantom{\frac{A}{A}} (1+2\phi)+2\chi\ S_12\right\}\ u^{\prime\prime}
 +\left(\vphantom{\frac{A}{A}} 2\phi'+2\chi'\ S_{12}\right)\ u' 
+\left[k_{cm}^2-2M_{red} V
\right.\nonumber\\ && \left.\vphantom{\frac{A}{A}}
 - \left\{\vphantom{\frac{A}{A}}(1+2\phi) + \chi\ S_{12} \right\}
 \frac{{\bf L}^2}{r^2} 
 -\frac{{\bf L}^2}{r^2} \chi\ S_{12} + \phi^{\prime\prime} +
 \chi^{\prime\prime}\ S_{12}\right]\ u = 0.
\label{app:NTC4} \end{eqnarray} 
Under the substitution $u = A^{-1/2}v$, where
\begin{equation}
  A \equiv (1+2\phi) + 2\chi\ S_{12},
\label{app:NTC5} \end{equation} 
over into the radial equation for $v(r)$
\begin{equation}
 v^{\prime\prime}(r) + \left[ k_{cm}^2-\frac{l(l+1)}{r^2} - 2M_{red} W\right]\
 v(r) = 0 
\label{app:NTC6} \end{equation} 
with the (pseudo) potential
\begin{eqnarray}
 2M_{red} W &=& 2M_{red} A^{-1/2} V\ A^{-1/2} 
 - A^{-2}\left( \phi' + \chi'\ S_{12}\right)^2 
 -\left(A^{-1}-1\right)\ k_{cm}^2 \nonumber\\ &&
+\left\{\vphantom{\frac{A}{A}} A^{1/2}\left[L^2,A^{-1/2}\right]
 + A^{-1/2}\left[L^2,A^{1/2}\right]\right\}/(2r^2).
\label{app:NTC7} \end{eqnarray} 
In passing we note that A and $S_{12}$ commute, and therefore
\begin{eqnarray*}
 A^{-2}\left( \phi' + \chi'\ S_{12}\right)^2 = 
 \left[A^{-1/2}\left( \phi' + \chi'\ S_{12}\right) A^{-1/2}\right]^2=
 \frac{1}{4}\left[A^{-1/2}\ A'\ A^{-1/2}\right]^2.
\end{eqnarray*}

Defining 
\begin{equation}
 X = (1+ 2\phi + 4\chi)^{1/2}\ \ , \ \ 
 Y = (1+ 2\phi - 8\chi)^{1/2},       
\label{app:NTC8} \end{equation} 
the transformation A is given as
\begin{eqnarray}
 A^{1/2} &=& \frac{1}{3}\left(2 X+Y\right) + \frac{1}{6}\left(X-Y\right)\ S_{12}
 \nonumber\\
 A^{-1/2} &=& \left\{\frac{1}{3}\left(X+2 Y\right) + 
 \frac{1}{6}\left(-X+Y\right)\ S_{12}\right\}/(XY). 
\label{app:NTC9} \end{eqnarray} 
Using (\ref{app:NTC10}) one readily derives 
\begin{eqnarray}
&& \left\{ A^{1/2}\left[L^2,A^{-1/2}\right]_{-} + A^{-1/2}
\left[L^2, A^{1/2}\right]_{-}\right\} = \nonumber\\ &&
 -2 \frac{(X-Y)^2}{XY}\ \frac{\sqrt{J(J+1)}}{2J+1}\
\left(\begin{array}{cc}
 2\sqrt{J(J+1)} & -1 \\ -1 & -2\sqrt{J(J+1)}\end{array}\right). 
\label{app:NTC10} \end{eqnarray} 
Writing 
 $A^{-1} = \alpha + \beta\ S_{12}$ one finds  
\begin{eqnarray}
 \alpha &=& +\left(1+2\phi-4\chi\right) 
 \left[\vphantom{\frac{A}{A}} \left(1+2\phi+4\chi\right)\left(1+2\phi-8\chi\right)\right]^{-1}, 
\nonumber\\
 \beta &=& -2\chi\ 
 \left[\vphantom{\frac{A}{A}} \left(1+2\phi+4\chi\right)\left(1+2\phi-8\chi\right)\right]^{-1}, 
\label{app:NTC11} \end{eqnarray} 
leading to
\begin{eqnarray}
 -\left(A^{-1}-1\right) &=& 
\left[ \left\{\vphantom{\frac{A}{A}} (2\phi-8\chi)(1+2\phi+4\chi)-8\chi\right\}
 +2\chi\ S_{12}\right]\cdot\nonumber\\ && \times
 \left[(1+2\phi+4\chi)(1+2\phi-8\chi)\right]^{-1}.
\label{app:NTC12} \end{eqnarray} 

\section{Gaussian Yukawa-Functions}            
\label{app:A}
The basic Fourier transforms for the soft-core potentials is Refs.~\cite{NRS78,MRS89}
\begin{equation}
 \int\frac{d^3k}{(2\pi)^3} \frac{e^{i{\bf k}\cdot{\bf r}}}{{\bf k}^2+m^2}
 ({\bf k}^2)^n \exp\left(-{\bf k}^2/\Lambda^2\right) \equiv
 \frac{m}{4\pi} (-m^2)^n \phi_C^n(r) = (-\mbox{\boldmath $\nabla$}^2)^n
 \frac{m}{4\pi} \phi_C^0(r),
\label{app:A.1}\end{equation}
and similar ones for the tensor-, spin-orbit-, and the quadratic-spin-orbit
potentials.
The basic central, tensor, and spin-orbit functions are         
\begin{enumerate}
\item[(i)] central potentials:
\begin{subequations}
\begin{eqnarray}
\hspace*{-4mm} \phi_C^0(r) &=& \exp(m^2/\Lambda^2)\left[
 e^{-mr} {\cal E}rfc\left(-\frac{\Lambda r}{2}+\frac{m}{\Lambda}\right)
 -e^{ mr}{\cal E}rfc\left(\frac{\Lambda r}{2}+\frac{m}{\Lambda}\right)\right]/2mr\ ,
 \nonumber\\ \\
\hspace*{-4mm} \phi_C^1(r) &=& \phi_C^0(r)-\frac{1}{2\sqrt{\pi}}
 \left(\frac{\Lambda}{m}\right)^3 \exp\left[-\left(\frac{\Lambda r}{2}\right)^2\right],
 \\
\hspace*{-4mm} \phi_C^2(r) &=& \phi_C^1(r)+\frac{1}{2\sqrt{\pi}}
 \left(\frac{\Lambda}{m}\right)^5 \left[\frac{3}{2}-\left(\frac{\Lambda r}{2}\right)^2
\right] \exp\left[-\left(\frac{\Lambda r}{2}\right)^2\right],
\label{app:A.2}\end{eqnarray}
\end{subequations}
\item[(ii)] tensor potentials:
\begin{subequations}
\begin{eqnarray}
\hspace*{-4mm} \phi_T^0(r) &=& 
 \frac{1}{3}\frac{1}{m^2}r\frac{\partial}{\partial r}\frac{1}{r}
\frac{\partial}{\partial r} \phi_C^0(r) =
 \left\{\exp(m^2/\Lambda^2)\left[ \vphantom{\frac{A}{A}}[1+mr+(mr)^2/3]
 e^{-mr} 
 \cdot\right.\right. \nonumber\\ && \left.\left. \times
{\cal E}rfc\left(-\frac{\Lambda r}{2}+\frac{m}{\Lambda}\right)
 -[1-mr+(mr)^2/3]
 e^{ mr}{\cal E}rfc\left(\frac{\Lambda r}{2}+\frac{m}{\Lambda}\right)\right]
 \right.\nonumber\\ && \left.
 -\frac{4}{\sqrt{\pi}}\left(\frac{\Lambda r}{2}\right)\left[1+\frac{2}{3}
\left(\frac{\Lambda r}{2}\right)^2\right] \exp\left[-\left(\frac{\Lambda r}{2}
\right)^2\right]\right\}/2(mr)^3\ , \\
\hspace*{-4mm} \phi_T^1(r) &=& \phi_T^0-\frac{1}{6\sqrt{\pi}}   
\left(\frac{\Lambda}{m}\right)^5\left(\frac{\Lambda r}{2}\right)^2 
\exp\left[-\left(\frac{\Lambda r}{2}\right)^2\right].
\label{app:A.3}\end{eqnarray}
\end{subequations}
\item[(iii)] spin-orbit potentials:
\begin{subequations}
\begin{eqnarray}
\hspace*{-4mm} \phi_{SO}^0(r) &=& 
 -\frac{1}{m^2}\frac{1}{r}\frac{\partial}{\partial r} \phi_C^0(r) =
 \left\{\exp(m^2/\Lambda^2)\left[\vphantom{\frac{A}{A}} [1+mr]
 e^{-mr} 
 \cdot\right.\right. \nonumber\\ && \left.\left. \times
{\cal E}rfc\left(-\frac{\Lambda r}{2}+\frac{m}{\Lambda}\right)
 -[1-mr]
 e^{ mr}{\cal E}rfc\left(\frac{\Lambda r}{2}+\frac{m}{\Lambda}\right)\right]
 \right.\nonumber\\ && \left.
 -\frac{4}{\sqrt{\pi}}\left(\frac{\Lambda r}{2}\right)
\left(\frac{\Lambda r}{2}\right) \exp\left[-\left(\frac{\Lambda r}{2}
\right)^2\right]\right\}/2(mr)^3\ , \\
\hspace*{-4mm} \phi_{SO}^1(r) &=& \phi_{SO}^0-\frac{1}{4\sqrt{\pi}}   
\left(\frac{\Lambda}{m}\right)^5\left(\frac{\Lambda r}{2}\right)^2 
\exp\left[-\left(\frac{\Lambda r}{2}\right)^2\right].
\label{app:A.4}\end{eqnarray}
\end{subequations}
\item[(iv)] quadratic-spin-orbit potentials:
\begin{eqnarray}
\hspace*{-4mm} \phi_{Q}^0(r) &=& 
 -\frac{m^5}{4\pi} \frac{3}{(mr)^2} \phi_T^0(r).
\label{app:A.5}\end{eqnarray}
\end{enumerate}
The Fourier transforms of the Pomeron-type of potentials are gaussian-integrals,
which can be obtained from the above formulas by the substitutions
\begin{equation}
 \frac{1}{2}\Lambda \equiv m_P,\ \ m=0,\ \ \phi^{P,n}_i = \phi_i^{n+1}.
\label{app:A.6}\end{equation}
For explicit formulas see Refs.~\cite{NRS78,MRS89}.

\section{New Version Quark-Pair-Creation model \cite{THAR11} }
\label{app:D}
\begin{figure}[hbt]
 \begin{center} \begin{picture}(400,100)(0,0)
 \SetPFont{Helvetica}{9}
 \SetScale{1.0} \SetWidth{1.5}
 \ArrowLine(15,50)(75,50)   
 \ArrowLine(15,70)(75,70)   
 \ArrowLine(15,90)(75,90)   
 \Vertex(75,50){3}
 \Vertex(105,50){3}
 \Gluon(75,50)(105,50){2}{3}
 \Line(75,70)(105,70)   
 \Line(75,90)(105,90)   
 \ArrowLine(105,50)(165,50)  
 \ArrowLine(105,70)(165,70)   
 \ArrowLine(105,90)(165,90)   
 \ArrowLine(75,50)(165,00)   
 \ArrowLine(165,20)(105,50)  

 \Text( 5, 90)[]{$q_1$}
 \Text( 5, 70)[]{$q_2$}
 \Text( 5, 50)[]{$q_3$}
 \Text(180,90)[]{$q_1'$}
 \Text(180,70)[]{$q_2'$}
 \Text(180,50)[]{$q_4$}
 \Text(180,20)[]{$q_5$}
 \Text(180,00)[]{$q_3'$}
 \Text(85,00)[]{(a) direct}

 \SetOffset(200,0)
 \ArrowLine(15,50)(75,50)   
 \ArrowLine(15,70)(75,70)   
 \ArrowLine(15,90)(75,90)   
 \Vertex(75,70){3}
 \Vertex(105,50){3}
 \Gluon(75,70)(105,50){2}{3}
 \Line(75,70)(105,70)   
 \Line(75,90)(105,90)   
 \ArrowLine(105,50)(165,50)  
 \ArrowLine(105,70)(165,70)   
 \ArrowLine(105,90)(165,90)   

 \ArrowLine(75,50)(165,00)   
 \ArrowLine(165,20)(105,50)  

 \Text( 5, 90)[]{$q_1$}
 \Text( 5, 70)[]{$q_2$}
 \Text( 5, 50)[]{$q_3$}
 \Text(180,90)[]{$q_1'$}
 \Text(180,70)[]{$q_2'$}
 \Text(180,50)[]{$q_4$}
 \Text(180,20)[]{$q_5$}
 \Text(180,00)[]{$q_3'$}
 \Text(85,00)[]{(b) exchange}

 \end{picture} 
  \end{center}
  \caption{\sl $^3P_0$- and $^3S_1$-quark-pair-creation (QPC) }
  \label{fig.ts10} 
  \end{figure}
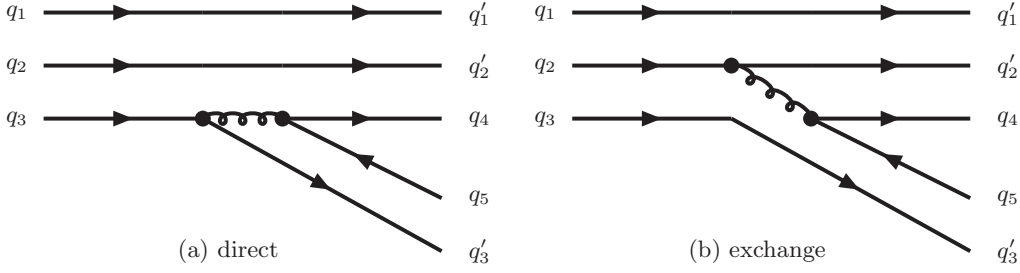

In this appendix we give a short description of the evaluation of the 
BBM coupling constants in the QPC-model using the Fierz-transformation 
technique. For details we refer to Ref.~\cite{THAR11}.
Here, apart from the Fierz-transformation, the techniques used are those 
of \cite{LeY73,LeY75,Chai80}.
In Fig.~\ref{fig.ts10} the two kind of processes, direct (a) and 
exchange (b), are shown.
The derivation of the BBM-couplings starts from the generalized $^3P_0$ (S) 
and $^3S_1$ (V) Pair-creation Hamiltonians          
\begin{eqnarray}
 {\cal H}_I^{(S)} &=& -4\gamma_{q\bar{q}}^{(S)} 
        \left(\sum_{i} \bar{q}_i q_i\right)\cdot
                     \left(\sum_{j} \bar{q}_j  q_j\right),   
\nonumber\\ 
 {\cal H}_I^{(V)} &=& -
 \gamma_{q\bar{q}}^{(V)} \left(\sum_{i} \bar{q}_{i,\alpha}(\mbox{\boldmath $\lambda$})^\alpha_{\ \beta}
 \gamma^\mu q_{i,\beta}\right)\otimes 
           \left(\sum_{j} \bar{q}_{j,\gamma}(\mbox{\boldmath $\lambda$})^\gamma_{\ \delta}
 \gamma_\mu q_{j,\delta}\right)
\label{eq:1.1}\end{eqnarray}
where $\gamma_{q\bar{q}}^{(V)}$ is a phenomenological constant, 
and the summations run as $i,j=u,d,s$.
In this QPC-model in the fundamental process there is a (confined) scalar or
gluon propagator. This implies, assuming a constant propagator, an extra factor
depending on a scalar or (massive) gluon exchange
  $(-i)^2.(\mp i/m_G^2) \sim \pm i/\Lambda_{QPC}^2$.
meaning $ \sim \pm i H_{int}$.\\
Rearrangement is supposed to take place when a quark-antiquark pair is created by some 
mechanism in a baryon, where one quark from the baryon combines into a mesonic 
state with the anti-quark from the pair. The quark from the pair recombines with the 
two remaining quarks of the baryon to make the baryon in the final state. This rearrangements
into mesons of different kind can be understood from a Fierz-transformation applied to 
(\ref{eq:1.1}). One has the identity \cite{Okun84}
\begin{eqnarray}
 && {\cal H}_I^{(S)} = \gamma_{q\bar{q}}^{(S)}\sum_{i,j} 
 \left[ \vphantom{\frac{A}{A}}
 \hspace*{0.0cm}  +\ \bar{q}_i\ q_j\cdot\bar{q}_i\ q_j              
 + \bar{q}_i\gamma_\mu q_j\cdot \bar{q}_j\gamma^\mu q_i 
 \right.\nonumber \\ && \left.
  -\frac{1}{2}\ \bar{q}_i\sigma_{\mu\nu} q_j\cdot\bar{q}_j\sigma^{\mu\nu} q_i 
  -\bar{q}_i\gamma_\mu\gamma_5 q_j\cdot\bar{q}_j\gamma^\mu\gamma^5 q_i 
  +\ \bar{q}_i\gamma_5 q_j\cdot 
     \bar{q}_j\gamma^5 q_i \vphantom{\frac{A}{A}} \right],
\nonumber\\
 && {\cal H}_I^{(V)} = +\gamma_{q\bar{q}}^{(V)}\sum_{i,j} \left[ \vphantom{\frac{A}{A}}
 \hspace*{0.0cm}  +\ \bar{q}_i\ q_j\cdot\bar{q}_i\ q_j              
 -\frac{1}{2}\ \bar{q}_i\gamma_\mu q_j\cdot \bar{q}_j\gamma^\mu q_i 
 \right.\nonumber \\ && \left.
  -\frac{1}{2}\ \bar{q}_i\gamma_\mu\gamma_5 q_j\cdot 
     \bar{q}_j\gamma^\mu\gamma^5 q_i 
  -\ \bar{q}_i\gamma_5 q_j\cdot 
     \bar{q}_j\gamma^5 q_i \vphantom{\frac{A}{A}} \right].
\label{eq:1.2}\end{eqnarray}
Here, we considered only the flavor-spin Fierzing.
\footnote{                
It should be noted that the terms for the couplings of the B-axial $J^{PC}=1^{+-}$-
and tensor $J^{PC}=2^{++}$mesons are missing on the r.h.s. of (\ref{eq:1.2}).
The same is true for the $^3P_0$-interaction (\ref{eq:1.1}).
}
The appropriate Fierzing 
of the color structure is different for diagram (a) and diagram (b) in 
Fig.~\ref{fig.ts10}:
(i)\ For diagram (a) we use the identity \cite{Okun84}
\begin{equation}
(\mbox{\boldmath $\lambda$})^\gamma_{\ \delta}\cdot
(\mbox{\boldmath $\lambda$})_\alpha^{\ \beta} =    
\frac{16}{9} \delta^\gamma_\alpha \delta^\beta_\delta
-\frac{1}{3} (\mbox{\boldmath $\lambda$})^\gamma_\alpha \cdot
 (\mbox{\boldmath $\lambda$})^\beta_\delta
\label{eq:1.3}\end{equation}
Since the mesons are colorless, the second term in (\ref{eq:1.3}) may be 
neglected, and color gives the simple factor $16/9$.\\
\noindent (ii)\ In diagram (b) there is in fact a sum over $q_1$ and $q_2$.
Because the baryons are colorless, we have
\begin{equation}
(\mbox{\boldmath $\lambda$}_1)_\alpha^{\ \beta} +    
(\mbox{\boldmath $\lambda$}_2)_\alpha^{\ \beta} =    
-(\mbox{\boldmath $\lambda$}_3)_\alpha^{\ \beta}.     
\label{eq:1.4}\end{equation}
Therefore, for this diagram we have, using (\ref{eq:1.3}), the identity
\begin{equation}
(\mbox{\boldmath $\lambda$}_5)^\gamma_{\ \delta}\cdot
\sum_{i=1,2}(\mbox{\boldmath $\lambda$}_i)_\alpha^{\ \beta} =    
-\frac{16}{9} \delta^\gamma_\alpha \delta^\beta_\delta
+\frac{1}{3} (\mbox{\boldmath $\lambda$}_5)^\gamma_\alpha \cdot
 (\mbox{\boldmath $\lambda$}_3)^\beta_\delta
\label{eq:1.5}\end{equation}
Again, for colorless mesons the second term in (\ref{eq:1.5}) may be 
neglected, and color gives the simple factor $-16/9$.\\
\noindent {\it We find that the direct (a) and exchange (b) diagram give different color
factors. Such a difference does not occur in the $^3P_0$-model. Now, it appears that
the momentum overlap for type (b) is usually much smaller than for type (a),
see \cite{THAR11}for details.
This can be traced back to our use of a constant propagator for the (confined) gluon.
Therefore, in the following we neglect processes described in diagram (b).
Then, the difference between the $^3P_0$- and $^3S_1$-model is, apart from an overall
constant, exclusively given by the different coefficients in the flavor-spin Fierz-identities
(\ref{eq:1.2}).}

In the $^3S_1$-model for the interaction Hamiltonian for the pair-creation 
one uses the one-gluon-exchange (OGE) model \cite{RGG75,HZ87}, see Fig.~\ref{fig.ts10}.
Considering one-gluon exchange, see Fig.~\ref{fig.ts10}, one derives the 
effective vertex \cite{RGG75,HZ87} by using a (confined) 
constant $P_g(ji)$ gluon propagator between quark line i and line j:
$P_g(ji) \sim \delta_{ji}/m_g^2$, where the (effective) gluon mass is taken to be 
$m_g \approx (0.8 fm^{-1}) \approx 250$ MeV \cite{HZ87}.
We notice that the color factor for the coupling of colorless mesons to colorless baryons
is always the same, and we can include this into an effective coupling $\gamma_S$, i.e.
\begin{equation}
 \frac{\pi \alpha_s (\mbox{\boldmath $\lambda$}_i\cdot\mbox{\boldmath $\lambda$}_j)}
 {m_G^2} \Rightarrow \gamma_{q\bar{q}}^{(V)}.
\label{eq:1.7}\end{equation}
Here we use for the gluon a constant (confined) propagator $P_g = 1/m_G^2$.
As is clear from (\ref{eq:1.1}) $\gamma_{q\bar{q}}$ has the dimension
[MeV]$^{-2}$. Also, we notice that $m_G \approx \Lambda_{QPC}$, therefore
$\gamma_{q\bar{q}} \longrightarrow \gamma_{q\bar{q}}/\Lambda_{QPC}^2$.
From the momentum conservation rules one now gets different dependences between the momenta
as compared to the version of the $^3P_0$-model in \cite{LeY73,Chai80}. 
Hence, we have different momentum overlap-integrals.

From the results for the couplings of the mesons in the $^3P_0$-model those for the
$^3S_1$--model meson-couplings can be read off by comparing the coefficients in
the Fierz-identities (\ref{eq:1.2}) and (\ref{eq:1.1}) for the corresponding operators.
Here, we assume that the effect of color in the $^3P_0$- and $^3S_1$-model 
can be absorbed into $\gamma_{q\bar{q}}^{(S,V)}$, see below.
For example, the prediction for the scalar-meson couplings will have the ratio 
$g_\epsilon(^3S_1) = \left[\gamma_{q\bar{q}}^{(V)}/\gamma_{q\bar{q}}^{(S)}\right] 
g_\epsilon(^3P_0)$. Apart from an overall constant, the couplings for the $^3S_1$-model
can be read off from those of the $^3P_0$-model.

 \subsection{Meson-states, Meson- and baryon wave-functions}
\label{app:Da}
We list the $\langle B,M| H_{int}| A\rangle$ matrix
elements for the different type of mesons. Restriction on the quark-level
to process (a) in Fig.~\ref{fig.ts10}, using the Fierzed form of the
interaction Hamiltonians in (\ref{eq:1.1}). So, below we will give the
results for the $^3P_0$-model.
Following \cite{RW67} we write the meson creation operators as
\begin{eqnarray}
 J^{PC}= 0^{-+}:\ \hspace{5mm}  
 d_{M,P}^\dagger({\bf k}) &=& i \sum_{r,s=\pm}\int d^3k_1 d^3k_2\ 
 \delta({\bf k}-{\bf k}_1-{\bf k}_2)\ 
 \cdot\nonumber\\ && \times 
 \widetilde{\psi}_M^{(L=0)}({\bf k}_1,{\bf k}_2)\
 \varphi^{(0)}(r,s)\ b^\dagger({\bf k}_1,r)\ d^\dagger({\bf k}_2,s), \\
 J^{PC}= 1^{--}:\ d_{M,V}^\dagger({\bf k},m) &=& \sum_{r,s=\pm}\int d^3k_1 d^3k_2\ 
 \delta({\bf k}-{\bf k}_1-{\bf k}_2)\ 
 \cdot\nonumber\\ && \times 
 \widetilde{\psi}_M^{(L=0)}({\bf k}_1,{\bf k}_2)\
 \varphi^{(1)}_m(r,s)\ b^\dagger({\bf k}_1,r)\ d^\dagger({\bf k}_2,s), \\
 J^{PC}= 0^{++}:\ d_{M,S}^\dagger({\bf k},m) &=& \sum_{r,s=\pm}\int d^3k_1 d^3k_2\ 
 \delta({\bf k}-{\bf k}_1-{\bf k}_2)\ (-)^m
 \cdot\nonumber\\ && \times 
 \widetilde{\psi}_{M,m}^{(L=1)}({\bf k}_1,{\bf k}_2)\
 \varphi^{(1)}_{-m}(r,s)\ b^\dagger({\bf k}_1,r)\ d^\dagger({\bf k}_2,s), \\
 J^{PC}= 1^{++}:\ d_{M,A}^\dagger({\bf k},m) &=& \sum_{r,s=\pm}\int d^3k_1 d^3k_2\ 
 \delta({\bf k}-{\bf k}_1-{\bf k}_2)\ 
 C(1,1,1;m_L,m_\sigma,m)\cdot\nonumber\\ && \times
 \widetilde{\psi}_{M,m_L}^{(L=1)}({\bf k}_1,{\bf k}_2)\
 \varphi^{(1)}_{m_\sigma}(r,s)\ 
 b^\dagger({\bf k}_1,r)\ d^\dagger({\bf k}_2,s), \\
 J^{PC}= 1^{+-}:\ d_{M,B}^\dagger({\bf k},m) &=& \sum_{r,s=\pm}\int d^3k_1 d^3k_2\ 
 \delta({\bf k}-{\bf k}_1-{\bf k}_2)\ 
 \cdot\nonumber\\ && \times 
 \widetilde{\psi}_{M,m}^{(L=1)}({\bf k}_1,{\bf k}_2)\
 \varphi^{(0)}(r,s)\ b^\dagger({\bf k}_1,r)\ d^\dagger({\bf k}_2,s), \\
 J^{PC}= 2^{++}:\ d_{M,T}^\dagger({\bf k},m) &=& \sum_{r,s=\pm}\int d^3k_1 d^3k_2\ 
 \delta({\bf k}-{\bf k}_1-{\bf k}_2)\ 
 C(1,1,2;m_L,m_\sigma,m)\cdot\nonumber\\ && \times
 \widetilde{\psi}_{M,m_L}^{(L=1)}({\bf k}_1,{\bf k}_2)\
 \varphi^{(1)}_{m_\sigma}(r,s)\ 
 b^\dagger({\bf k}_1,r)\ d^\dagger({\bf k}_2,s), 
\label{eq:2.1a}\end{eqnarray}
for respectively the pseudoscalar-, vector-, scalar-, axial-vector mesons 
of the first ($A_1$ etc.) and second kind  ($B_1$ etc.)\cite{tinv}, and
tensor mesons.
The baryon and meson wave , harmonic oscillator, functions are
\begin{eqnarray*}
\widetilde{\psi}_N({\bf k}_1,{\bf k}_2,{\bf k}_3) &=& 
\left(\frac{\sqrt{3} R_A^2}{\pi}\right)^{3/2} \exp\left[-\frac{R_A^2}{6}
 \sum_{i<j}({\bf k}_i-{\bf k}_j)^2\right]\ , \nonumber\\
\widetilde{\psi}_M^{(L=0)}({\bf k}_1,{\bf k}_2) &=& 
\left(\frac{R_M^2}{\pi}\right)^{3/4} \exp\left[-\frac{R_M^2}{8}
 ({\bf k}_1-{\bf k}_2)^2\right]\ , \nonumber\\
\widetilde{\psi}_{M,m}^{(L=1)}({\bf k}_1,{\bf k}_2) &=& 
 \frac{R_M}{\sqrt{2}}\left(\frac{R_M^2}{\pi}\right)^{3/4}
\left[- \mbox{\boldmath $\epsilon$}_{m}\cdot({\bf k}_1-{\bf k}_2) \right]\ . 
 \exp\left[-\frac{R_M^2}{8}
 ({\bf k}_1-{\bf k}_2)^2\right].  
\label{eq:2.1b}\end{eqnarray*}
Here we used the spherical unit vectors
$\mbox{\boldmath $\epsilon$}_{\pm 1} = 
\mp\frac{1}{\sqrt{2}}\left({\bf e}_1 \pm i {\bf e}_2\right)\ \ ,\ \ 
\mbox{\boldmath $\epsilon$}_{0} = {\bf e}_3$.  
 \subsection{Coupling-constant Formulas}
\label{app:Db}
The matrix elements 
$\langle B-f({\bf p}')\ M({\bf k})|{\cal H}_I^{(S),(V)}|B_i({\bf p})\rangle$ 
involve the momentum space overlap integrals, which can be performed in a 
straightforward manner \cite{THAR11}.
\noindent The summary of the derived formulas in \cite{THAR11}, in the case of the 
 $^3P_0$-model, for the divers (I=1)-couplings is:
\begin{eqnarray*}
  g_P &=& +\pi^{-3/4}\ \gamma_{q\bar{q}}\
 \frac{ (m_P R_P)^{1/2}}{(\Lambda_{QPC}R_P)^2}\cdot(6\sqrt{2})\ , \nonumber\\
  g_V &=& +\pi^{-3/4}\ \gamma_{q\bar{q}}\
 \frac{ (m_V R_V)^{1/2}}{(\Lambda_{QPC}R_V)^2}\cdot(3/\sqrt{2})\ , \nonumber\\
  g_S &=& +\pi^{-3/4}\ \gamma_{q\bar{q}}\
 \frac{ (m_S R_S)^{-1/2}}{(\Lambda_{QPC}R_S)^2}\cdot\frac{9 m_S}{M_B}\ , \nonumber\\
  g_A &=& -\pi^{-3/4}\ \gamma_{q\bar{q}}\
 \frac{ (m_A R_A)^{-1/2}}{(\Lambda_{QPC}R_A)^2}\cdot\frac{6 m_A}{M_B},   
\end{eqnarray*}
 with $\Lambda_{QPC} \approx 600$ MeV, and $R_M \approx 0.66$.




\begin{thebibliography}{99}
\bibitem{dercopax}      
Note that in this paper we suppose that $f_A$ does not contain the 
one-pion-pole etc. In momentum space $\widetilde{f}_A({\bf k}^2)$ is a 
smooth function of ${\bf k}^2$.
\end{thebibliography}
\end{document}